\documentclass[aps,prl,twocolumn,longbibliography,nofootinbib,nobibnotes]{revtex4-2}
\usepackage{amsmath,amsthm,amsfonts,amssymb,braket,algorithm}
\usepackage[noend]{algpseudocode}

\usepackage{standalone}
\usepackage{tikz}
\usetikzlibrary{shapes.geometric, arrows}
\usetikzlibrary{positioning}
\usepackage{setspace}

\usepackage{graphicx}
\usepackage{subfigure}
\usepackage[colorlinks=true,citecolor=blue,linkcolor=blue]{hyperref}

\newcommand{\secref}[1]{Section~\hyperref[#1]{{\ref*{#1}}}}
\newcommand{\figref}[1]{Fig.~\hyperref[#1]{{\ref*{#1}}}}
\newcommand{\eqnref}[1]{Eq.~\hyperref[#1]{{(\ref*{#1})}}}
\newtheorem{theorem}{Theorem}
\newtheorem{example}[theorem]{Example}
\newtheorem{remark}[theorem]{Remark}

\makeatletter
\renewcommand{\paragraph}{%
  \@startsection{paragraph}{4}%
  {\z@}{3.25ex \@plus 1ex \@minus .2ex}{-1em}%
  {\normalfont\normalsize\bfseries}%
}
\makeatother

\newcommand{\PL}{p_{\mathrm{L}}}
\newcommand{\Var}{\mathop{\rm Var}}
\newcommand{\GS}{\textrm{GS}}

\begin{document}

\title{Scalable surface code decoders with parallelization in time}
\author{Xinyu Tan}
\thanks{These two authors contributed equally.}

\author{Fang Zhang}
\thanks{These two authors contributed equally.}

\author{Rui Chao}

\author{Yaoyun Shi}

\author{Jianxin Chen}

\affiliation{Alibaba Quantum Laboratory, Alibaba Group USA, Bellevue, Washington 98004, USA}

\begin{abstract}
Fast classical processing is essential for most quantum fault-tolerance architectures.
We introduce a sliding-window decoding scheme that provides fast classical processing for the surface code through parallelism. 
Our scheme divides the syndromes in spacetime into overlapping windows along the time direction, which can be decoded in parallel with any inner decoder. 
With this parallelism, our scheme can solve the decoding throughput problem as the code scales up, even if the inner decoder is slow.
When using min-weight perfect matching and union-find as the inner decoders, we observe circuit-level thresholds of $0.68\%$ and $0.55\%$, respectively, which are almost identical to $0.70\%$ and $0.55\%$ for the batch decoding. 
\end{abstract}

\maketitle

Fault-tolerance theory allows scalable and universal quantum computation provided that the physical error rates are below a threshold.
Aharonov and Ben-Or in their seminal work~\cite{aharonov2008fault} give a fault-tolerance scheme with no classical operations. 
However, the resulting threshold is far from feasible.  
Multiple fault-tolerance architectures have since been proposed~\cite{knill2005quantum,aliferis2005quantum,gottesman2013fault}, but their estimations for the threshold and resource overhead mostly assume instantaneous classical computations.

One prominent architecture~\cite{dennis2002topological,fowler2012surface} uses the surface code~\cite{kitaev2003fault} and achieves universality via magic-state distillation~\cite{bravyi2005universal}. 
In particular, the implementation of a non-Clifford gate using a magic state typically involves a classically controlled Clifford correction.
Therefore, the error correction between consecutive non-Clifford gates should be fast enough to keep up with the rapidly decohering quantum hardware, so that the error syndromes do not backlog~\cite{terhal2015quantum}. 
This requires an adequately high decoding throughput---the amount of error syndromes that can be processed by a decoder in unit time.

Many decoding schemes for the surface code have high thresholds~\cite{raussendorf2007fault,fowler2009high}, yet they do not address the inadequate decoding throughput problem.
On the other hand, local decoding schemes~\cite{harrington2004analysis,duclos2010fast,bravyi2013quantum,fujii2014measurement,herold2015cellular,torlai2017neural,holmes2020nisq+,ueno2022qulatis} are fast and scalable to a certain degree, but their speed comes at the expense of accuracy.
The accuracy of local decoders can be improved by appending a global decoder~\cite{delfosse2020hierarchical,meinerz2022scalable,gicev2021scalable,chamberland2022techniques,smith2022local,ueno2022neo} while still pursuing relatively high decoding throughputs. 
Other schemes based on specialized hardware~\cite{fowler2012towards,fowler2013minimum,das2022afs,overwater2022neural} have also been proposed. 
However, to the best of our knowledge, none of the approaches mentioned above have demonstrated adequate accuracy, throughput, and scalability simultaneously.

In this work, we introduce the \emph{sandwich decoder} for the surface code, which solves the throughput problem using parallelism.
Our work is inspired by the idea of ``overlapping recovery'' in~\cite{dennis2002topological} (later rediscovered in~\cite{das2021lilliput}), which we reformulate as the \emph{forward decoder}. 
Both the sandwich and forward decoders are sliding-window decoders, \emph{i.e.}, they divide the error syndromes in spacetime into overlapping windows in the time direction.
However, the forward decoder needs to process these windows sequentially which results in a limited throughput. 
Meanwhile, our sandwich decoder removes the dependency between the windows so that they can be handled in parallel, \emph{e.g.}, using separate classical processing units~\cite{bartolucci2021fusion}. 
Adjacent sandwich windows may diagnose differently upon the same syndromes, which would compromise the fault-tolerance property of the decoding scheme. 
We reconcile such inconsistency by decoding the controversial syndromes in a further subroutine. 

The parallelism of the sandwich decoder is a great advantage for scalability. 
An inherently sequential algorithm like the forward decoder can hardly take advantage of parallel computational resources, and thus will have difficulty maintaining adequate throughput when the code distance increases. 
Meanwhile, the sandwich decoder can solve the throughput problem as the code scales up, as long as it is given enough parallel processing units. 
Little communication is needed between processing units as there is no dependency between windows. 
Thus the throughput requirement can be easily satisfied by adding more cores or processors, which is much easier than pushing the processor clock speed. 
Furthermore, the number of parallel processing units needed only scales with the speed of the quantum hardware and the code distance, not with the length of the quantum computation.

We benchmark the sandwich decoder with the memory experiment for the distance-$d$ rotated surface code~\cite{tomitasvore} under circuit-level noise. 
In particular, we decode each window using the min-weight perfect matching~\cite{dennis2002topological} or union-find decoder~\cite{delfosse2021almost}, and observe numerical thresholds of $0.68\%$ and $0.55\%$, respectively, for the logical error rate per $d$ cycles of syndrome extraction. 
These values are almost identical to the corresponding thresholds for the batch decoders. 
It is reasonable to expect similar preservation of accuracy when using other inner decoders. 
In consequence, our sandwich decoder may allow one to prioritize the accuracy of the inner decoder, as throughput is assured simply with adequate computational resources.  

\begin{figure*}[!t]
\centering
\begin{tabular}{ccc}
\subfigure[\label{fig:qubit_layout_mt}]{
\raisebox{.6cm}{\includegraphics[scale=.21]{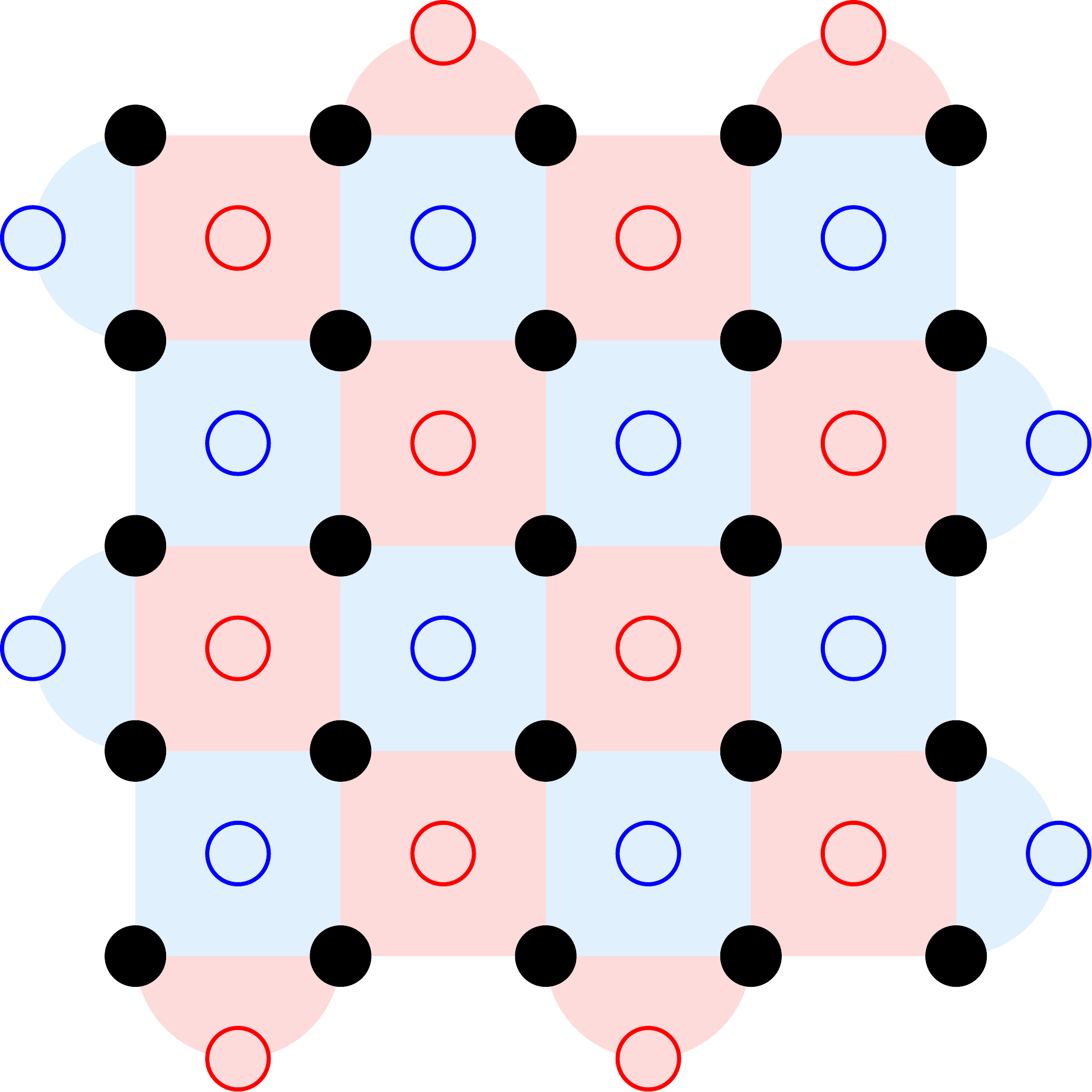}}}
&\quad
\subfigure[\label{fig:z_decoder_graph_mt}]{
\raisebox{.2cm}{\includegraphics[scale=.205]{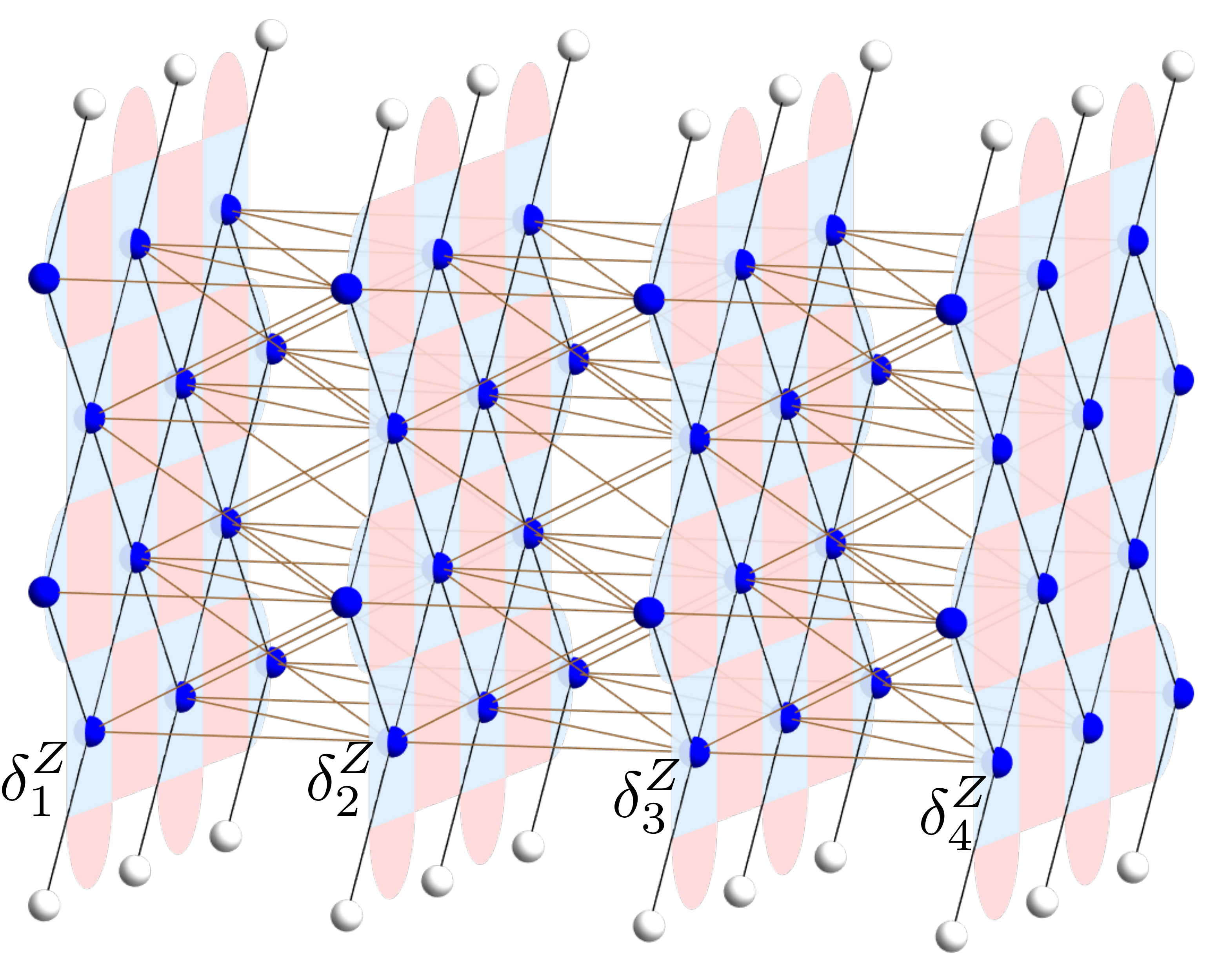}}}
&
\subfigure[\label{fig:x_decoder_graph_mt}]{
\raisebox{.2cm}{\includegraphics[scale=.2]{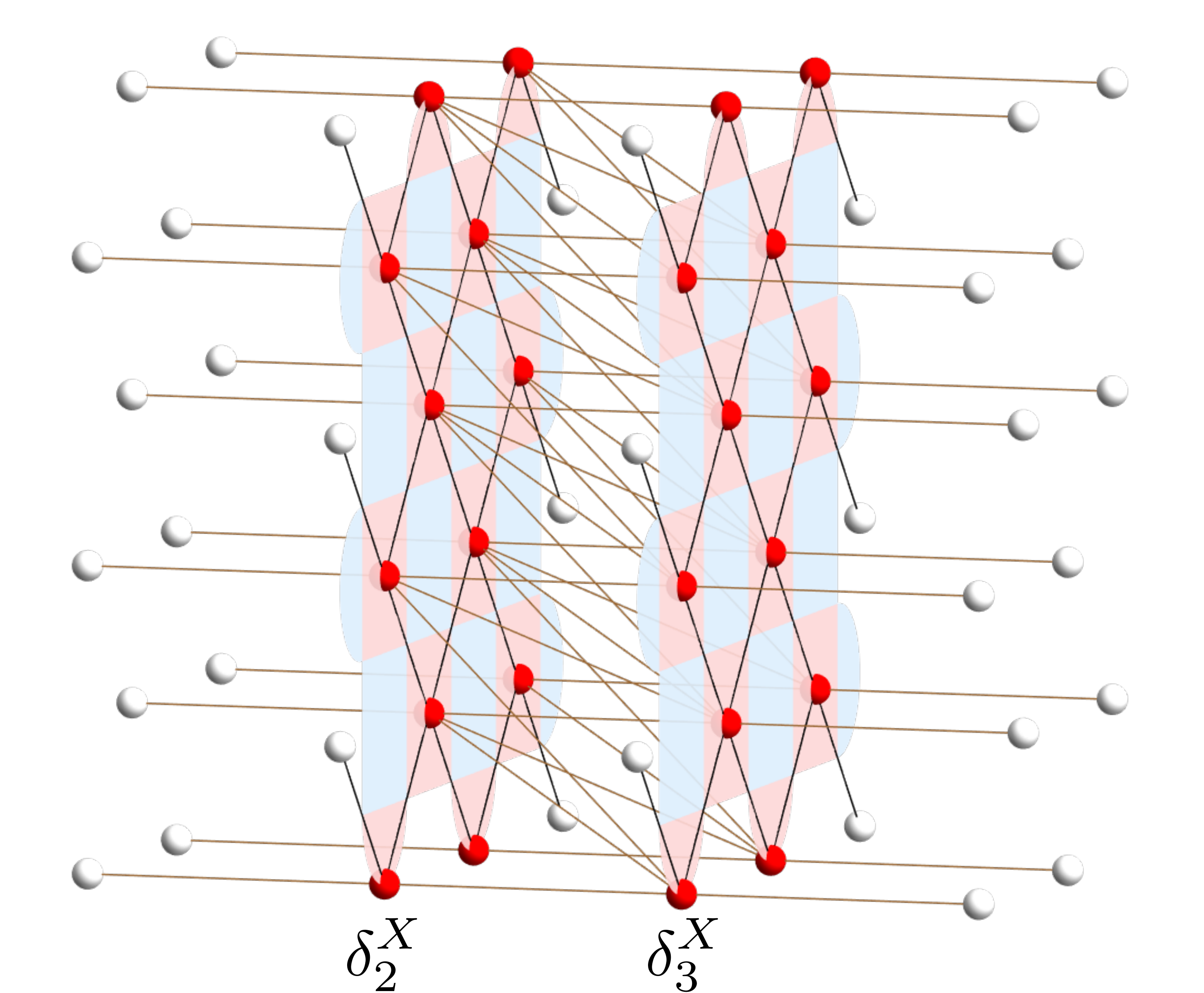}}}
\end{tabular}
\caption{
Decoder graphs of $Z$-type (b) and $X$-type (c) for a memory experiment with $3$ syndrome-extraction cycles that preserves $|\overline0\rangle$ of a distance-$5$ rotated surface code (a). 
Data qubits (black) reside on the plaquette corners.
Check operators of $Z$-type (blue) and $X$-type (red) are measured with ancilla qubits (empty circles) on the plaquette centers.
Blue and red vertices denote $Z$- and $X$-type real detectors, respectively; white vertices denote imaginary detectors.
Each edge represents the set of faults that flip the incident detectors.
Each decoder graph has two open space boundaries, whereas the $X$-type decoder graph also has two open time boundaries.
}\end{figure*}

\emph{Decoder graphs with boundaries.}---
We focus on the memory experiment that preserves the logical state~$|\overline0\rangle$ for the $[\![d^2,1,d]\!]$ rotated surface code; the argument for other variants of the surface code or logical basis states proceeds analogously. 
Specifically, we first initialize all the data qubits into the state~$|0\rangle$.
Then, we repeatedly apply a syndrome-extraction circuit for $n$ cycles and obtain syndromes $\sigma_i^X,\sigma_i^Z\in\{0,1\}^{\frac{d^2-1}{2}}$ of the $X$- and $Z$-type check operators, respectively, for $i=1,\cdots,n$. 
Finally, we measure all the data qubits in the $Z$ basis and obtain outcomes $\mu\in\{0,1\}^{d^2}$.

For the surface code, each cycle of \emph{detectors} is the XOR of two consecutive cycles of syndromes.
More precisely, let $\sigma^Z_{n+1}(\mu)\in\{0,1\}^{\frac{d^2-1}{2}}$ be the syndromes of the $Z$-type check operators evaluated from the data qubit measurement outcomes~$\mu$.
Define
\begin{align}
\begin{split}
&\delta_1^Z:=\sigma_1^Z, \\ 
&\delta_i^P:=\sigma_i^P\oplus\sigma_{i-1}^P,\;\; P\in\{X,Z\}, \; i=2,3,\cdots,n,\; \textrm{and} \\
&\delta_{n+1}^Z:=\sigma^Z_{n+1}(\mu)\oplus\sigma^Z_n.
\end{split}
\end{align}
We further assume that the syndrome-extraction circuit is fault-tolerant~\cite{tomitasvore,sm} and the whole circuit of the memory experiment is afflicted with stochastic Pauli noises. 
Specifically, each gate, qubit idling, and initialization (resp., measurement) is modeled as the ideal operation followed (resp., preceded) by a random Pauli, referred to as a \emph{fault}, supported on the involved qubit(s).

Under our assumptions about the circuit and noise model, detectors are $0$ in the absence of faults; thus, any \emph{defects}---detectors with value $1$---indicate the presence of faults.
Furthermore, the occurrence of each fault flips at most two detectors of each type ($X$ or $Z$)~\cite{tomitasvore}. 
We define a detector to be \emph{open} if there is a fault which flips that detector but no other detector of the same type; otherwise, it is \emph{closed}.
See~\cite{sm} for examples.

\begin{figure*}[!t]
\centering
\begin{tabular}{ccc}
\subfigure[\label{fig:window_graph_mt}]{
\raisebox{.1cm}{\includegraphics[scale=.15]{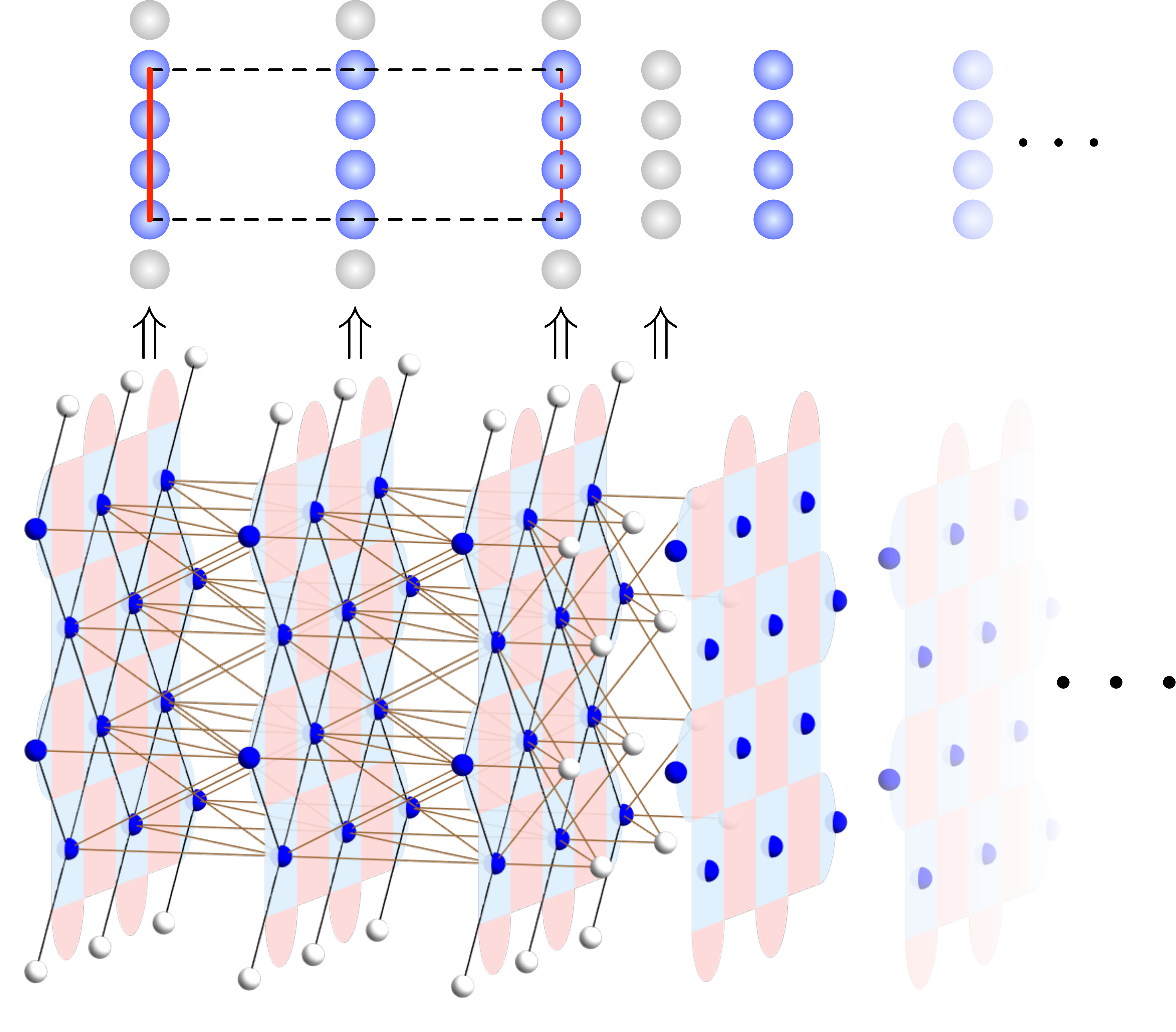}}}
&\quad
\subfigure[\label{fig:forward_window_mt}]{\raisebox{.3cm}{\includegraphics[scale=.33]{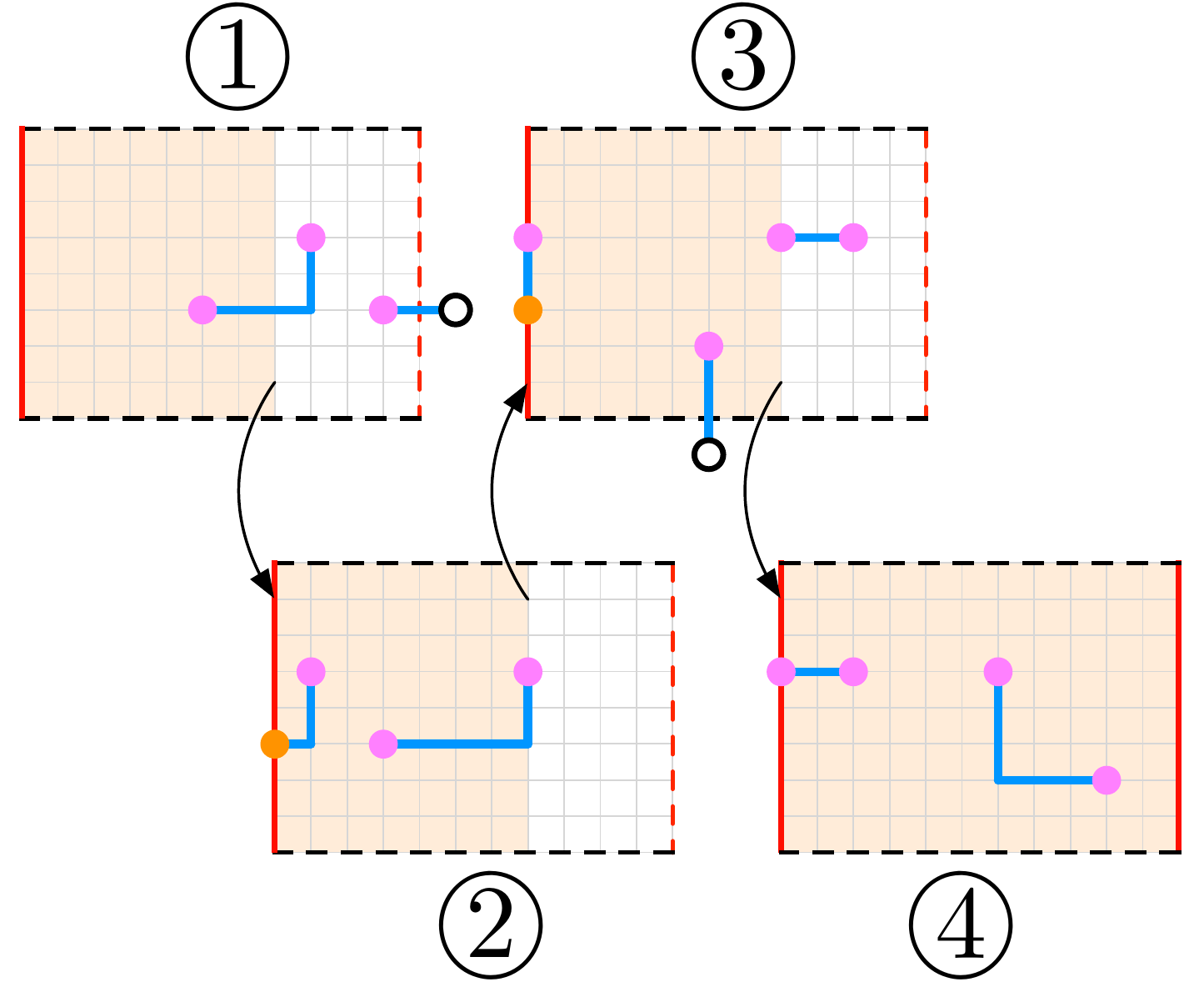}}}
&
\subfigure[\label{fig:sandwich_window_mt}]{
\raisebox{.3cm}{\includegraphics[scale=.33]{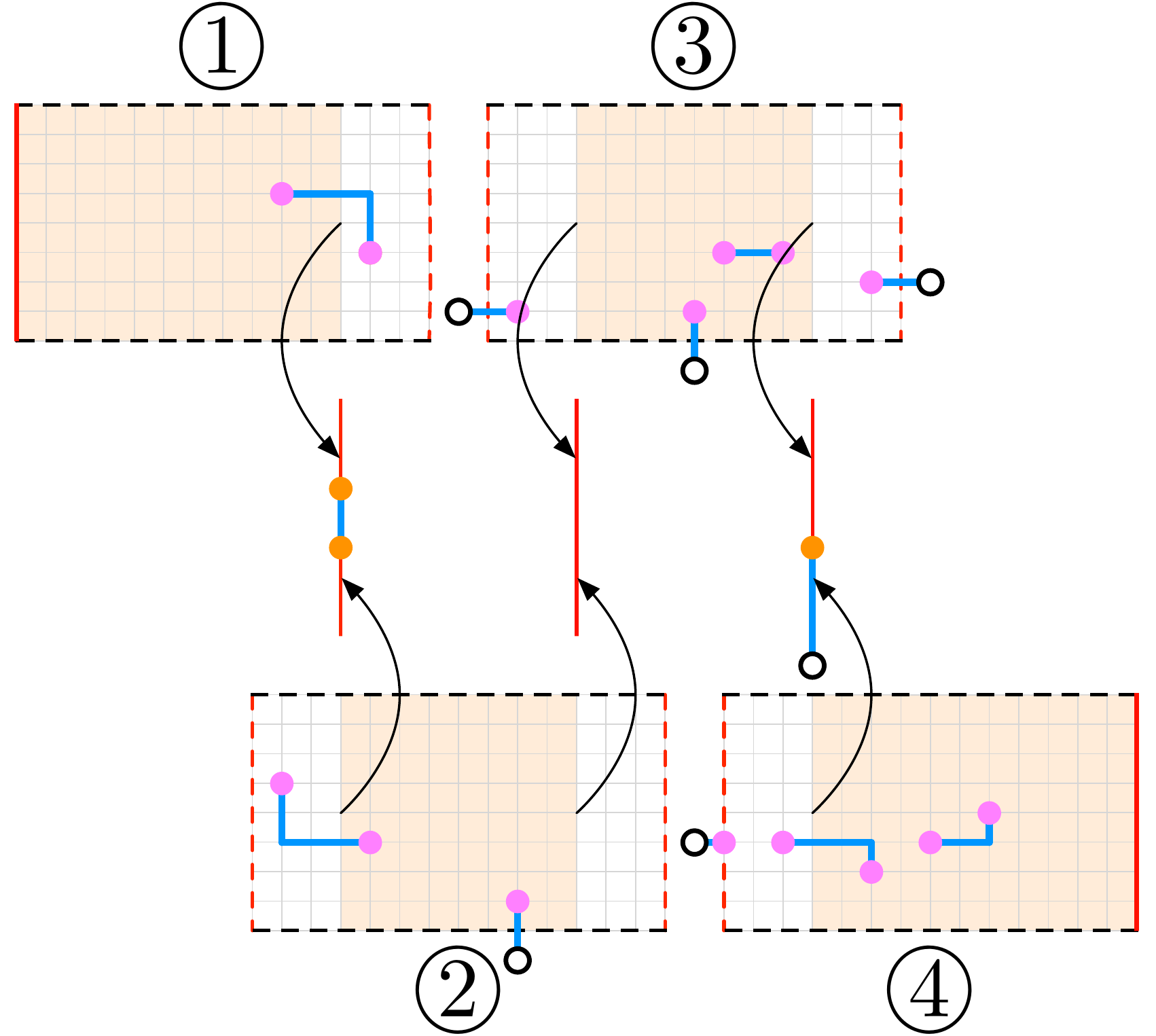}}}
\end{tabular}
\caption{
(a) Decoder graph of a first window of length $w=3$.
As in Fig.~\ref{fig:z_decoder_graph_mt}, blue and white vertices denote $Z$-type real and imaginary detectors, respectively.
Each edge represents the set of faults that flip the incident detectors.
The past time boundary (red line) is closed; the top and bottom space boundaries (black dashed line) and the future time boundary (red dashed line) are open.
Imaginary detectors near the future time boundary are equivalent to vertices in $\delta_4^Z$.
Forward decoder (b) and sandwich decoder (c).
Observed defects (pink dot) are annihilated by the corrections (blue line).
Corrections in the cores (brown region) are retained, whereas those in the buffers (transparent region) are discarded. 
Retained corrections can create updated defects (orange dot) on the seams between adjacent cores.
Imaginary detectors reside near the open boundaries (dashed line), and only those incident to corrections are shown (empty circle).
The forward decoder must decode the windows sequentially by the order labeled, whereas the sandwich decoder can in principle decode the windows in parallel.
}\end{figure*}

Figure~\ref{fig:z_decoder_graph_mt} illustrates a \emph{$Z$-type decoder graph} constructed as follows.
First, add one vertex for each $Z$-type detector. 
Then, add an edge between two vertices (detectors) if there is a fault that flips them both.
Finally, for each open detector, add an \emph{imaginary detector} and an edge connecting them.
We also assign each imaginary detector a binary value, such that each fault flips either zero or two $Z$-type detectors.
Each edge in the decoder graph thus represents an equivalence class of faults that flip the same two  detectors.

The notions ``open'' and ``closed'' can be extended to the boundaries of a decoder graph.
For example, an $X$ fault on any of the data qubits on the top and bottom boundaries of the lattice in Fig.~\ref{fig:qubit_layout_mt} flips only one $Z$-type real detector (\emph{i.e.}, a bit in $\delta^Z$) and by definition it is open. 
Hence, the top and bottom space boundaries of the 3D decoder graph in Fig.~\ref{fig:z_decoder_graph_mt} can be referred to as open.
Contrarily, there exists no fault that flips only one detector in $\delta_1^Z$ or $\delta_{n+1}^Z$ (unless on a space boundary).
Thus, the past (left) and future (right) time boundaries in Fig.~\ref{fig:z_decoder_graph_mt} are closed.

One can similarly construct an $X$-type decoder graph; see Fig.~\ref{fig:x_decoder_graph_mt}.
During the first (resp., last) cycle of syndrome extraction, a measurement fault on the ancilla of any $X$-type check operator flips only one detector in $\delta_2^X$ (resp., $\delta_n^X$).
Hence, both time boundaries are open. 
When a fault flips only one real detector on a intersection of space and time boundaries, it might help to distinguish if the fault afflicts, for instance, an ancilla measurement or a boundary data qubit.
Thus, in Fig.~\ref{fig:x_decoder_graph_mt} we connect such open real detectors to two imaginary detectors rather than one.

Given that we prepare the logical state $|\overline{0}\rangle$, we focus on the $Z$-type decoder graph as depicted in Fig.~\ref{fig:z_decoder_graph_mt}.

\emph{Sliding-window decoders.}---
A \emph{window} consists of a number of consecutive cycles of detectors in the decoder graph; see Fig.~\ref{fig:window_graph_mt}. 
Both the forward and sandwich decoders work on one window at a time. 
Within each window, an \emph{inner decoder} finds a set of edges from the decoder graph that can \emph{annihilate} all defects.
(Formally, a defect is annihilated if it is incident to an odd number of edges in the set.)
These edges are regarded as \emph{corrections}.
The corrections assembled from all windows should collectively annihilate every observed defect in the whole decoder graph.
Besides the inner decoders, the efficacy of our sliding-window decoders also relies on a crucial feature that any two consecutive windows overlap.
These overlaps allow each window to only retain a relatively trustworthy subset of the corrections for later assembly and discard the rest.

Hereafter, we associate each window with a subgraph of Fig.~\ref{fig:z_decoder_graph_mt} consisting of the real detectors enclosed in the window and certain imaginary detectors.
Concretely, the first window contains detectors from $\delta_1^Z$ to $\delta_w^Z$, the second window from $\delta_{1+s}^Z$ to $\delta_{w+s}^Z$, and so forth.
That is, windows, each with length $w$, proceed rightwards with step size $s<w$.
(The final window contains detectors up to $\delta_{n+1}^Z$.)
Analogous to the construction of Figs.~\ref{fig:z_decoder_graph_mt} and~\ref{fig:x_decoder_graph_mt}, here imaginary detectors are used such that each fault flips zero or two detectors in a window graph.
For instance, the top and bottom space boundaries of each window are open.
Importantly, in the initial window of both decoders, the past time boundary is closed; however, the vertices on the future time boundary are open and are connected with imaginary detectors to account for the faults that flip one detector in $\delta_w^Z$ and the other in $\delta_{w+1}^Z$.
See Fig.~\ref{fig:window_graph_mt}.
A key distinction between the two decoders is the time-boundary conditions of the remaining windows, which we now discuss separately. 

\emph{Forward decoder.}---
Figure~\ref{fig:forward_window_mt} illustrates the forward decoder. 
For each window before the final one that spans the detectors from $\delta_{1+is}^Z$ to $\delta_{w+is}^Z$, let its past boundary be closed and future boundary be open. 
Also, let the \emph{core} region denote the set of edges in the window that are incident to at least one vertex ranging from $\delta_{1+is}^Z$ to $\delta_{(i+1)s}^Z$ and the \emph{buffer} region denote the remaining edges in the window. 
For the final window, both time boundaries are closed and all edges belong to the core.
The core regions in two adjacent windows overlap in exactly the vertices on the past boundary of the later window.

The forward decoder processes the windows in temporal order.
Within each window starting from $\delta_{1+is}^Z$, the inner decoder first finds a set of corrections that can annihilate the existing defects but only retains the corrections in the core region. 
If the current window is the final one, all observed defects will have been annihilated and the decoder will terminate.
Otherwise, only the defects from $\delta_{1+is}^Z$ to $\delta_{(i+1)s}^Z$ will be annihilated, and the detectors on $\delta_{1+(i+1)s}$ will be updated and deferred to the next window.

Intuitively, the corrections found in the core become more reliable with larger buffer region, as the future faults outside the window are less likely to affect the core. 
A window needs no buffer preceding the core because all past defects have been reliably annihilated, rendering its past time boundary closed.
As far as we know, the idea of forward decoder first appears in~\cite{dennis2002topological} with $w=2s$, and is rediscovered in~\cite{das2021lilliput} with $s=1$ but without specifying the future-boundary conditions of windows. 
A limitation for the forward decoder is that windows must be processed sequentially due to the defect updates. 

\emph{Sandwich decoder.}---
Next, we introduce the sandwich decoder whose windows can be processed effectively in parallel; see Fig.~\ref{fig:sandwich_window_mt}.
Without loss of generality, assume $w+s$ is even and $s\ge2$.

Except for the past boundary of the initial window and the future boundary of the final window, all the time boundaries of windows are open.
For each window that spans the detectors from $\delta_{1+is}^Z$ to $\delta_{w+is}^Z$, the core region contains all edges incident to at least one vertex ranging from $\delta_{(w-s)/2+1+is}^Z$ ($\delta_1$ if initial window) to $\delta_{(w+s)/2-1+is}^Z$ ($\delta_{n+1}^Z$ if final window).
The core regions in two adjacent windows overlap in exactly one cycle of detectors, which we refer to as a \emph{seam}.

The sandwich decoder has two subroutines.
The first subroutine decodes all the windows. 
That is, each inner decoder finds corrections that can annihilate all defects within the window but only retains the ones in the core.
Observe that the retained corrections collectively annihilate all defects in the whole decoder graph except on $\delta^Z_{(w\pm s)/2+is}$---the seams between adjacent core regions.
The second subroutine decodes all the seams. 
Within each seam, it first updates the existing defects using the corrections retained in adjacent cores and then finds another set of corrections to annihilate the updated defects using the same heuristic as that of the inner decoder.
Note that the second subroutine can always find valid corrections for each seam since the top and bottom space boundaries of the two-dimensional decoder graph (\emph{i.e.}, a three-dimensional decoder graph with only one layer of detectors and both time boundaries closed) are open.
The output of the sandwich decoder consists of corrections from the cores in the first subroutine and the seams in the second subroutine.

Each window can be decoded independently, and each seam is ready to be decoded once the two adjacent windows have been decoded. 
Therefore, these windows and seams are naturally parallelizable.
The seams can in fact be generalized to three-dimensional windows; see~\cite{sm}.
We also discuss in~\cite{sm} other generalizations of the sandwich decoder, especially the potential to adapt it to the stability experiment~\cite{gidney2022stability} and lattice surgery~\cite{chamberland2022universal}.

\begin{figure}[!t]
\centering
\begin{tabular}{c}
\hspace{1cm}
\subfigure{\hspace{-1cm}
\raisebox{.0cm}{\includegraphics[scale=.45]{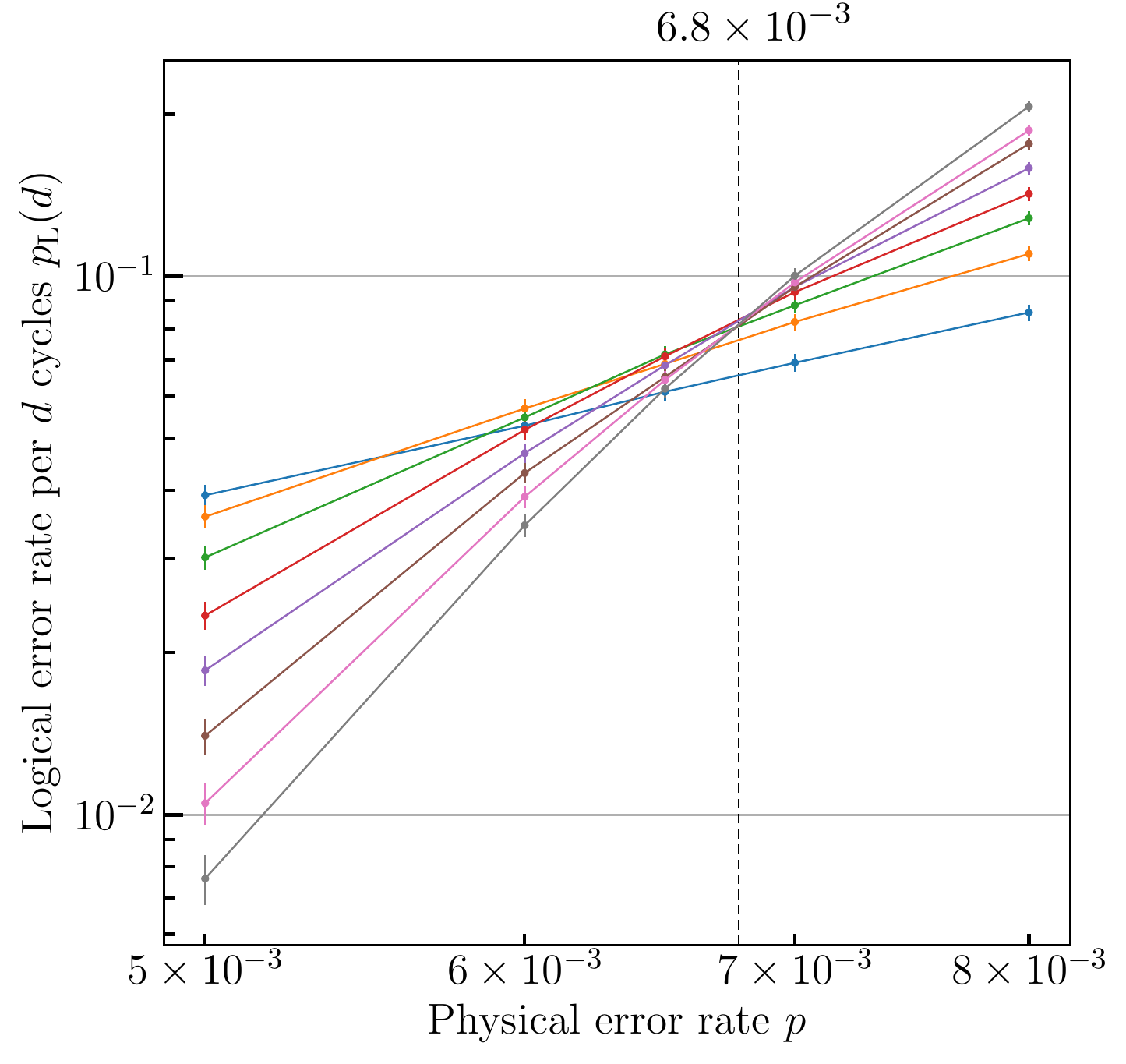}}}
\vspace{-.2cm}
\\
\hspace{1cm}
\subfigure{\hspace{-1cm}
\raisebox{.0cm}{\includegraphics[scale=.45]{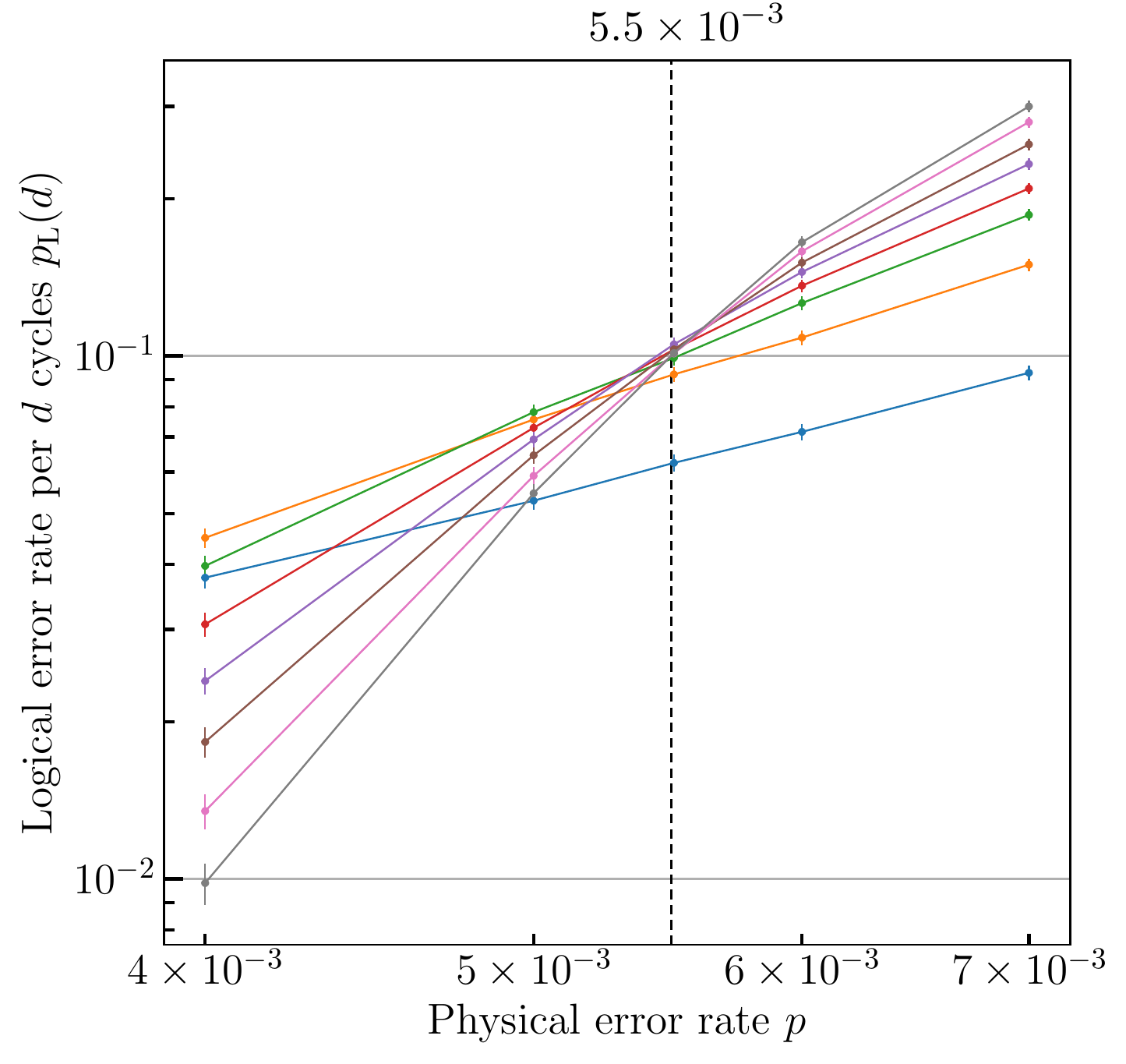}}}
\vspace{-.2cm}
\end{tabular}
\caption{
Threshold plots for the sandwich decoder with min-weight perfect matching (a) and union-find (b) decoders as the inner decoder. 
For fixed code distance $d=3,5,\cdots,17$ and physical error rate $p$, we first vary the number of cycles of syndrome extraction $n$ and simulate each memory experiment for $10^5$ shots. 
Then, we collect the estimated logical error rates per shot for varying $n$ and calculate the logical error rate per $d$ cycles $\PL(d)$, depicted as dot. 
Error bars indicate $95\%$ statistical confidence and dashed lines indicate the thresholds.
}\label{fig:thresholds_mt}
\end{figure}

\emph{Performance.}---
We benchmark our sandwich decoder for odd code distances $d=3,5,\cdots,17$ with step size $s=(d+1)/2$ and window size $w=3s$.
We employ the independent depolarizing noise model with rate $p$ for the entire circuit.

We take the ansatz that each cycle of syndrome extraction---after applying the corrections---flips the logical qubit independently with probability $\PL(1)$.
Assuming perfect data qubits initializations and final measurements, the probability of flipping the logical qubit after $i$ cycles of syndrome extraction $\PL(i)$ satisfies
\begin{align}
1-2\PL(i) = \left[1-2\PL(1)\right]^i.
\end{align}
We estimate the logical error rate per $d$ cycles $\PL(d)$ via Monte Carlo simulations and observe thresholds of physical error rates $p=0.68\%$ and $0.55\%$ for the min-weight perfect matching and union-find decoders, respectively as the inner decoder. 
See Fig.~\ref{fig:thresholds_mt}.
We also estimate $\PL(d)$ for batch decoding, \emph{i.e.}, decode all the syndromes of a memory experiment as a single window, and observe thresholds of $0.70\%$ and $0.55\%$ for the respective inner decoder. 

See~\cite{sm} for simulation details and numerical analysis such as varying step size and window size, comparison with the forward decoder, \emph{etc}. 

\emph{Discussion.}---
We have validated the sandwich decoder for the surface code with memory experiment. 
In Supplementary Material~\cite{sm}, we provide a parallel divide-and-conquer formalism for our sandwich decoder, which applies to general stabilizer codes and possibly general logical operations. 
For example, for lattice surgery one may need to divide syndromes into windows along the space direction, as well as the time direction. 
However, further theoretical justification or numerical validation is needed to fully evaluate our sandwich decoder in real-time decoding for fault-tolerant computation. 

\medskip

We would like to thank Xiaotong Ni for the insightful discussion and Hui-Hai Zhao for providing additional computational resource to speed-up the simulation. This work was supported by Alibaba Group through Alibaba Research Intern Program, and conducted when X.T. was a research intern at Alibaba Group USA.

\medskip

\emph{Note added.}---
An independent work~\cite{skoric2022parallel} was available to the public concurrently with ours. 


\let\oldaddcontentsline\addcontentsline
\renewcommand{\addcontentsline}[3]{}

\let\addcontentsline\oldaddcontentsline

\newpage
\onecolumngrid
\clearpage

\newpage
\null
\onecolumngrid

\begin{center}
\large \textbf{Supplementary Material for `Scalable surface code decoders with parallelization in time'}
\end{center}

\renewcommand{\thepage}{S\arabic{page}} 
\renewcommand{\thesection}{S\arabic{section}}  
\renewcommand{\thetable}{S\arabic{table}}  
\renewcommand{\theequation}{S\arabic{equation}}  
\renewcommand{\thefigure}{S\arabic{figure}}
\renewcommand{\thesubfigure}{(\alph{subfigure})}
\setcounter{figure}{0}
\setcounter{page}{1}
\setcounter{equation}{0}

\tableofcontents

\section{Decoder graphs with boundaries}

We focus on the memory experiment for the $[\![d^2,1,d]\!]$ rotated surface code~\cite{sm_tomitasvore} that preserves logical state $|\overline0\rangle$; the argument for other variants of the surface code or logical basis states proceeds analogously. 
Specifically, we first initialize all the data qubits into $|0\rangle$ states.
Then, we repeatedly apply a syndrome-extraction circuit for $n$ cycles and obtain syndromes $\sigma_i^X,\sigma_i^Z\in\{0,1\}^{\frac{d^2-1}{2}}$ of the $X$- and $Z$-type check operators, respectively, for $i=1,\ldots,n$. 
Finally, we measure all the data qubits in the $Z$ basis and obtain outcomes $\mu\in\{0,1\}^{d^2}$.

For the surface code, each cycle of \emph{detectors} is the XOR of two consecutive cycles of syndromes.
More precisely, 
\begin{align}
\begin{split}
\delta_1^Z&:=\sigma_1^Z, \\ 
\delta_i^P&:=\sigma_i^P\oplus\sigma_{i-1}^P,\;\; P\in\{X,Z\}, \; i=2,3,\ldots,n, \\
\delta_{n+1}^Z&:=\sigma^Z_{n+1}(\mu)\oplus\sigma^Z_n.
\label{eqn:detection_event}
\end{split}
\end{align}
Here, $\sigma^Z_{n+1}(\mu)\in\{0,1\}^{\frac{d^2-1}{2}}$ are syndromes of the $Z$-type check operators evaluated from the outcomes~$\mu$ of final measurements on the data qubits.
We further assume that the syndrome-extraction circuit is fault-tolerant (\emph{e.g.}, see~Fig.~S1
) and the whole circuit of the memory experiment is afflicted with stochastic Pauli noises. 
Specifically, each qubit preparation, idle qubit, gate, and qubit measurement is modeled as the ideal operation followed or preceded by a random Pauli fault supported on the involved qubit(s).

Under our assumptions about the circuit and noise model:
\begin{enumerate}
    \item All detectors are $0$ in the absence of faults. We regard the detectors with value $1$ as \emph{defects}. 
    \item The occurrence of each fault flips at most two detectors of each type ($X$ or $Z$). We define a detector to be \emph{open} if there is a fault that flips that detector but no other detector of the same type; otherwise, it is \emph{closed}.
\end{enumerate}

\begin{figure}[!ht]
\centering
\begin{tabular}{ccc}
\subfigure[\label{fig:layout_1_sm}]{
\raisebox{.2cm}{\includegraphics[scale=.3]{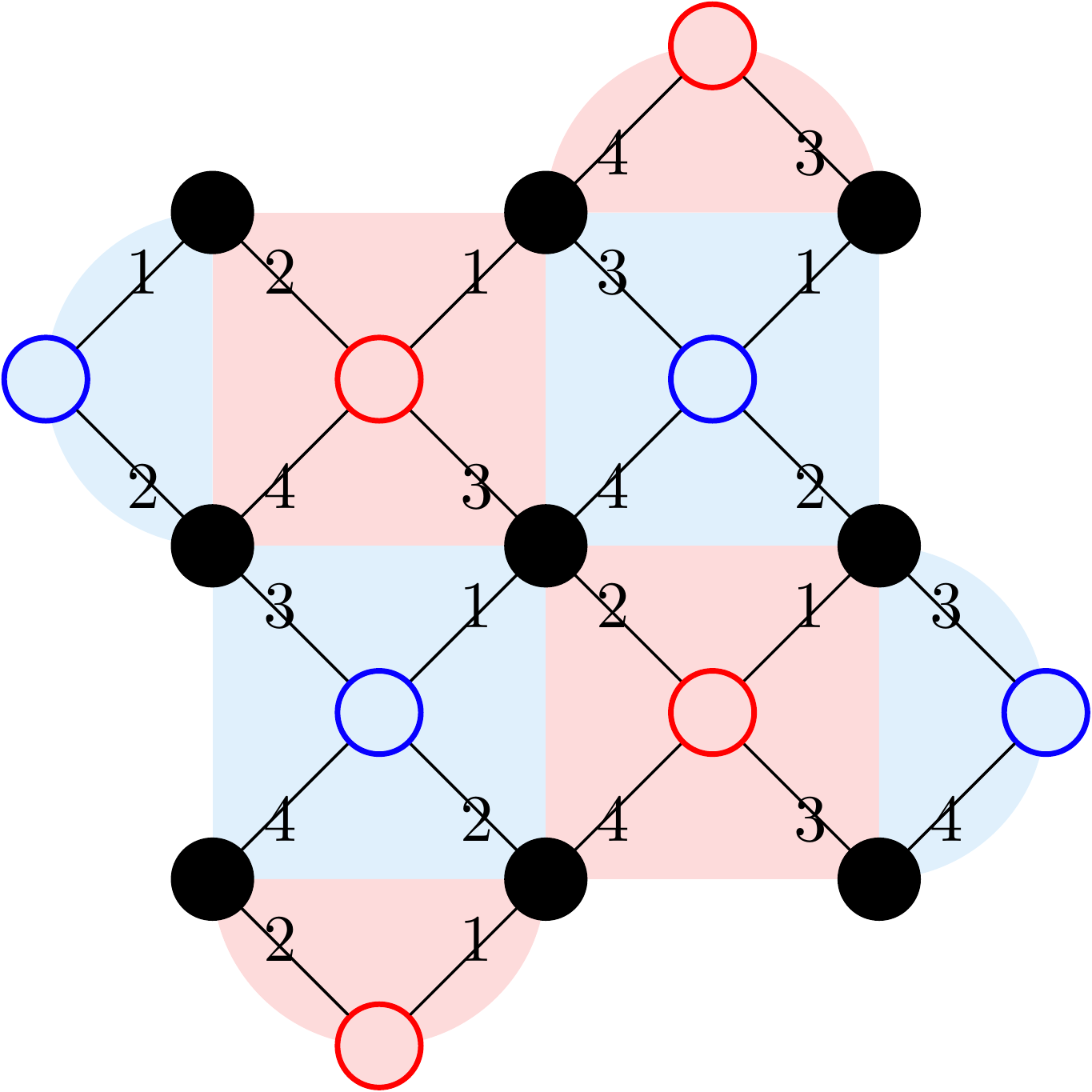}}}
&\qquad
\subfigure[\label{fig:blue_check_sm}]{
\raisebox{.4cm}{\includegraphics[scale=.2]{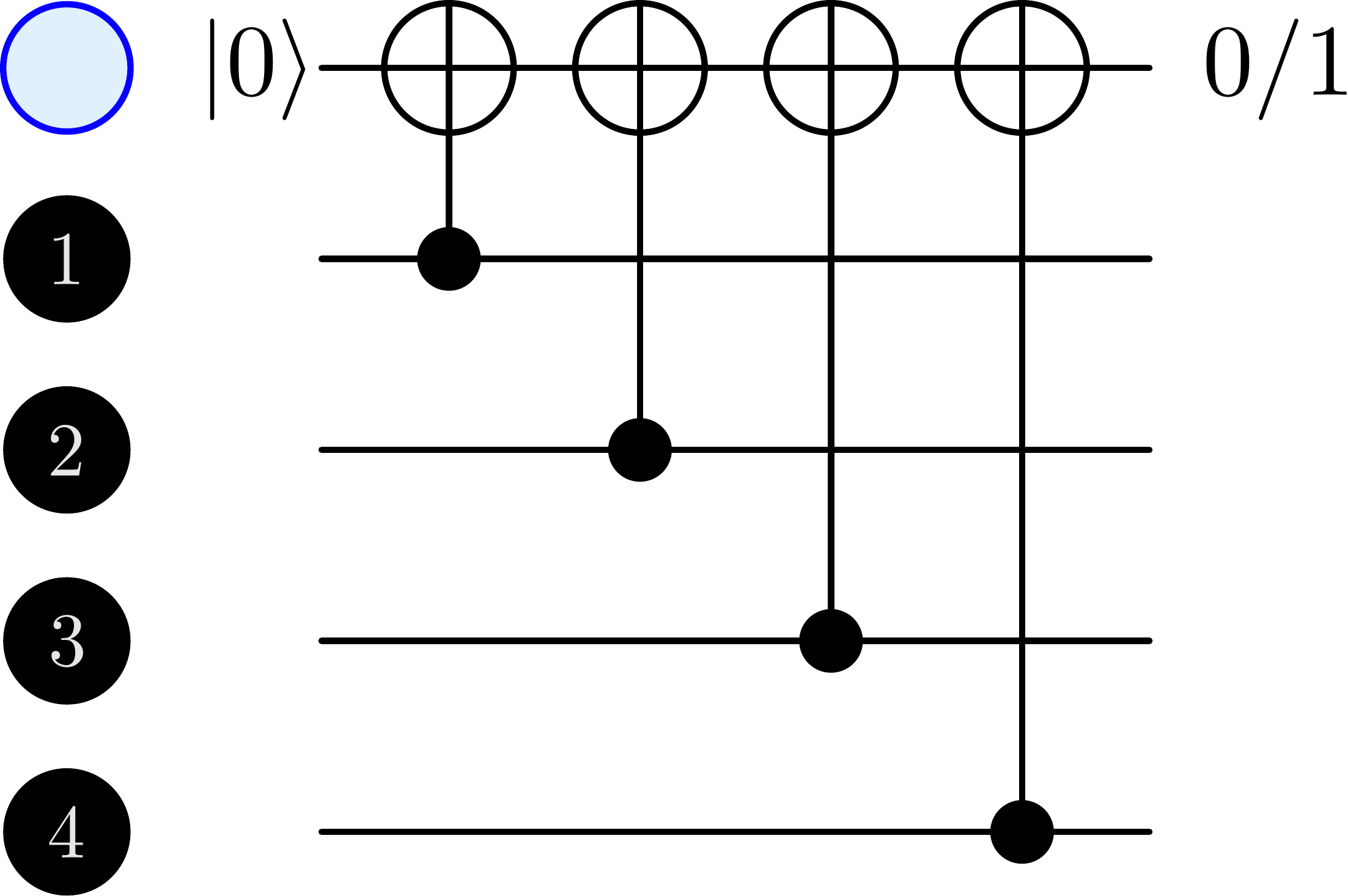}}}
&\qquad
\subfigure[\label{fig:red_check_sm}]{
\raisebox{.4cm}{\includegraphics[scale=.2]{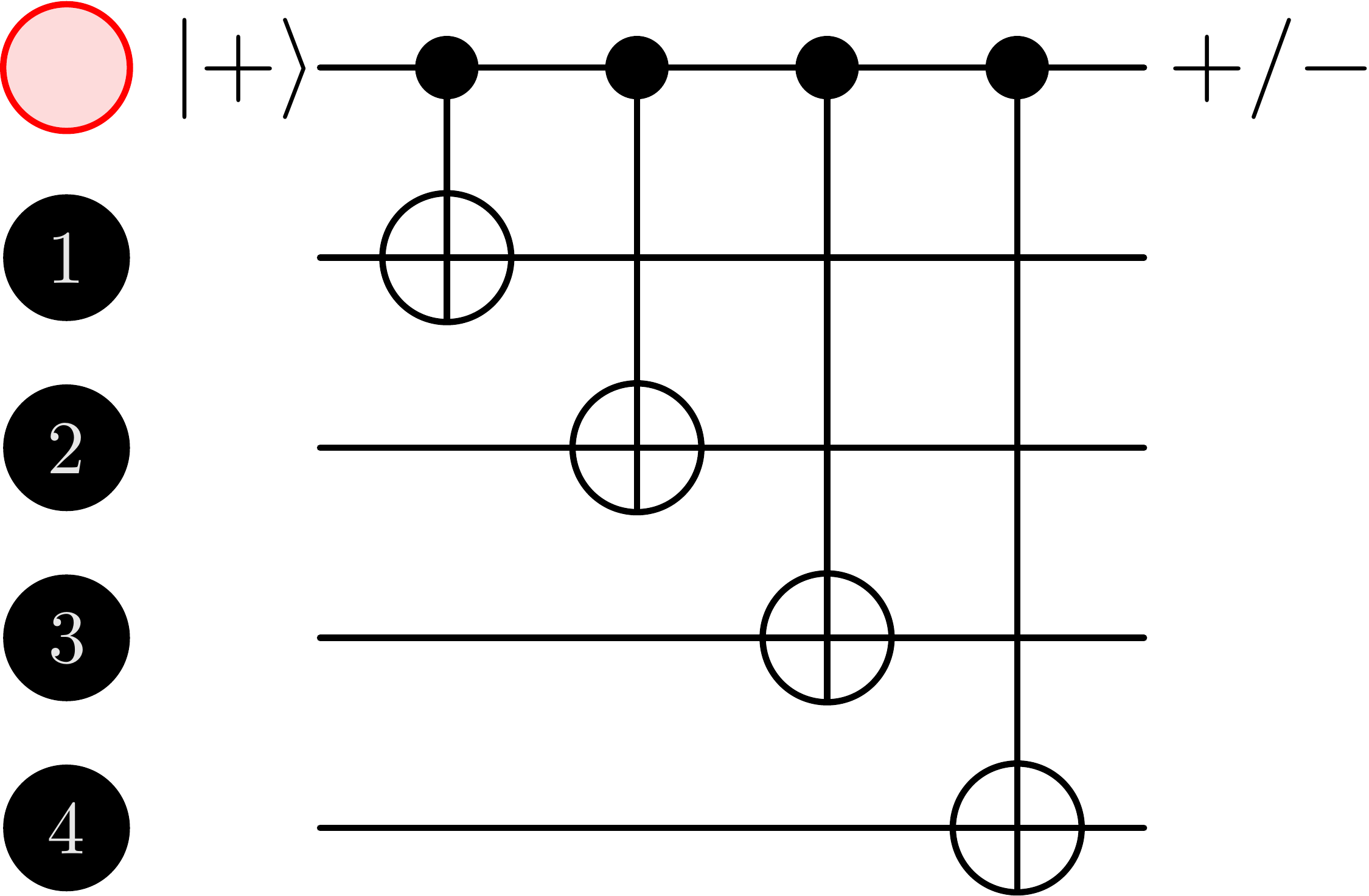}}}
\end{tabular}
\caption{
Rotated surface code $[\![d^2,1,d]\!]$ with $d=3$ (a).
Data qubits (black) reside on the plaquette corners.
Check operators of $Z$-type (blue) and $X$-type (red) are measured using the circuits in (b) and (c), respectively, with ancilla qubits (empty circles) on the plaquette centers.
First, prepare each ancilla in the $|0\rangle$ or $|+\rangle$ state; then, apply CNOT gates on qubit pairs connected by black links, in the order specified by the numbers on the plaquette corners; finally, measure each ancilla in the $Z$ or $X$ basis.
}\label{fig:circuits_sm}
\end{figure}

\begin{example}
In Fig.~S2(a), 
an $X$ fault on data qubit $A$ flips detectors $\alpha$ and $\beta$; whereas an $X$ fault on data qubit $B$ only flips detector $\gamma$. $\gamma$ is an open detector and $\alpha$ and $\beta$ are closed detectors. 
\end{example}

Figure~S2(b) illustrates a \emph{$Z$-type decoder graph} constructed as follows.
First, add one vertex for each $Z$-type detector. 
Then, add an edge between two vertices (detectors) if there is a fault that flips both. 
Finally, for each open detector, add an \emph{imaginary detector} and an edge connecting them. 
We also assign each imaginary detector a binary value, such that each fault flips either zero or two $Z$-type detectors.
Each edge in the decoder graph thus represents an equivalence class of faults which flip the same two detectors.

The goal of a decoding procedure is to \emph{annihilate} all defects by finding a proper set of \emph{corrections}---edges which give rise to the exact same defects. Formally, a defect is annihilated if it is incident to an odd number of edges in the set.

\begin{figure}[!ht]
\centering
\begin{tabular}{ccc}
\subfigure[\label{fig:layout_sm}]{
\raisebox{.4cm}{\includegraphics[scale=.24]{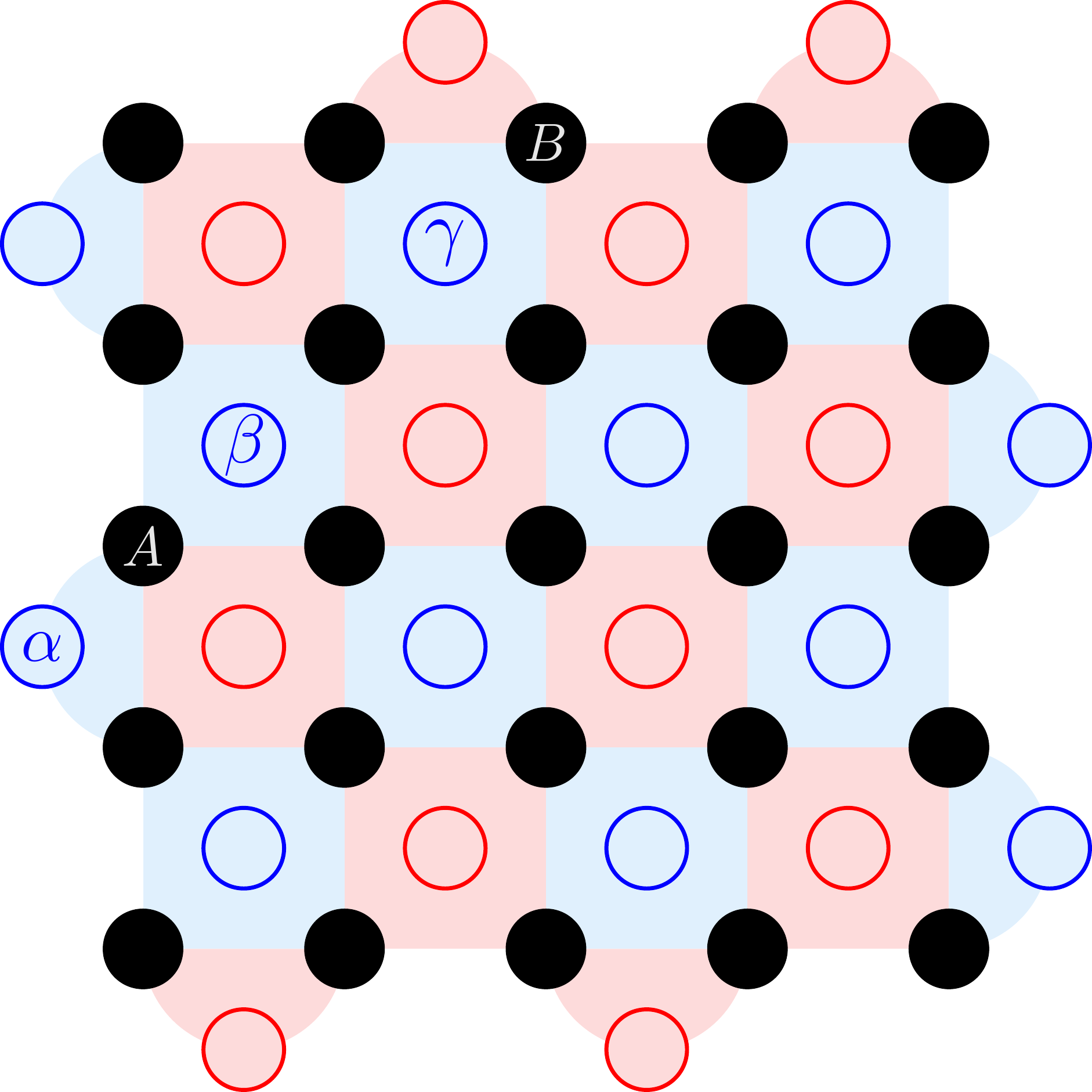}}}
&\quad
\subfigure[\label{fig:blue_graph_sm}]{
\raisebox{.2cm}{\includegraphics[scale=.19]{fig/z_decoder_graph}}}
&
\subfigure[\label{fig:red_graph_sm}]{
\raisebox{.2cm}{\includegraphics[scale=.18]{fig/x_decoder_graph}}}
\end{tabular}
\caption{
Decoder graphs of $Z$-type (b) and $X$-type (c) for a memory experiment with $3$ syndrome-extraction cycles that preserves $|\overline0\rangle$ of a distance-$5$ rotated surface code (a). 
Blue and red vertices denote $Z$- and $X$-type real detectors, respectively; white vertices denote imaginary detectors. 
Vertices which represent detectors from the same cycle constitute a \emph{layer}. These layers are arranged in temporal order from left to right. 
Each edge represents the set of faults that flip the incident detectors.
Each decoder graph has two open space boundaries. In addition, the $X$-type decoder graph has two open time boundaries.
}\label{fig:graphs_sm}
\end{figure}

\subsection{Open and closed boundary conditions}

In our memory experiment, each three-dimensional decoder graph has six boundaries: four \emph{space boundaries} and two \emph{time boundaries}. 
Each space boundary consists of all the real detectors adjacent to the top-, bottom-, left-, or right-most (in the directions of Fig.~S2(a)
)\footnote{Hereafter whenever we say ``top,'' ``bottom,'' ``left,'' or ``right'' in the context of space boundaries, we refer to the directions in Fig.~S2(a)
.} data qubits of each layer respectively. 
The first and last layers of real detectors are the time boundaries, representing the detectors at the time of data qubit initialization and final data qubit measurements. 

A boundary in a decoder graph is called open if every detector on this boundary is open. A boundary is closed if it is not open.

\begin{example}To understand open and closed space boundaries, let us consider the $Z$-type decoder graph with the layout specified in Fig.~S2(b). 
Every $X$ error on the data qubits of the top row flips only one real detector on the top space boundary. By our definition, it is not hard to see that each detector on the top space boundary is open and is thus connected to an imaginary detector. 
Therefore, the top space boundary is an open boundary. Similarly, the bottom space boundary is also open. Meanwhile, since every non-corner detector on the left and right space boundaries are closed, the left and right space boundaries are closed. 
\end{example}

\begin{remark}
One way to intuitively justify the words ``open'' and ``closed'' is by looking at the forms of undetectable errors. For codes without space boundaries (such as the toric code), an undetectable error always looks like a cycle or a combination of cycles, either topologically trivial (in which case it will never cause a logical error) or not (in which case it may be a logical operator). For codes with space boundaries, an undetectable error can also be a path with both ends at the open boundaries, as if the path goes into and out of the code patch through those boundaries.
\end{remark}

To understand open and closed time boundaries, let us continue with the quantum memory experiment for a logical qubit $|\overline0\rangle$. For the $Z$-type decoder graph, both time boundaries are closed since every fault flips exactly two $Z$ detectors (including the imaginary detectors on the open space boundaries). 
\begin{example}
Suppose there is only a $Z$ stabilizer measurement error during the last cycle of syndrome extraction, \emph{i.e.}, $\sigma_i^Z = 0$ for $i\in \{1, \cdots, n+1\} \setminus \{n\}$ and $\sigma^Z_n = 0\cdots 010\cdots 0$. It follows from \eqnref{eqn:detection_event} that there are only two defects in $\delta_n^Z$ and $\delta_{n+1}^Z$ respectively. The edge connecting these two defects indicate the $Z$ stabilizer measurement error.
\end{example}
\begin{example}
Suppose there is only a data qubit measurement error at the end of the memory experiment. We have $\sigma_i^Z = 0$ for $i\in \{1, \cdots, n\}$. As we calculate the final set of $Z$ syndromes $\sigma_{n+1}^Z$ based on the data qubit measurement results $\mu$, one flipped data qubit affects all the check operators that it involves. Therefore, $\sigma_{n+1}^Z$ has $1$ or $2$ non-trivial syndromes and $\delta_{n+1}^Z$ has $1$ or $2$ defects. With the imaginary detectors on the open space boundaries, any single data qubit measurement fault flips exactly two detectors.
\end{example}
However, for the $X$-type decoder graph, the outcome of the first cycle of $X$ stabilizer extraction is a random binary string even if there are no errors. Therefore, we need to make the initial time boundary open (\emph{i.e.}, allowing the bottom detection events to connect to some virtual vertices) in order to explain those non-trivial syndrome measurement results. Similarly, the ending time boundary also needs to be open. Since in the end after we measure the data qubits in the $Z$ basis, the outcome $\mu$ does not provide any information about the $X$ syndromes.

We emphasize that the open and closed boundary conditions are not just some mathematical tricks that marginally improve the performance of the decoder; on the contrary, to get meaningful results from the quantum memory experiment, one must correctly close or open the boundaries according to the context (\emph{e.g.}, the code or the specific type of errors). 

As we focus on the memory experiment of preserving logical $|\overline0\rangle$, our goal is to prevent the logical $Z$ operator from being flipped (an odd number of times). 
But if one of the time boundaries in the $Z$-type decoder is open, there will be low-weight (\emph{i.e.}\ short) undetectable $X$ errors with both endpoints on that time boundary. Such an undetectable $X$ error can easily flip the logical $Z$ operator, violating the principle that only at least $\lceil d/2\rceil$ physical errors can cause a failure. 
On the other hand, when both time boundaries are closed, everything makes sense: The only open boundaries are the top and bottom space boundaries, and low-weight $X$ errors starting and ending at one of those boundaries can only flip the logical $Z$ operator an even number of times. To flip the logical $Z$ operator, an error must cross from one open space boundary to the other open space boundary, but then it is a logical operator with weight $\ge d$, and all is well.

We later provide another way to understand the open time boundary condition in \secref{subsec:preliminaries} when we introduce ``sliding windows'' and show with numerical evidence in \secref{subsec:open_vs_closed} that misusing closed and open time boundary conditions greatly increase the logical error rate. 

\section{Sliding-window decoders}
\subsection{Motivations}
Among most of the existing implementations and simulations of quantum memory experiments, decoding is usually an offline process. This means that the decoding will not start until the entire quantum circuit finishes execution and produces a total of $n$ rounds of error syndromes. Since this type of decoders takes the entire batch of those $n$ rounds of syndromes as input, we refer to them as the \emph{batch decoders}.

Batch decoders are conceptually simple and they work well as a ``proof of concept'' showing that a QEC can indeed protect quantum information (in a certain parameter regime) as well as demonstrating general ideas (MWPM, UF, \emph{etc.}.) to recover this information. However, there are a few problems with batch decoders. First, it is unclear how they can scale up temporally. Most current demonstrations of the quantum memory experiment do not have too many surface code cycles, especially when the code distance $d$ is large; usually only $n \sim d$ surface code cycles are simulated. An ideal ``quantum memory'' should last for an indefinitely large number of surface code cycles, as long as the logical error rate per cycle (which should decrease exponentially with the code distance) allows. But an approach based on batch decoders needs $O(pd^2n)$ space just to store the syndromes and thus will quickly run out of memory. Furthermore, \emph{after} the logical qubit is measured, decoding will take an additional amount of time that scales at least linearly with $n$. For some decoding methods it may take even more time (\emph{e.g.}, a naive implementation of the MWPM decoder would generate a complete graph with $O(p^2d^4n^2)$ edges and thus scale quadratically with $n$).

Moreover, a quantum memory which only supports single-qubit measurements as logical operations is obviously not useful in pursuing fault-tolerance and universality. As far as we know, it is unclear how to perform (especially non-Clifford) logical operations using batch decoders in the context of magic state injections, due to the complicated intermediate measurements. 

\begin{figure}[!ht]
    \centering
    \includegraphics[width=.5\textwidth]{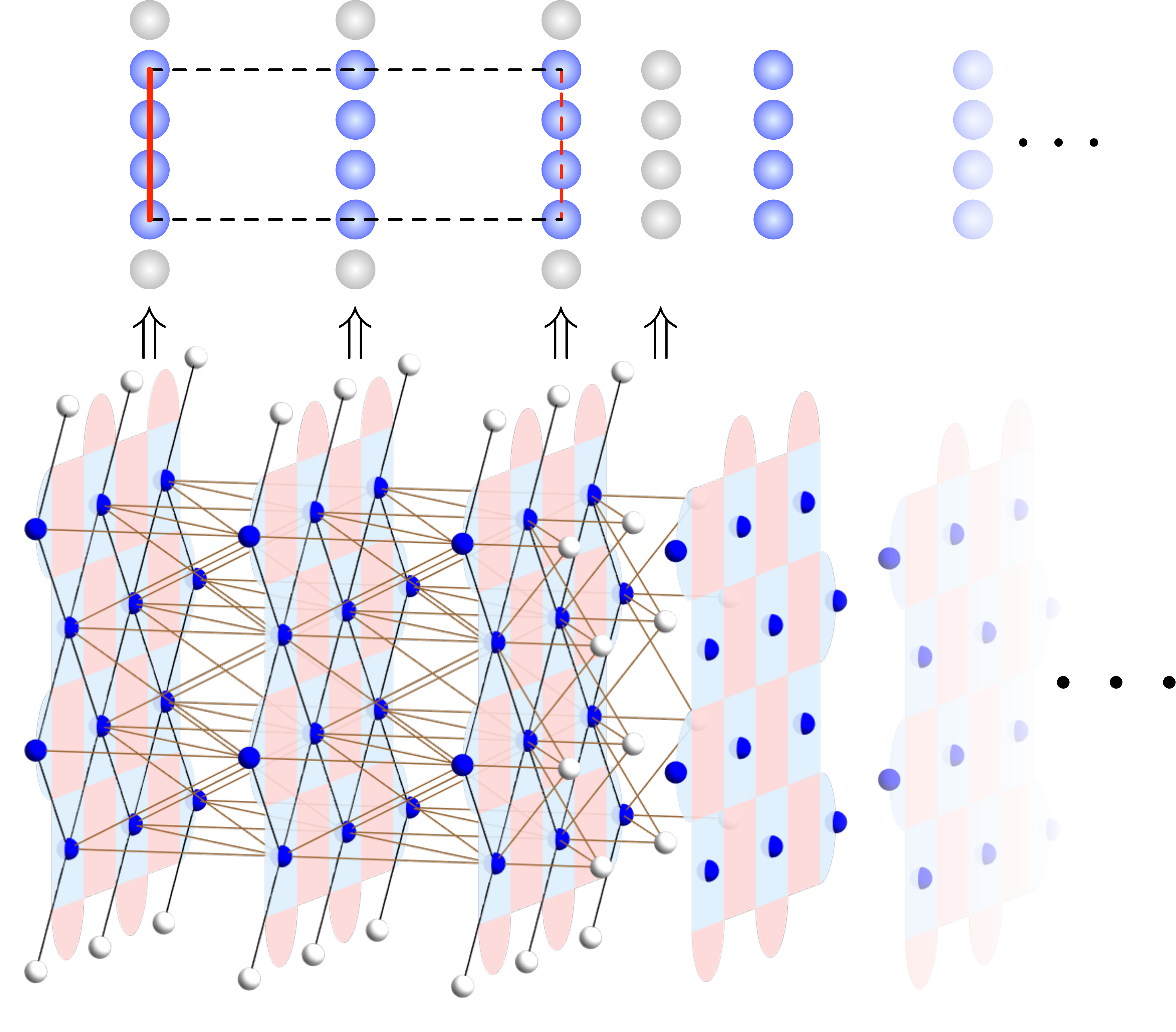}
    \caption{Visual representations of a sliding window of size $3$ in 3D (bottom) and 2D (top) for $d=5$. Time flows from left to right. The detectors within a cycle, \emph{i.e.}, on each layer, is abstracted as a column of nodes. The decoding window, which includes $3$ layers of detectors, is abstracted as a rectangle. The past (left) time boundary marked in a red solid line is closed and the future (right) time boundary marked in a dashed line is open. Since the decoder graph also has two open space boundaries, the top and bottom edges of the rectangle are marked in dashed lines as well. }
    \label{fig:decoder_graph_window}
\end{figure}

One way to deal with long sequences of QEC cycles is to use \emph{sliding-window decoders}. 
In this paper, we define a decoding window of size $w$ as a collection of $w$ consecutive cycles of detectors (Fig.~S3). 
Two consecutive windows will be offset by a certain number of layers, which we refer to as the \emph{step size}. 
The following three steps for a general sliding-window decoder can give readers an intuitive understanding of how it works:
\begin{enumerate}
    \item allocate all the detectors to multiple windows,
    \item decode each window, and
    \item combine all the individual corrections such that the overall corrections are still valid.
\end{enumerate}

The idea of sliding-window decoders seems valid based on the intuition that to decode any part of the entire decoder graph in spacetime, we only need a relatively small amount of local information on errors, \emph{e.g.}, detectors in a window of size $O(d)$. However, there exist very few implementations which can properly handle all three steps for sliding-window decoders (particularly, the last two) and nicely preserve the logical information throughout the quantum circuit as time increases. 

In this paper, we rigorously define and analyze the sliding-window decoder framework, including the insights and limitations of an existing implementation~\cite{sm_das2021lilliput}, and propose two variants: the \emph{forward decoder} (\secref{subsec:forward}) and the scalable \emph{sandwich decoder} (\secref{subsec:sandwich}). They not only can address the aforementioned problems of a batch decoder, but also have many other desirable advantages. In summary:
\begin{itemize}
    \item (forward, sandwich) scale nicely in code distance, with almost equivalent threshold as that of a batch decoder,
    \item (forward, sandwich) scale nicely in time as the number of QEC cycles increases, with almost equivalent performance regarding the logical error rate as that of a batch decoder,
    \item (sandwich) each window can be decoded in perfect parallel, allowing multiple classical execution units to effectively increase the decoding throughput,
    \item (sandwich) has a clear potential to perform logical operations on multiple qubits (in \secref{sec:generalizations}, we briefly discuss how one may approach the problem of generalizing the sandwich  decoder to the case of lattice surgery). 
\end{itemize}

\subsection{Preliminaries}\label{subsec:preliminaries}
Before revealing more details of the sandwich decoder, we shall introduce some important concepts.

\paragraph{Inner decoder} In a sliding-window decoding scheme, each window is treated as an independent decoding task with its own decoder graph. Thus, each window is fed into an \emph{inner decoder}, a subroutine that generates corrections for that window only. Therefore, designing a sliding-window decoder includes choosing an inner decoder (\emph{e.g.}, UF or MWPM) and an appropriate decoder graph (\emph{e.g.}, open and closed boundary conditions) for each window. We describe our design choice of the inner decoder for our experiments in \secref{subsec:inner_decoder}.

\paragraph{Artificial time boundaries}
The idea of allocating detectors in a decoder graph into windows naturally creates two \emph{artificial time boundaries} for each window (except for the first and the last ones, each of which has one real time boundary and one artificial time boundary). 
Note that these artificial time boundaries do not represent any real initialization or termination of the memory experiment.  Therefore, they should naturally be open, indicating that there may still be detectors at the other side of each artificial time boundary unknown to the current window.

From the perspective of the current window, an isolated defect on such a time boundary is more likely to be caused by some faults from the future or the past. However, if the inner decoder regards this boundary as closed, it will be forced to generate corrections of a higher weight within the current window, which is more likely to cause a logical error in the final result.

However, with some modifications, an artificial time boundary of a window may also become closed. 
For an example, see the syndrome propagation procedure described in \secref{subsec:forward}.

\paragraph{Core regions and buffer regions}
In both forward and sandwich decoders, each inner decoder takes all the detectors within a window as input and return corrections for the entire window. However, since adjacent windows generally have an overlap, we do not need to apply all these corrections. Instead, each window only accepts a part of the assignments that are relatively reliable and disregards the rest. 
The former region (where corrections are accepted) is called the \emph{core region} and the latter region (the rest of the window) is called the \emph{buffer region}. 
The size of a core region in general equals the step size.
The buffer regions are usually close to the open artificial time boundaries since their corrections are less trustworthy. We will be more precise about these two regions later.

Intuitively, the buffer regions between windows let them share precious contextual information on errors with each other. Therefore, having a large buffer is beneficial when we merge individual corrections generated by each inner decoder back to the entire decoder graph.
The corrections accepted in the core region will be reliable enough only when each buffer region is large enough.

\paragraph{Correction consistency}
One important principle of surface codes is that the ``correct'' corrections are not unique: Any two sets of corrections that differ by one or more stabilizers are logically equivalent. This fact poses a challenge for sliding-window decoders: Even if each decoder window individually finds a ``correct'' set of corrections, there may not be an obvious method to combine them into a consistent set of corrections for the entire decoder graph.

An example of this is illustrated in \figref{fig:sandwich_window}(b): The windows labeled \raisebox{.5pt}{\textcircled{\raisebox{-.9pt} {1}}} and \raisebox{.5pt}{\textcircled{\raisebox{-.9pt} {2}}} return inconsistent corrections along the ``seam'' where we want to merge. The possibility of such an inconsistency means that the combined corrections may not annihilate all defects and that even a low-weight fault may cause a logical error.

\subsection{Forward decoder}\label{subsec:forward}
Our forward decoder is a generalization of the ``overlapping recovery'' method introduced in \cite{sm_dennis2002topological}
as well as the sliding-window scheme used in LILLIPUT (a Lightweight Low Latency Look-Up Table decoder) proposed by Das \emph{et al.} in \cite{sm_das2021lilliput}. The main idea is to sequentially decode each window one at a time, propagating necessary information from the current window to the next window with the \emph{syndrome propagation} procedure implicitly introduced in \cite{sm_dennis2002topological} and more explicitly described in \cite{sm_das2021lilliput}.

\paragraph{Syndrome propagation}
The syndrome propagation procedure solves the inconsistency problem by forcing the next window to output consistent corrections with the current window. The detectors on the oldest layer of the next window are updated according to the decoding results for the current window, which only accepts corrections old enough to be outside of the next window.

\begin{figure}[!ht]
\centering
\begin{tabular}{ccc}
\subfigure[\label{fig:ttliliput_sm}]{\raisebox{.3cm}{\includegraphics[scale=.21]{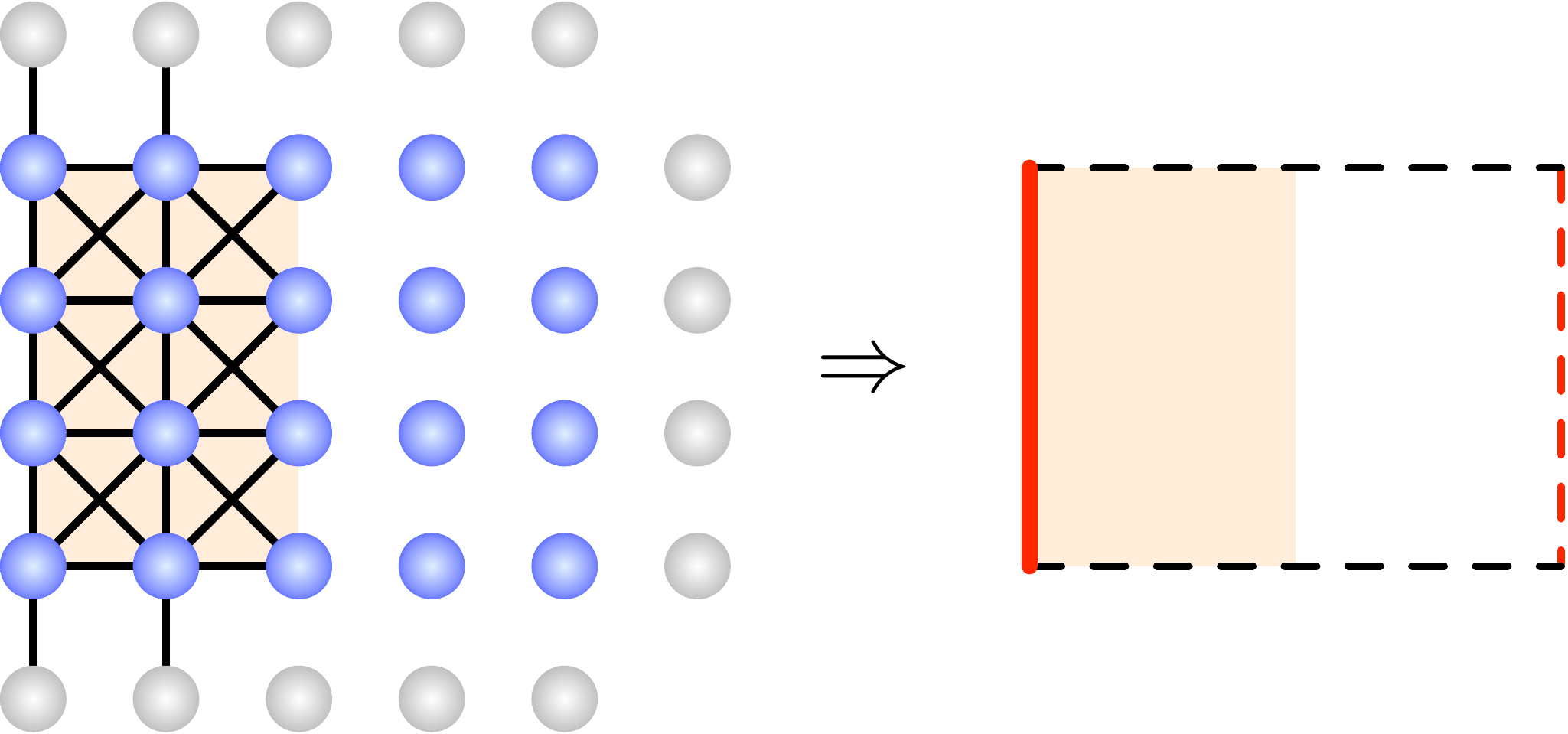}}} 
&\qquad
\subfigure[\label{fig:ttsandwich_sm}]{
\raisebox{.3cm}{\includegraphics[scale=.21]{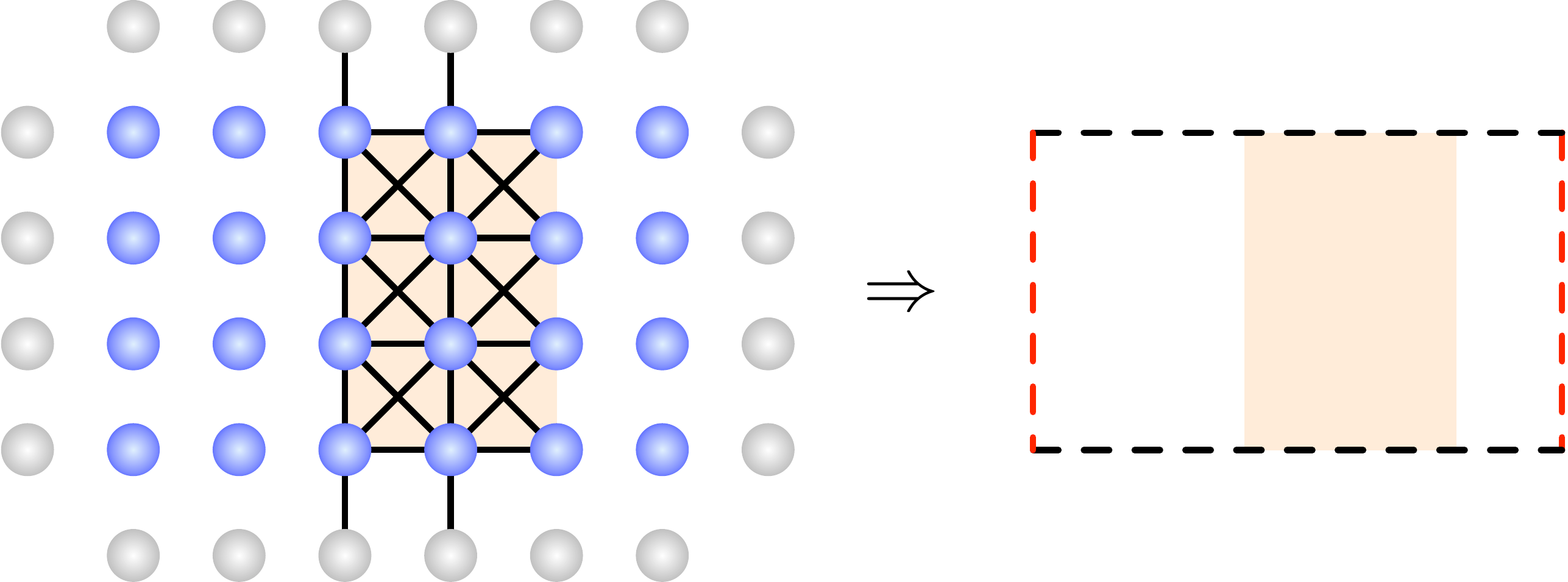}}} 
& \qquad
\subfigure[\label{fig:window_maintext}]{
\raisebox{.3cm}{\includegraphics[scale=.21]{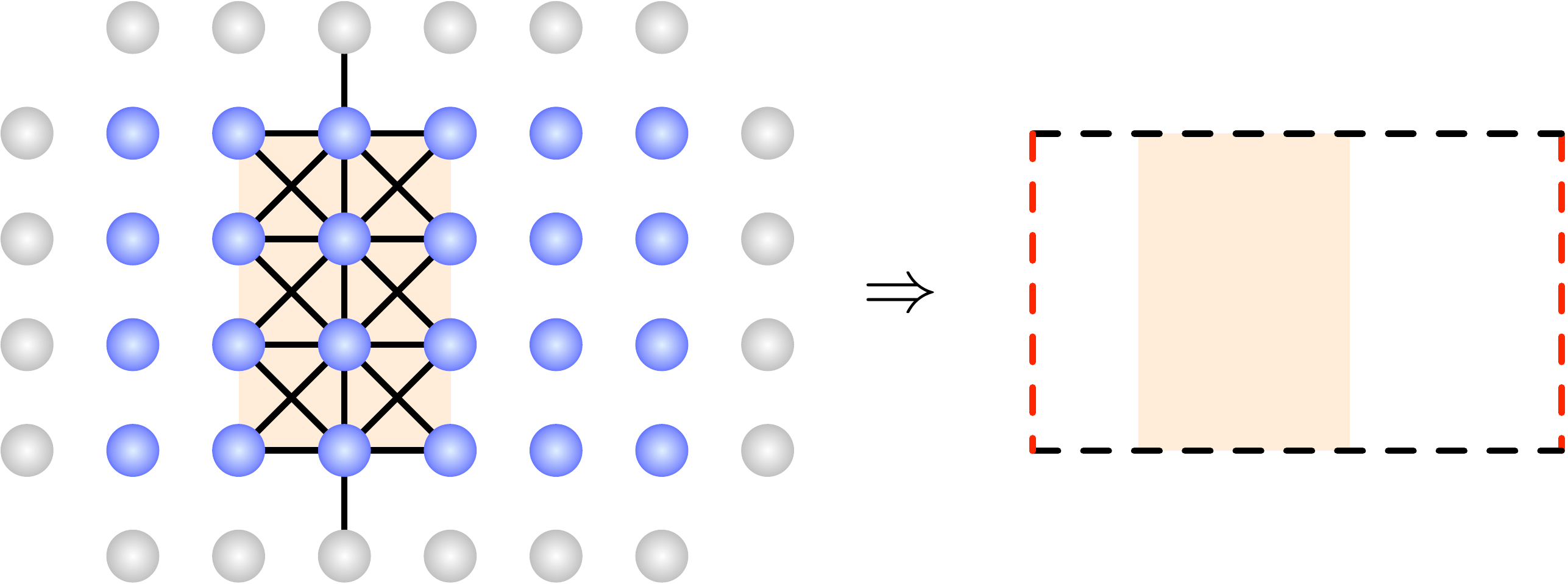}}}
\end{tabular}
\caption{Illustrations of the core region and buffer region(s) for a window. Only edges in the core region (shaded in brown) are drawn. (a) A forward window with size $w=5$ and step size $s=2$. (b,c) A type-$1$ sandwich window with size $w=6$ and step size $s=2$. Each buffer region has size $b=2$. The sandwich decoder simulations in this paper adopt the design in (b) and the core and buffer design described in the main text corresponds to (c). Both (b) and (c) are valid and the core regions across different windows in each design are disjoint edge sets. }
\label{fig:ttsandwich_window}
\end{figure}

Let us ignore it for now and let $s$ be the step size. The core region of each forward window includes the edges within the oldest $s$ layers and the edges connecting the $s$-th oldest layer and the $(s+1)$-th oldest layer. In this case, we say that the core region for a forward window has size $s$. What remains is the buffer region which exactly consists of all its overlapping edges with the next window. See \figref{fig:ttsandwich_window}(a) for an example. 
Hence, all the defects within the oldest $s$ layers will be annihilated. The detectors in the $(s+1)$-th oldest layer will be updated if some accepted corrections connect from the $s$-th oldest layer to the $(s+1)$-th oldest layer. All other detectors in the remaining layers stay the same. Therefore, when the next window starts from the $(s+1)$-th layer of the current window, only this layer is updated (\figref{fig:sandwich_window}(a)). 

This way, for each window, since all the defects prior to it have been annihilated by previously accepted corrections, it has a closed past boundary and an open future boundary. The closed boundary means that the inner decoder will not change any corrections already accepted in the past, and thus consistency between windows is ensured.

Note that in LILLIPUT, it is unclear whether the future time boundary for each window is open or closed. But according to the arguments in \secref{subsec:preliminaries}, this boundary should be open by its nature.

\paragraph{Handling the last window}
The size of the last window depends on the total number of cycles in the memory experiment, meaning that it could be smaller than the regular intermediate windows. For the $Z$-type decoder graph in the memory experiment preserving logical $|\overline{0}\rangle$, both time boundaries of the last windows are closed. All the generated corrections are accepted, \emph{i.e.}, the entire window is the core region.

\paragraph{Step size}
Compared with LILLIPUT which fixes the step size to be $1$, our implementation of the forward decoder allows flexible step size.
The advantage of having a step size of $1$ is that for any fixed window size $w$, the length of the buffer region $b = w-1$ is maximized.
However, moving forward only by one QEC cycle at a time has an obvious disadvantage in terms of the total time complexity of decoding all windows. 
In an experiment with a total of $n$ surface code cycles, the number of sliding windows needed is $O(n)$, and each window needs at least $O(pd^2w)$ time to decode, making the total time complexity $O(pd^2nw)$. The extra factor of $O(w)$ severely limits the possible code distance that can be implemented in practice and may force an implementation to use much more classical computational resource to achieve the desired throughput.

This problem can be solved by moving forward $s = O(w)$ cycles at a time, \emph{while still preserving the same amount of buffer $b = w-s$}. Importantly, when $b$ is fixed, increasing the size of each window $w$ will increase the step size $s$ and thus \emph{decrease} the total time complexity (assuming a linear time complexity for the underlying decoder). For example, if we set $s = b$ and $w = 2b$, then the total time complexity becomes
\begin{align}
    O(pd^2w)\cdot O\left(\frac{n}{s}\right) = O(pd^2n),
\end{align}
\emph{i.e.}, only a constant overhead compared to the batch decoder.

\paragraph{Problem with parallelization}
A remarkable disadvantage of the forward approach is the strict dependency among all windows: One cannot start decoding the next window until the current window is finished.
Therefore, once the decoding time of each window fails to keep up with $s$ cycles of newly extracted syndromes where $s$ is the step size, the latency accumulates. This is one of the reasons why LILLIPUT employs look-up tables (LUTs) as their inner decoders. On one hand, since the LUTs are generated off-line (or in advance) with MWPM, they are fast and accurate while decoding each window; however, the huge amount of memory needed to store these LUTs also restricts LILLIPUT to code distance up to $5$ and window size up to $3$.

For sliding-window decoders, it will be a huge boost to the throughput if we can decode all windows in parallel. However, this is impossible for a forward decoder since this data dependency causes the critical path\footnote{The longest series of sequential operations in a parallel computation.} to run through all windows.

\subsection{Sandwich decoder}\label{subsec:sandwich}

We propose an alternative approach to sliding-window decoders that also allows adequate buffer regions and solves the correction consistency problem, but has a much shorter critical path that does not increase with the total number of windows\footnote{Technically, we still need to add up the logical corrections from all the windows, but this has negligible computational cost in practice, and theoretically it can be done in $O(\log n)$ parallel time anyway.} and thus enables parallelism between windows. We name this approach the \emph{sandwich} approach, which comes with two interpretations, as we will mention below.

We shall introduce two types of sandwich windows. 
If we do not specify the type of a window, it should be clear from the context. 

\paragraph{Type-$1$ windows: buffer regions in both directions}
Recall that syndrome propagation is the key ingredient in solving the correction consistency problem in the forward decoders, but it is also what causes the long critical path since it only proceeds in the forward direction.
On the other hand, since the decoder graph formalism is symmetric with respect to the direction of time, each window should be capable of propagating syndromes in the backward direction as well.

It follows naturally that the core region of each window (we shall ignore the first and the last windows for now) can be ``sandwiched'' by two buffer regions, resulting in the definition of a type-$1$ sandwich window. 
If the step size is $s$, then the core region includes all edges within the middle $s$ layers. 
There is also some freedom regarding the specific design of the core region for a type-$1$ sandwich window. For example, in the experiments readers shall see in this paper, we add the edges connecting across the latest layer in the middle $s$ layers and its next layer to the core region as well. We say that the core region has size $s$. What remains are two separate buffer regions, each of which has size $b$ and the window size is thus $w=s + 2b$. 
See \figref{fig:ttsandwich_window}(b) for an example. 
Each type-$1$ sandwich window overlaps with its next window and its previous window in $2b$ layers of detectors respectively. Both artificial time boundaries are open. 

Same with the forward decoders, the sandwich decoders also have flexible step size. 

Most importantly, type-$1$ windows are not dependent on any other window, so all of them can be decoded in parallel.

\paragraph{Type-$2$ windows: merging corrections}

Since type-$1$ windows propagate syndrome information forwards and backwards, there must be another type of windows that receive syndrome information from both ends which we define as type-$2$ windows. Since syndrome propagation closes the corresponding time boundary of the receiving window, each type-$2$ sandwich window has \emph{both} time boundaries closed, and the entire window is the core region. 

Since each type-$2$ window is sandwiched between two type-$1$ windows, it is dependent on the decoding results of the two independent type-$1$ windows. But since all type-$2$ windows are independent from each other, they can be decoded in parallel. 

\begin{figure}[!ht]
    \centering
    \includegraphics[width=.7\textwidth]{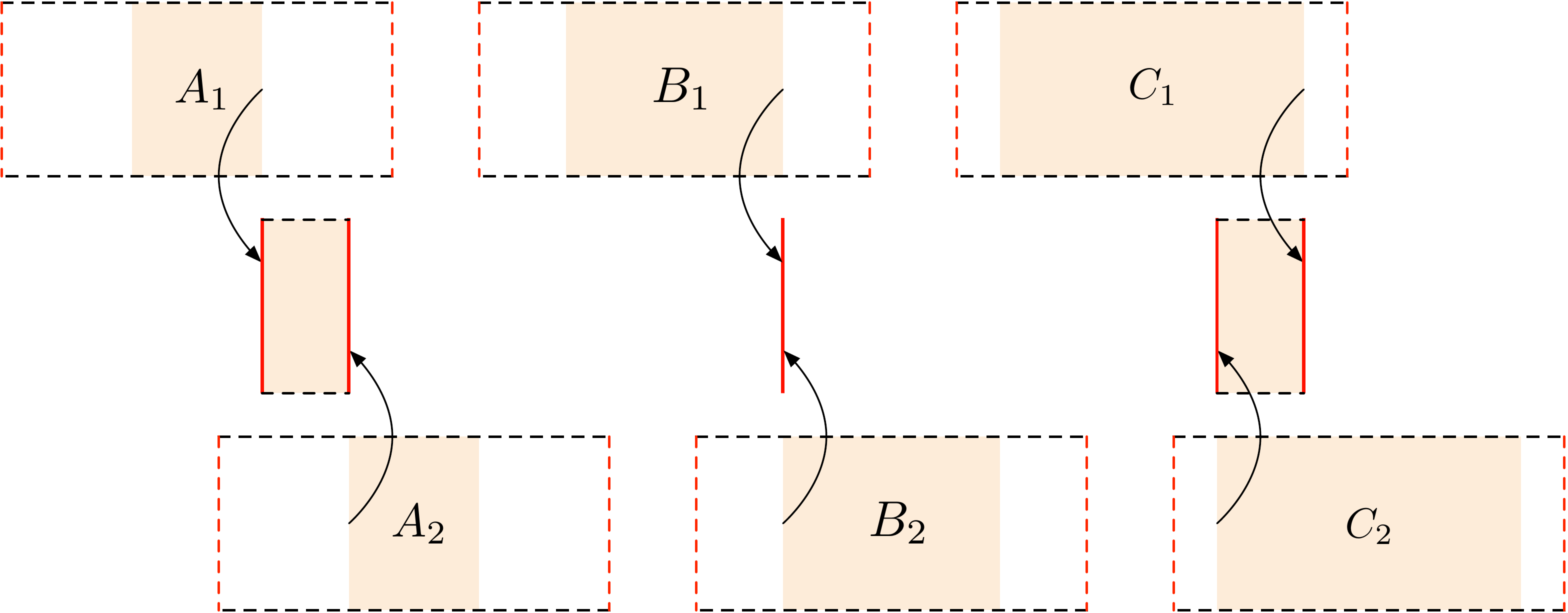}
    \caption{A. positive seam offset; B. zero seam offset; C. negative seam offset. }
    \label{fig:seam_offset_ref}
\end{figure}

After we apply the corrections from type-$1$ windows, there may exist defects not annihilated yet. 
The role of the type-$2$ windows is to reconcile this inconsistency by neutralizing \emph{all} remaining defects.
To illustrate this more clearly, we call the latest layer of detectors in a core region the \emph{right seam} and the oldest layer the \emph{left seam}. 
The core regions of two adjacent type-$1$ windows can have three patterns (\figref{fig:seam_offset_ref}) and we refer to this difference using a parameter called \emph{seam offset}. 

\begin{itemize}
    \item When the seam offset is $0$, meaning that the right seam of the former core overlaps with the left seam of the latter core. 
    The type-$2$ windows thus become two-dimensional and include the updated detectors in these pairs of seams. 
    \item When the seam offset is positive, the core regions do not have any overlap. The 3D type-$2$ windows thus include the detectors that are in neither type-$1$ windows and the updated detectors in the right and left seams pairs; 
    \item When the seam offset is negative, the core regions overlap in more than $1$ layers. The 3D type-$2$ windows thus include the corrections for the overlapping defects from type-$1$ windows. 
\end{itemize}

In each case, the size of a type-$2$ window is $|t|+1$ where $t$ is the seam offset value. After decoding all type-$2$ windows and apply every correction, all defects will be annihilated. 

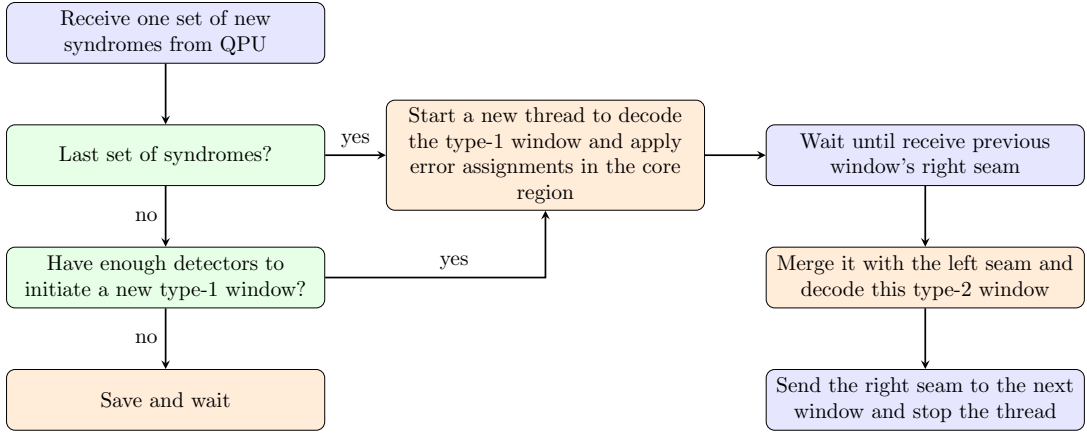
\begin{figure}[!ht]
\begin{singlespace}
\begin{tikzpicture}[node distance=1cm]
\tikzstyle{startstop} = [rectangle, rounded corners, minimum width=4cm, minimum height=1cm,text centered, text width=5cm, draw=black, fill=blue!10, align=flush center]
\tikzstyle{process} = [rectangle, rounded corners, minimum width=4cm, minimum height=1cm, text centered, text width=5cm, draw=black, fill=orange!15, align=flush center]
\tikzstyle{decision} = [rectangle, rounded corners, minimum width=4cm, minimum height=1cm, text centered, text width=5cm, draw=black, fill=green!10, align=flush center]
\tikzstyle{arrow} = [thick,->,>=stealth]

\node (start) [startstop] {Receive one set of new syndromes from QPU};
\node (dec_last) [decision, below=of start] {Last set of syndromes?};
\node (dec_window) [decision, below=of dec_last] {Have enough detectors to initiate a new type-$1$ window?};
\node (wait) [process, below=of dec_window] {Save and wait};
\node (decode) [process, right=of dec_last] {Start a new thread to decode the type-$1$ window and apply error assignments in the core region};
\node (lower_seam) [startstop, right=of decode] {Wait until receive previous window's right seam};
\node (decode_seam) [process, below=of lower_seam] {Merge it with the left seam and decode this type-$2$ window};
\node (upper_seam) [startstop, below=of decode_seam] {Send the right seam to the next window and stop the thread};

\draw [arrow] (start) -- (dec_last);
\draw [arrow] (dec_last) -- node[anchor=south] {yes} (decode);
\draw [arrow] (dec_last) -- node[anchor=east] {no} (dec_window);
\draw [arrow] (dec_window) -| node[anchor=south, xshift=-1.5cm] {yes} (decode);
\draw [arrow] (dec_window) -- node[anchor=east] {no} (wait);
\draw [arrow] (decode) -- (lower_seam);
\draw [arrow] (lower_seam) -- (decode_seam);
\draw [arrow] (decode_seam) -- (upper_seam);

\end{tikzpicture}
\end{singlespace}
\caption{The proposed workflow of a sandwich decoder in a real quantum computing system. }
\label{fig:sandwich_flow}
\end{figure}

\figref{fig:sandwich_flow} describes a general work-flow for implementing a sandwich decoder with multiprocessing units.

\paragraph{Handling the first and the last windows}
The first window should correct not only the middle $s$ layers, but also all the layers before, since it does not have a previous window. Therefore, the first window has only one buffer of size $b$ and the core region has size $s+b$. 
Similarly, the last window corrects the middle $s$ layers as well as all the layers after them. It also has only one buffer of size $b$ and the core region has size $s+b$. 

\begin{figure}[!ht]
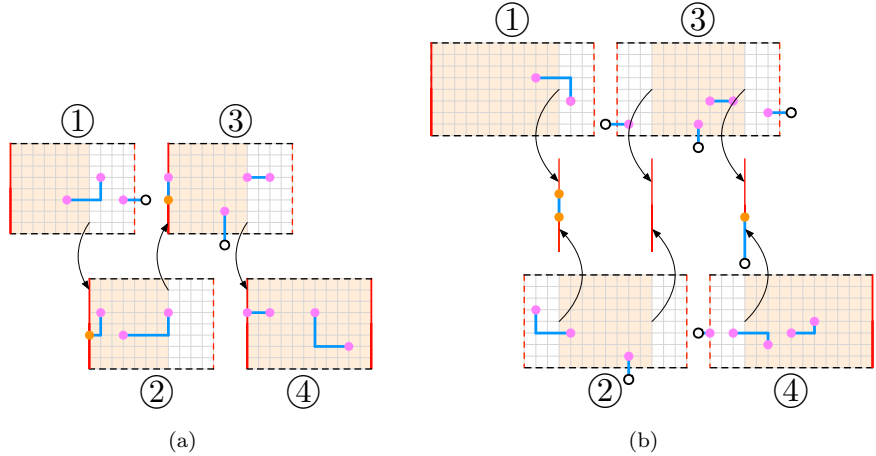

\centering
\begin{tabular}{cc}
\subfigure[\label{fig:liliput_sm}]{
\raisebox{.3cm}{\includegraphics[scale=.34]{fig/forward_window}}}
&\quad
\subfigure[\label{fig:sandwich_sm}]{
\raisebox{.3cm}{\includegraphics[scale=.35]{fig/sandwich_window}}}
\end{tabular}
\caption{Comparison between the forward decoder (a) and the sandwich decoder (b). The red vertical lines represent closed time boundaries, the dashed vertical lines represent open time boundaries, and the dashed horizontal lines represent open space boundaries. The core region of each window is marked with brown color. The buffer regions are the remaining white areas. }
\label{fig:sandwich_window}
\end{figure}

\paragraph{Comparison with the forward decoder}
\figref{fig:sandwich_window} visualizes the difference between the forward approach and the sandwich approach. Note that, for simplicity of illustration and implementation, we set the seam offset to be $0$, causing the type-$2$ windows to degenerate into 2D decoder graphs. 

\section{Monte Carlo simulation of the threshold}

\subsection{Union-find inner decoders}\label{subsec:inner_decoder}
In most of our experiments, we use a union-find (UF) decoder, as proposed in \cite{sm_delfosse2021almost}, to decode each individual window (including the forward windows and both types of the sandwich windows). In our implementation, we use the \emph{weighted growth} version of the decoder described in Section~5 of \cite{sm_delfosse2021almost}, although due to implementation considerations, our definition of the ``boundary size'' may be slightly different compared to the definition used in that paper.

We chose the UF decoder due to its low time complexity both in theory and in practice, but it was unclear whether the UF decoder is the best fit for a sliding-window scheme. The reason is that, in terms of its theoretical foundation, the UF decoder does not try to approximate the minimum-weight correction; instead, it tries to find an equivalence class that is likely to contain the actual error, and then it chooses an arbitrary correction in that equivalence class with a simple peeling decoder. This means that the updated detectors at the right and left seams obtained by applying only part of the correction output by the UF decoder may be misleading, although our experimental results indicates that this does not noticeably affect the performance in practice. In any case, we also conduct some experiments with a minimum-weight perfect matching (MWPM) decoder as an alternative inner decoder to validate the universality of the sandwich scheme.

\subsection{Sampling errors}

We employ the circuit-level depolarizing noise model with a single parameter $p$. 

More precisely, we assume that the preparation and measurement errors exist on all data and ancilla qubits, where a qubit is initialized to an orthogonal state with probability $p$ and a measurement result is flipped with probability $p$. Each single-qubit, two-qubit, and idle gate is implemented as a perfect gate followed by a depolarizing channel. With probability $p$, the perfect gate is afflicted by a non-trivial Pauli error chosen uniformly at random. 

The errors attached to different elementary operations are applied independently.

\subsection{Monte Carlo simulations}\label{sec:monte_carlo}

Given each code distance $d\in \{3, 5, \cdots, 17\}$, we choose a sandwich decoder with step size $s_d=(d+1)/2$ and window size $w_d=3s_d$. We use the 3D MWPM and UF decoder respectively as the inner decoder to handle each type-$1$ window and the 2D MWPM and UF decoder respectively to handle each type-$2$ window (\emph{i.e.}, the seam offset is $0$). 

\begin{figure}[!ht]
\centering
\begin{tabular}{cc}
\subfigure[\label{fig:mw_sandwich_threshold}]{
\raisebox{.2cm}{\includegraphics[scale=.6]{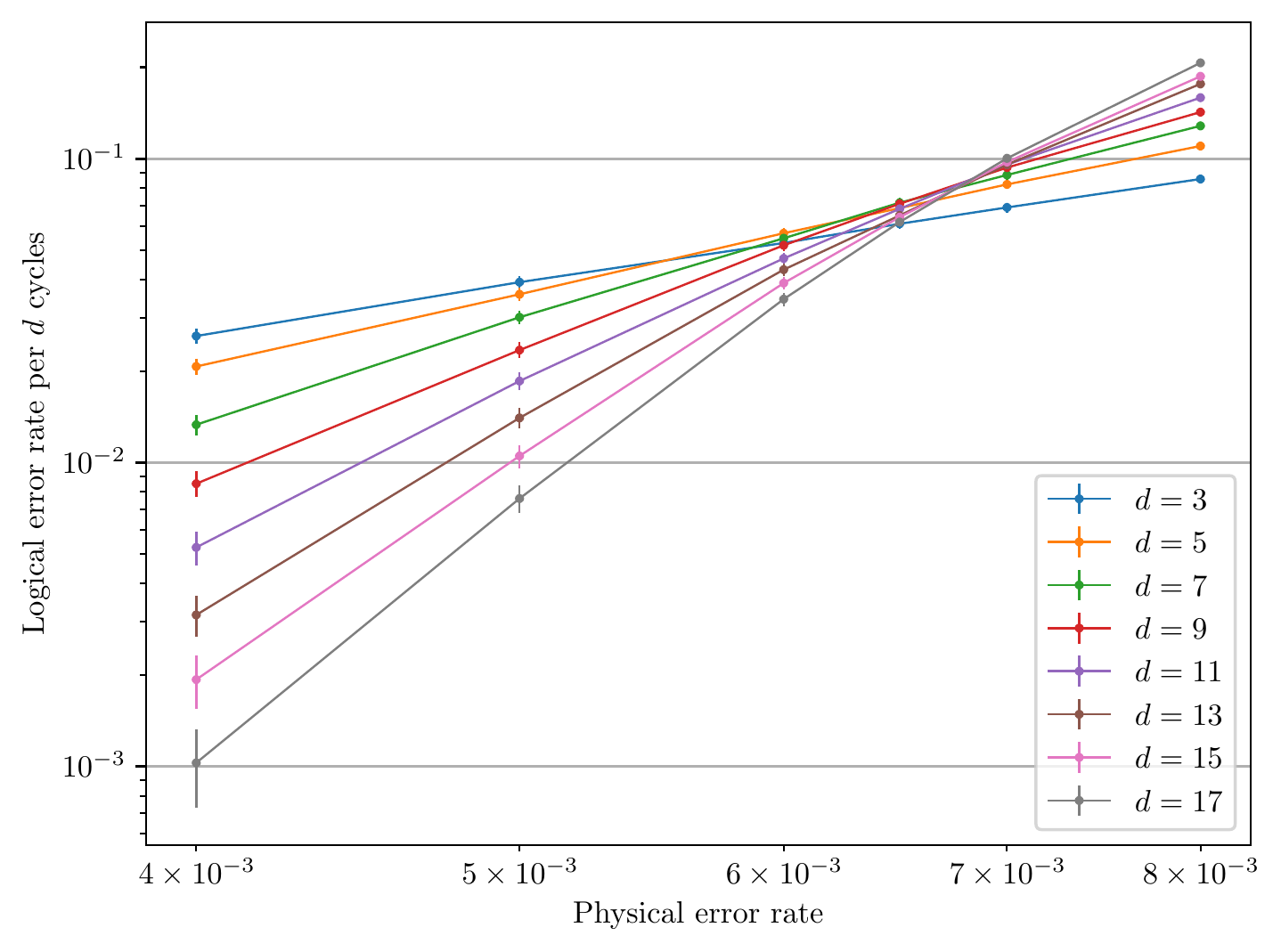}}}
&
\subfigure[\label{fig:mw_batch_threshold}]{
\raisebox{.2cm}{\includegraphics[scale=.6]{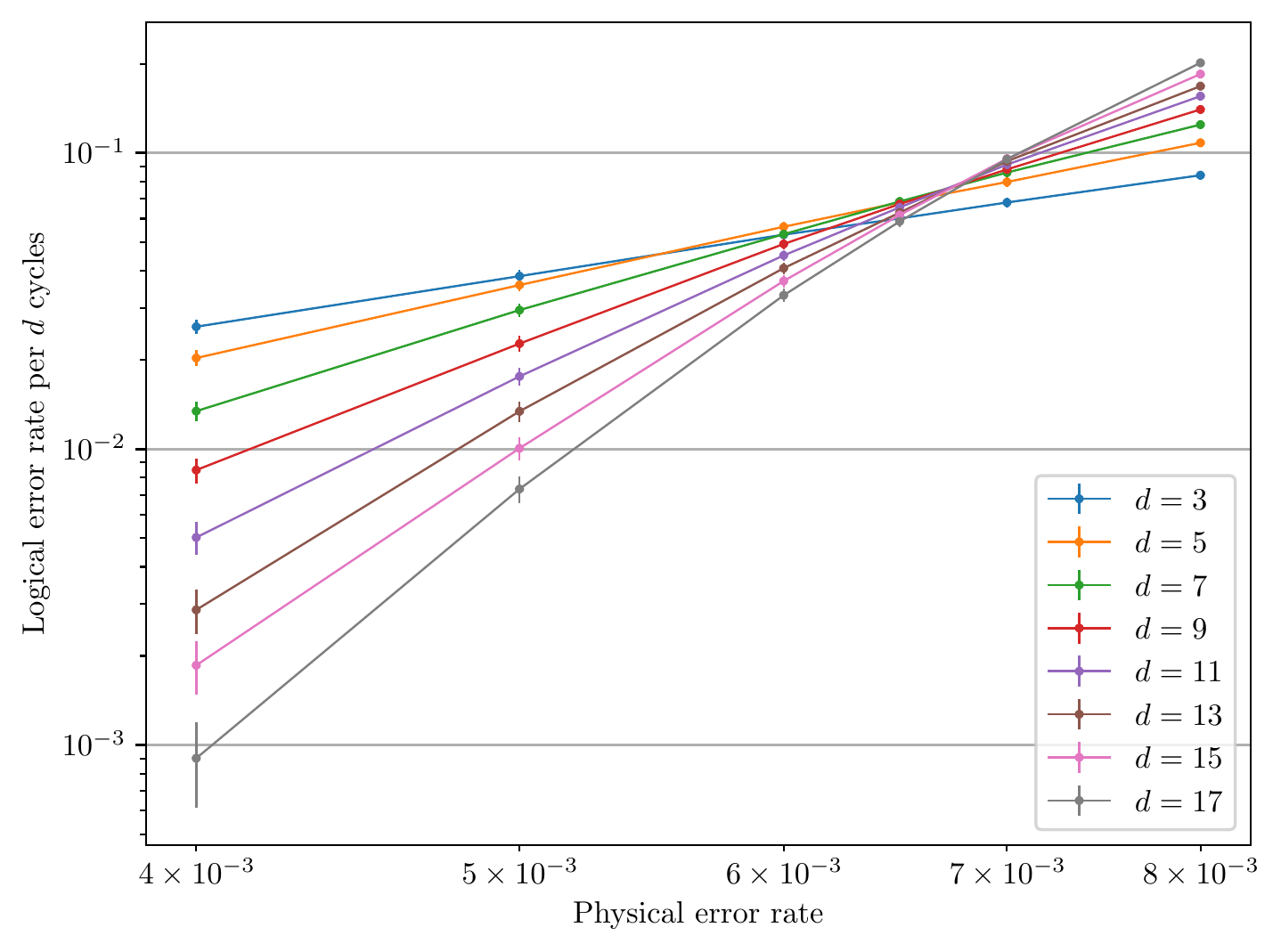}}} \\
\subfigure[\label{fig:sandwich_threshold}]{
\raisebox{.2cm}{\includegraphics[scale=.6]{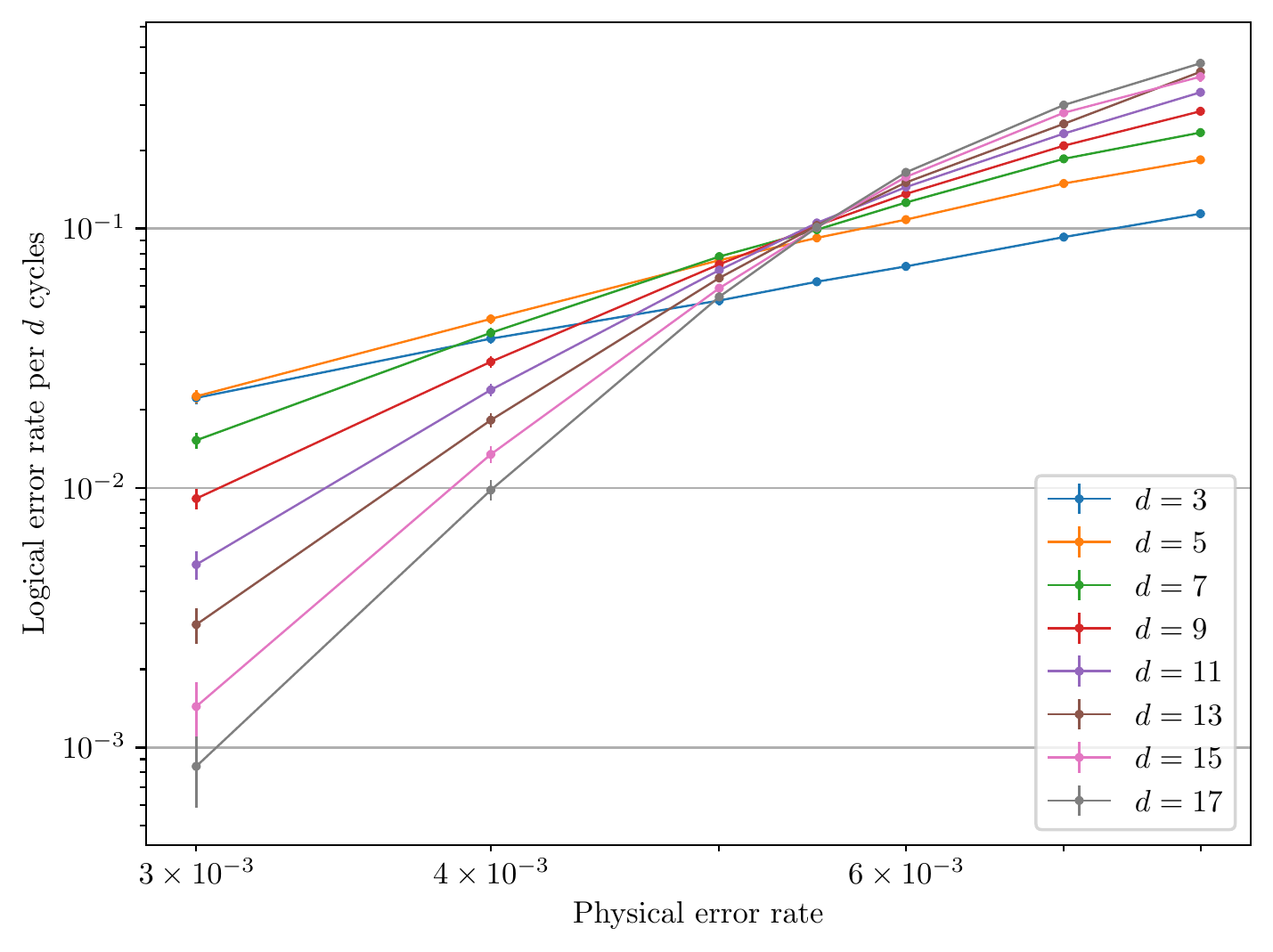}}}
&
\subfigure[\label{fig:uf_batch_threshold}]{
\raisebox{.2cm}{\includegraphics[scale=.6]{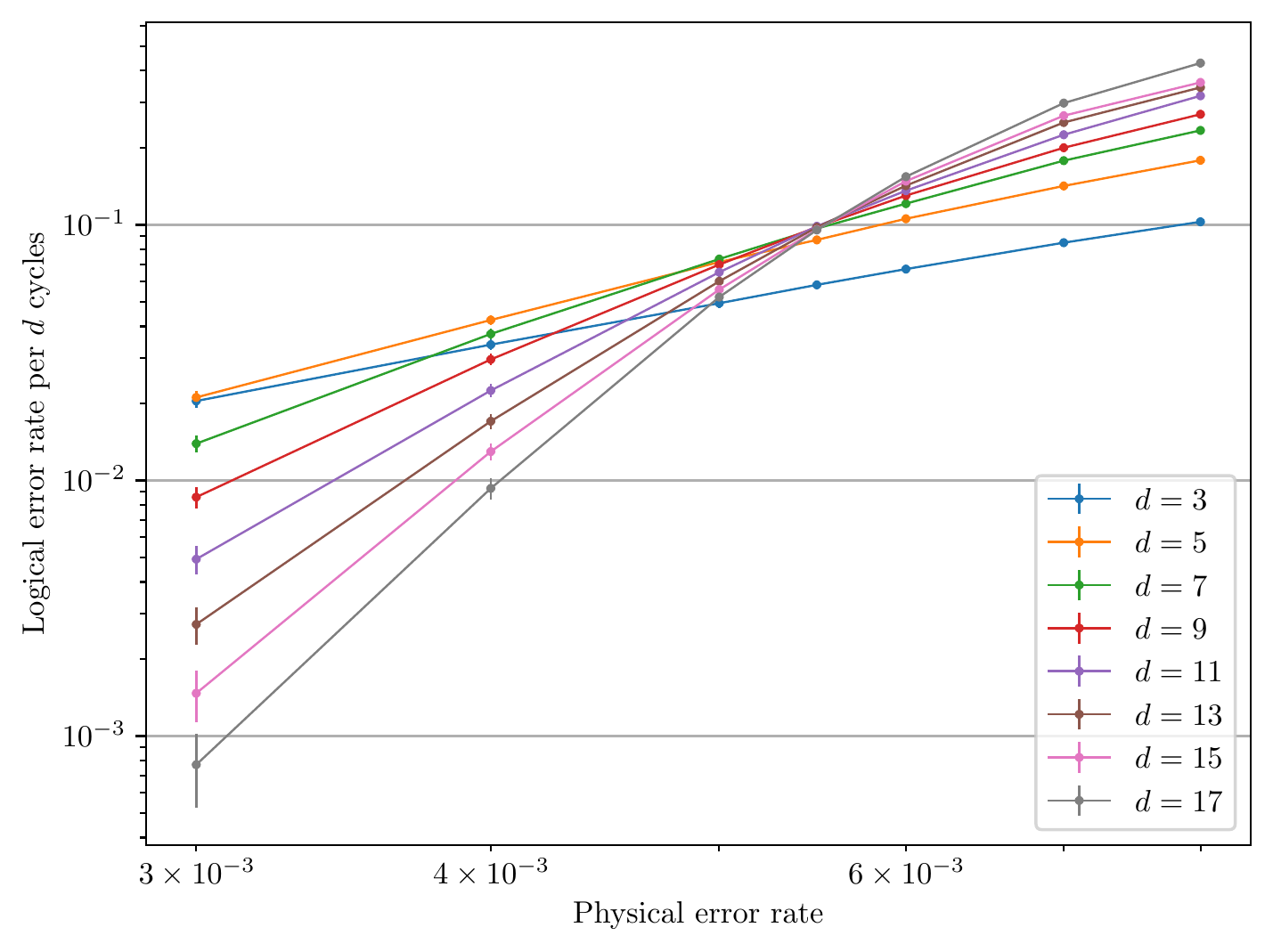}}}
\end{tabular}
\caption{Logical error rates for (a) the MWPM sandwich decoder ($0.68\%$), (b) the MWPM batch decoder ($0.70\%$), (c) the UF sandwich decoder ($0.55\%$), and (d) the UF batch decoder ($0.554\%$) for code distances $d=3,5,\ldots,17$. Logical error rates per $d$ cycles $\PL(d)$ are estimated with weighted least squares, as explained in \secref{sec:monte_carlo}. Error bars indicate $95\%$ statistical confidence according to a conservative estimate (\emph{i.e.}, overestimate) of the variance of $\hat{p}_L(d)$, as explained in \secref{sec:error_estimation}. 
The sandwich decoder has almost the same performance as the corresponding batch decoder.}
\label{fig:thresholds}
\end{figure}

For the experiments with the UF inner decoder, we consider physical error rates $p\in \{0.3\%$, $0.4\%$, $0.5\%$, $0.55\%$, $0.6\%$, $0.7\%$, $0.8\%\}$; for the experiments with the MWPM inner decoder, we consider physical error rates $p\in \{0.4\%$, $0.5\%$, $0.6\%$, $0.65\%$, $0.7\%$, $0.8\%\}$.
For each $p$, we run Monte Carlo simulation to find the logical error rate per $d$ cycles for $100,000$ shots. We obtain an estimated threshold of $0.55\%$ for UF sandwich decoder as shown in \figref{fig:thresholds}(c). We also run the same experiments using UF batch decoders and obtain a similar threshold as shown in \figref{fig:thresholds}(d). The thresholds for MWPM sandwich decoder and batch decoder are $0.68\%$ in \figref{fig:thresholds}(a) and $0.70\%$ in \figref{fig:thresholds}(b). 

To define the concept of ``the logical error rate per $d$ cycles'' $\PL(d)$, we make the ansatz that each cycle of syndrome extraction independently flips the logical qubit with a fixed probability $\PL(1)$. 
If we exclude the data qubits initialization and final measurement faults, the probability of flipping the logical qubit after $i$ cycles of syndrome extraction  $\PL(i)$ satisfies
\begin{align}
1-2\PL(i) = \left[1-2\PL(1)\right]^i.
\end{align}
Define $q$ to be the probability that the data qubit initialization and measurement collectively flip the logical qubit.
Then, the probability $p_{\mathrm{L},n}$ of logical error for an $n$-cycle memory experiment satisfies
\begin{align}\label{eq:logical_error_rate_recursion}
1-2p_{\mathrm{L},n} = (1-2q)\cdot (1-2\PL(1))^n.
\end{align}
We can thus calculate the logical error rate per $d$ cycles $\PL(d)$ from the estimated logical error rate per shot $\hat{p}_{\mathrm{L},n}$ using the weighted least squares estimator
\begin{align}\label{eq:weighted_least_squares}
\log(1-2\hat{p}_{\mathrm L}(d)) = \mathbf x^\top \mathbf{diag}(\mathbf w) \, \mathbf y \bigg/ \mathbf x^\top \mathbf{diag}(\mathbf w) \,\mathbf x
\end{align}
where 
\begin{align}
x_n = (n-\overline{n})/d, \qquad w_n =  1/\hat{\Var}(y_n), \qquad y_n = \log(1-2\hat{p}_{\mathrm L, n}) - \overline{\log(1-2\hat{p}_{\mathrm L, n})}.
\end{align}
The explicit form of the estimator $\hat{\Var}(y_n)$ will be given in \secref{sec:error_estimation} below.
In our experiments, we simulate with different overall numbers of cycles $n = \lfloor ks_d/2 \rfloor$ where $k\in \{8, 9, \cdots, 20\}$.

To more efficiently simulate the behavior of our scheme for different numbers of cycles, we conduct those simulation experiments simultaneously, reusing the sampled errors and decoder outputs for early cycles. 
That is, for each $d$ and $p$, we only construct one decoder graph with $n=10s_d$ and sample errors on it. Then, within the same decoder graph, we calculate the logical error rates per shot for all $n=\lfloor ks_d/2 \rfloor$ where $k\in \{8, 9, \cdots, 20\}$.
This causes the results of those experiments to be correlated, but over the $100,000$ independent shots, the effect of this correlation should be minor. 

More specifically, each simulation proceeds as follows:
\begin{enumerate}
    \item Before starting any sliding, sample edges on the entire decoder graph according to pre-calculated probabilities, except for the final data qubit measurement errors. 
    \item Decode each type-$1$ window with defects generated from the sampled edges, and identify the core region. More specifically, 
        \begin{enumerate}
            \item when the window has reached the last cycle (\emph{i.e.}, the last layer in the decoder graph), sample edges resulted from final data qubit measurement errors, decode window and apply corrections in the core region. Finally, calculate the logical error rate per $d$ cycles; 
            \item when the window has not reached the last cycle but would be the last window if the number of cycles equals $\lfloor ks_d/2 \rfloor$ for $k \in \{8, 9, \cdots, 19\}$, make two copies of the current situations. On one copy, treat this window as the last window in case (a). On the other copy, proceed as in case (c);
            \item when the window is not the last window for any numbers of cycles requested, decode, apply correction in the core region, and slide to the next window. 
        \end{enumerate}
    \item For each type-$2$ window (\emph{i.e.}, the overlapping seam between two consecutive decoded type-$1$ windows), put all inconsistent detectors into a 2D decoder and apply all the generated corrections. 
\end{enumerate}

\subsection{Error estimation}\label{sec:error_estimation}

For each value of $n$ (and combination of other parameters), our Monte Carlo simulation gives an estimated logical error rate per shot $\hat{p}_{\mathrm L,n}$ with variance
\begin{align}
\Var(\hat{p}_{\mathrm L,n}) = \frac{p_{\mathrm L,n} \cdot (1-p_{\mathrm L,n})}{N}
\end{align}
where $N=10^5$ denotes the number of shots.
The weights $w_i$ used in the least squares estimator \eqnref{eq:weighted_least_squares} are derived from the approximate variance
\begin{align}
\hat{\Var}(y_n) = \hat{\Var}\left(\log(1-2\hat{p}_{\mathrm L,n})\right) \approx \left(\frac{2}{2\hat{p}_{\mathrm L,n}-1}\right)^2\cdot \hat{\Var}(\hat{p}_{\mathrm L,n})
\end{align}

As mentioned above, our estimations $\hat{p}_{\mathrm L,n}$ for different values of $n$ are correlated. Therefore, we cannot use the usual variance estimator for weighted least squares. Instead, we take a conservative estimate of the variance 
\begin{align}
\hat{\Var}\left(\log(1-2\hat{p}_{\mathrm L}(d)) \right) \lessapprox \left(\sum_n |x_n| w_n \sqrt{\hat{\Var}(y_n)} \Bigg/\sum_n w_n x_n^2\right)^2.
\end{align}

\section{Numerical analysis}\label{num_analysis}

\subsection{Step size and window size}
When we evaluate the performance of sandwich decoders, the step size and window size are two natural features to consider. 
\begin{align}
    \text{window size} \ (w) = \text{step size} \ (s) + 2\times \text{buffer size} \ (b)
\end{align}

\begin{figure}
\centering
\begin{tabular}{cc}
\subfigure[\label{fig:vary_step}]{
\raisebox{.2cm}{\includegraphics[scale=.6]{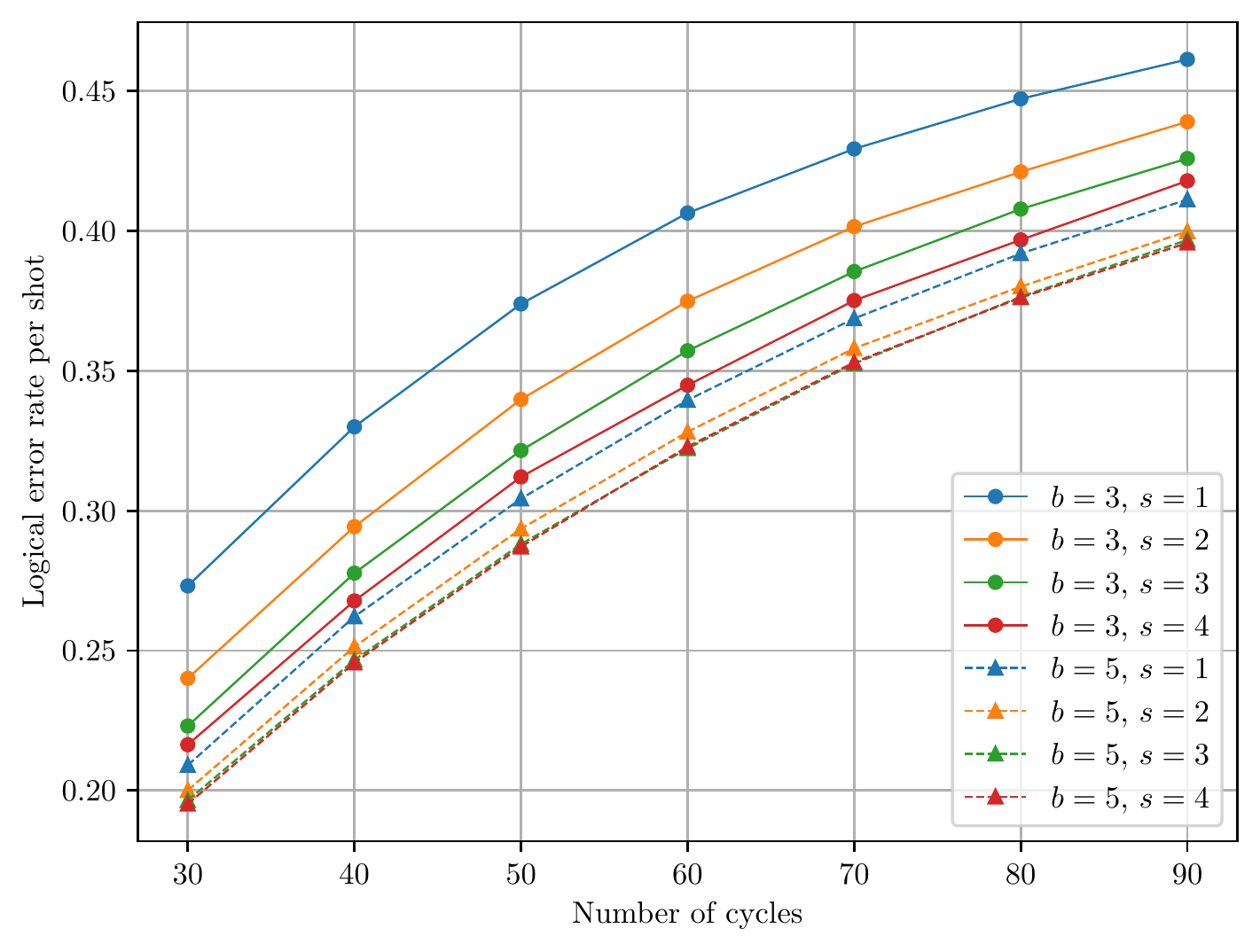}}}
&
\subfigure[\label{fig:vary_step_avg}]{
\raisebox{.2cm}{\includegraphics[scale=.6]{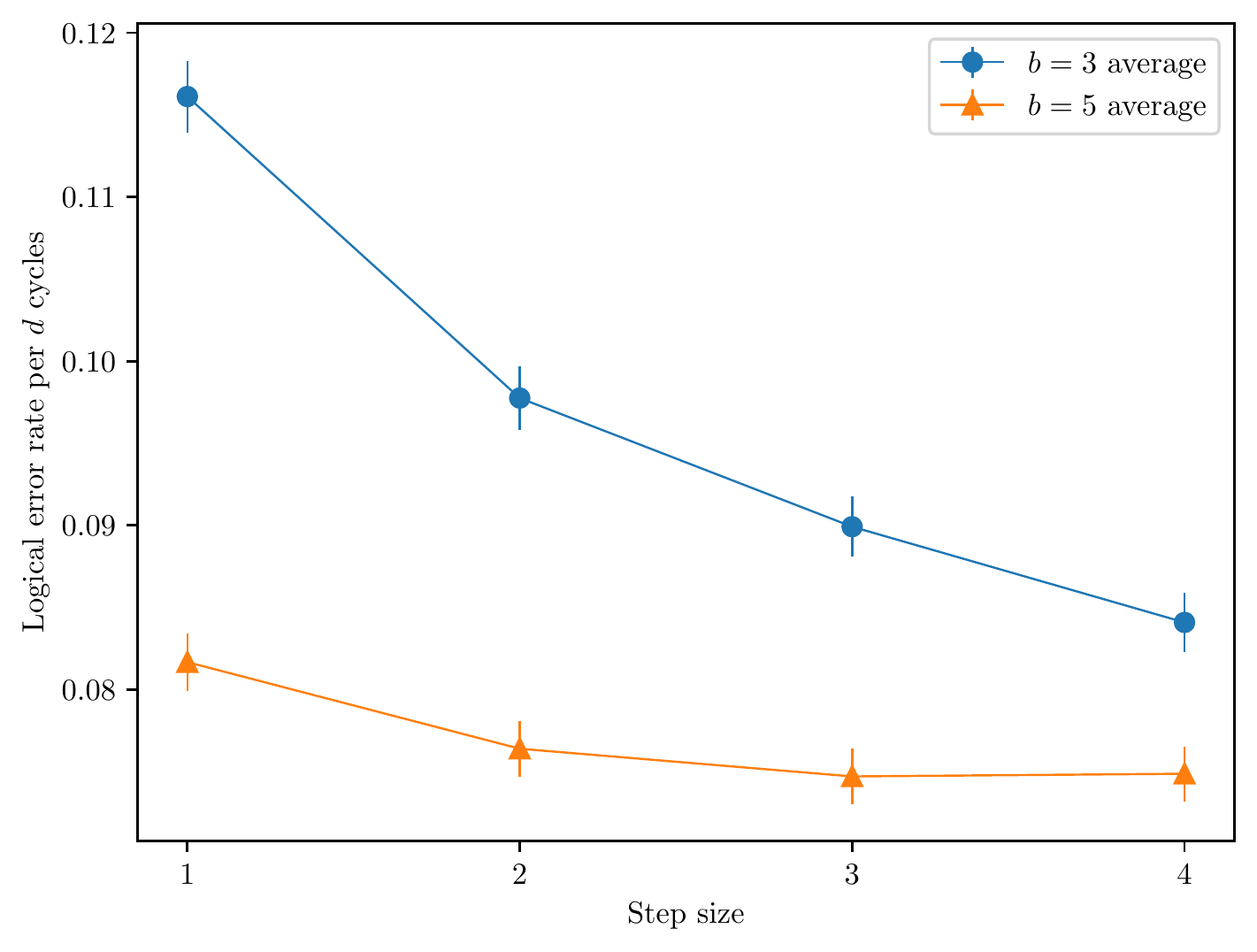}}}
\end{tabular}
\caption{Basic analysis on step size $s$ and buffer size $b$ when $d=9$, $p=0.005$, and numbers of cycles $n \in \{30, 40, 50, 60, 70, 80, 90\}$. Recall that the sandwich window size $w$ is $s + 2b$. We fix the buffer size to $3$ and $5$ and vary the step size. 
(a) The $x$-axis is the number of cycles and the $y$-axis is the logical error rate per shot. 
(b) Converting logical error rates per shot to logical error rates per $d$ cycles and changing the $x$-axis to the step size.}
\label{fig:basic}
\end{figure}

As the number of cycles increases, the logical error rate per shot also increases and gradually converges to $0.5$, following from \eqnref{eq:logical_error_rate_recursion}. \figref{fig:basic} gives an example for $d=9$ and $p=0.005$. Generally speaking, a larger buffer size, as well as a larger step size when the buffer size is fixed, results in lower logical error rates. 

\subsection{Sandwich vs. forward windows}

During each sandwich window, since we only accept the corrections in the middle, the decoder is given information of detection events from both the future and the past. The forward window (also LILLIPUT) can prevent premature matchings by taking account into its most recent future detection events. However, it sometimes fails to prevent problematic matchings from the past because the decoder's knowledge of its most recent past events is limited to a single layer of propagated syndromes.

\begin{figure}[!ht]
\centering
\begin{tabular}{cc}
    \subfigure[\label{fig:sandwich_forward_s_4}]{
    \raisebox{.2cm}{\includegraphics[scale=.46]{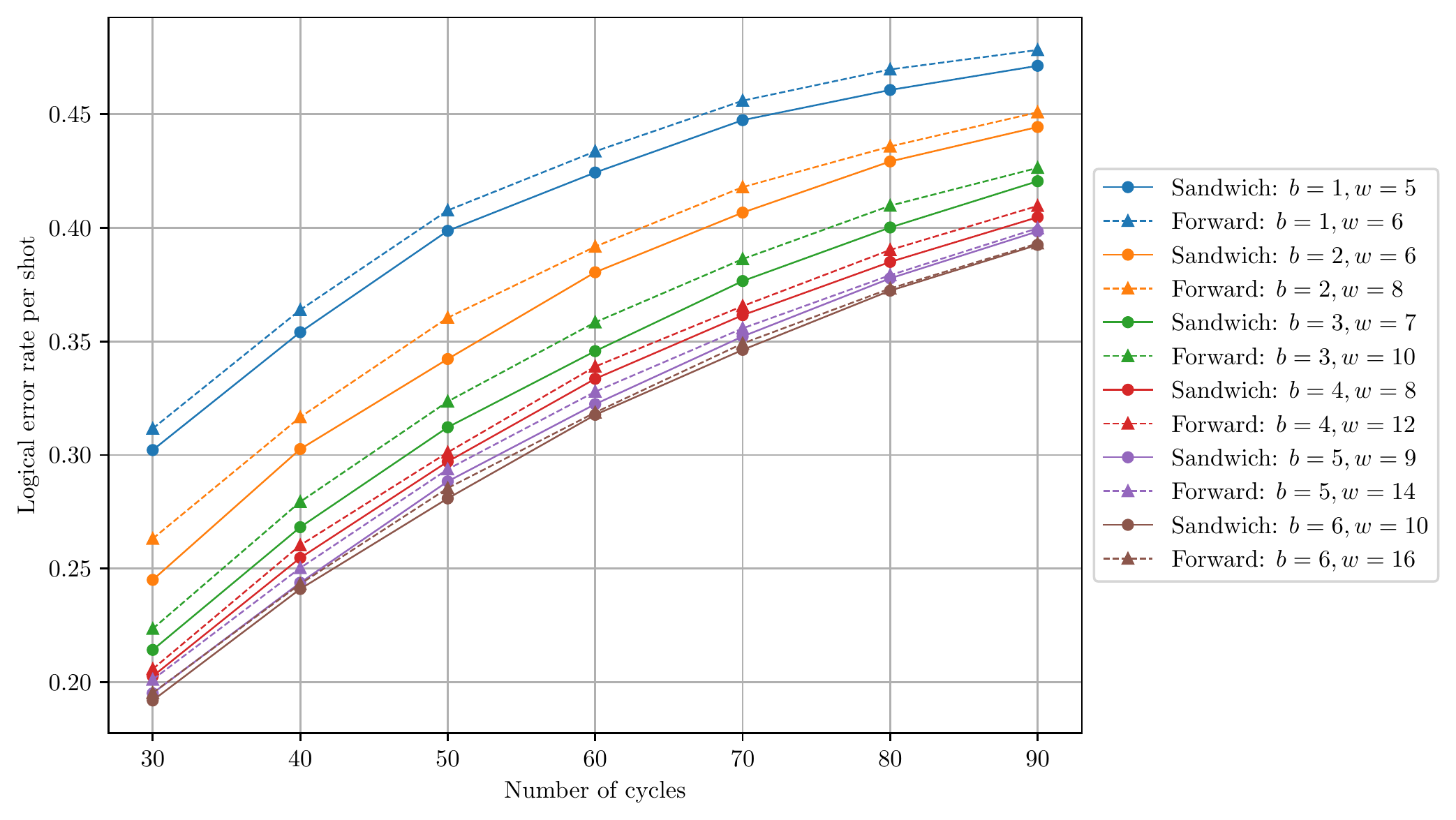}}}
    \subfigure[\label{fig:sandwich_forward_s_4_avg}]{
    \raisebox{.2cm}{\includegraphics[scale=.46]{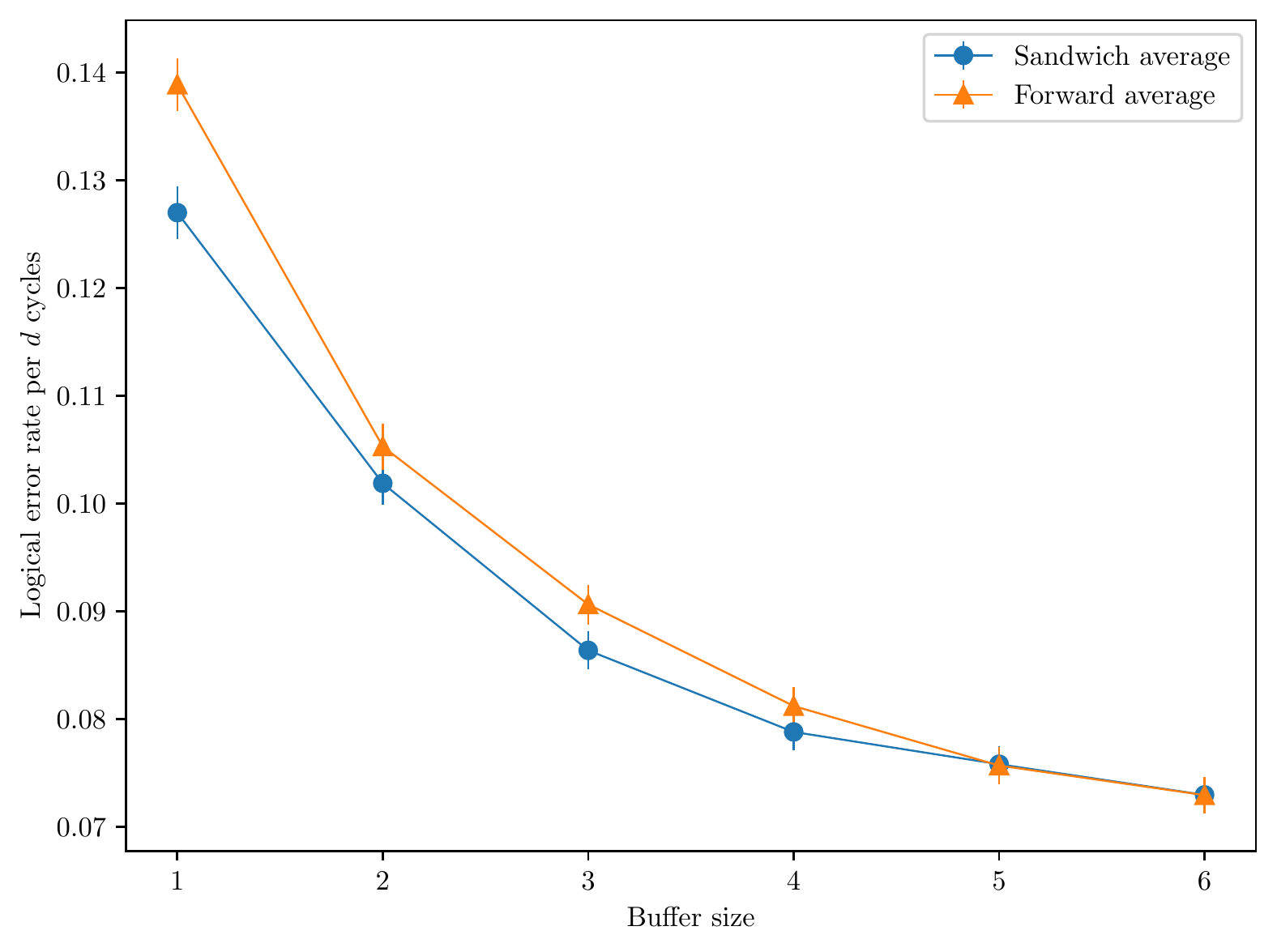}}}
\end{tabular}
\caption{Comparison between forward windows and sandwich windows. We fix $d=9$, $p=0.005$, and step size $s=4$. The forward window size is $s + b$ and the sandwich window size is $s+2b$. }
\label{fig:sandwich_forward}
\end{figure}

Indeed, we show in \figref{fig:sandwich_forward} that when we release the same amount of future events to both sandwich window and forward window, \emph{i.e.}, fix the step size and the size of buffer region(s), the sandwich windows produce lower logical error rates. Moreover, it means that given any forward decoder, we can construct a sandwich decoder such that it performs comparatively good or even better. 

\subsection{Open vs. closed time boundaries}\label{subsec:open_vs_closed}
As mentioned in \secref{subsec:forward}, it is unclear in LILLIPUT whether the future boundaries of windows are generally open or closed. Therefore we do a series of simulation experiments where we close all the time boundaries in windows and compare the performance with the normal case where only the past (left) time boundary of the first window and the future (right) time boundary of the last window are closed. Open boundaries have obvious advantage over closed boundaries. As we increase the buffer size, the advantage becomes less obvious. 

\begin{figure}[!ht]
    \centering
    \begin{tabular}{cc}
    \subfigure[\label{fig:forward_closed}]{
    \raisebox{.2cm}{\includegraphics[scale=.47]{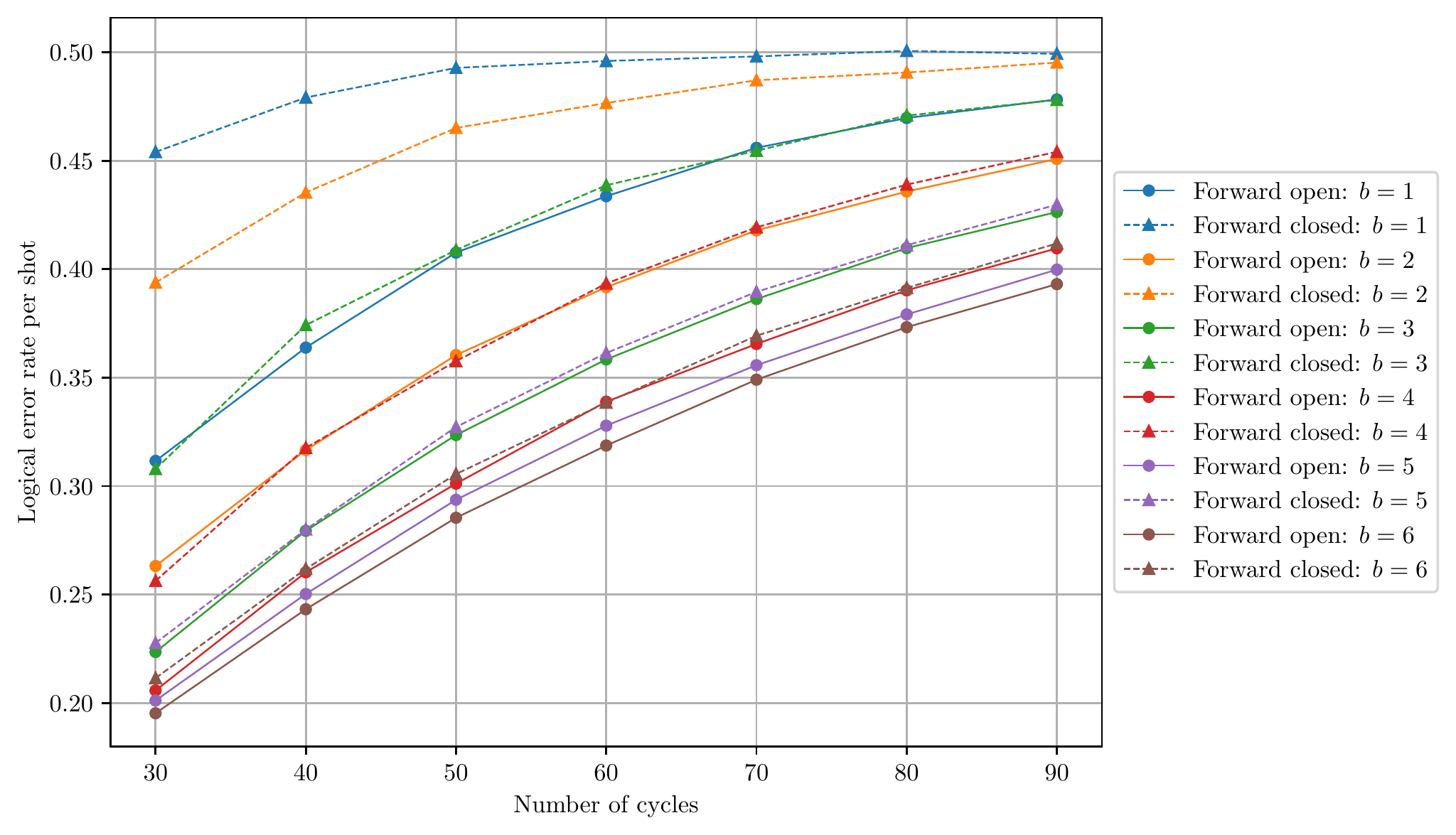}}}

    \subfigure[\label{fig:forward_closed_avg}]{
    \raisebox{.2cm}{\includegraphics[scale=.47]{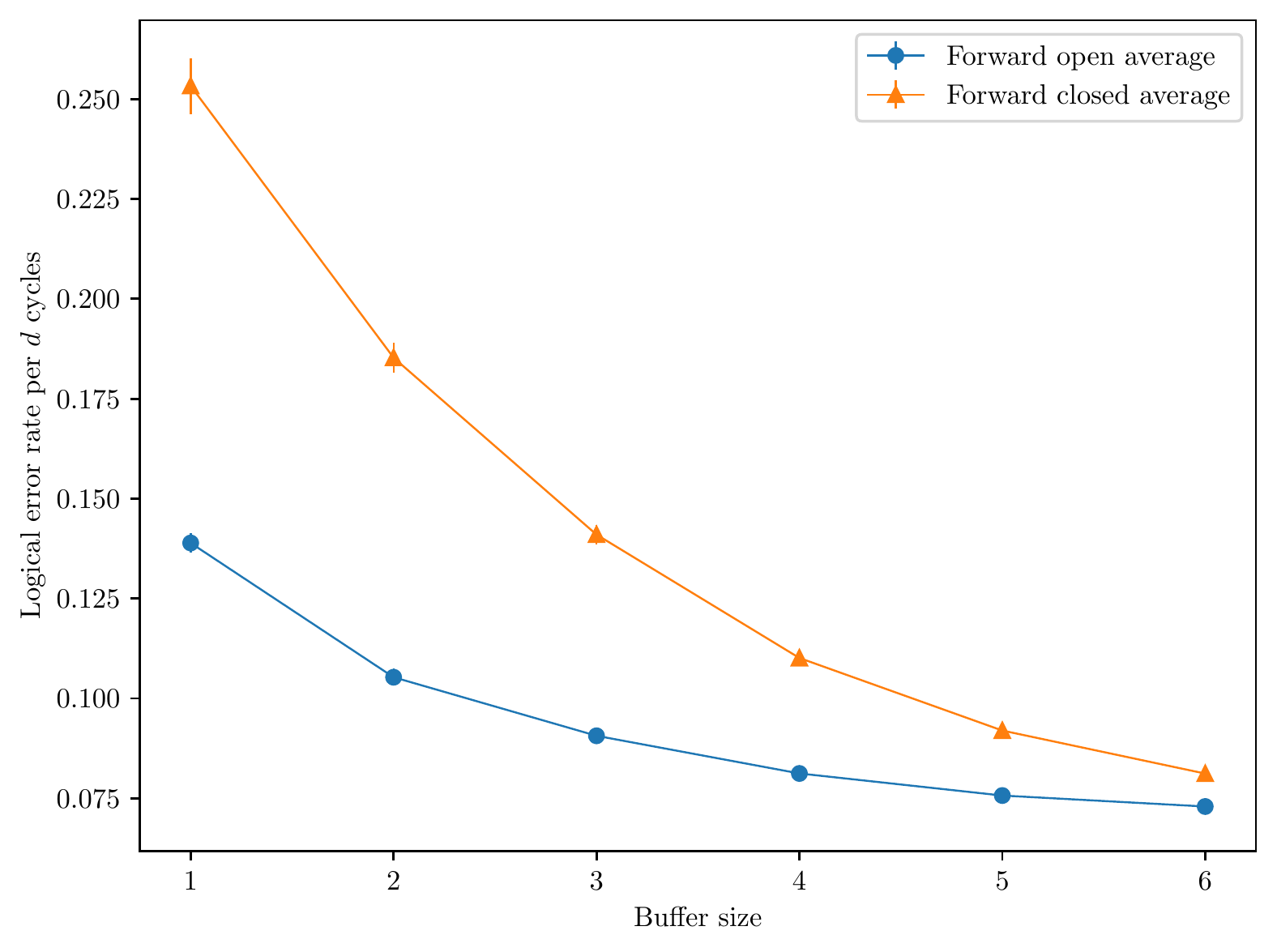}}}
    \\
    \subfigure[\label{fig:sandwich_closed}]{
    \raisebox{.2cm}{\includegraphics[scale=.47]{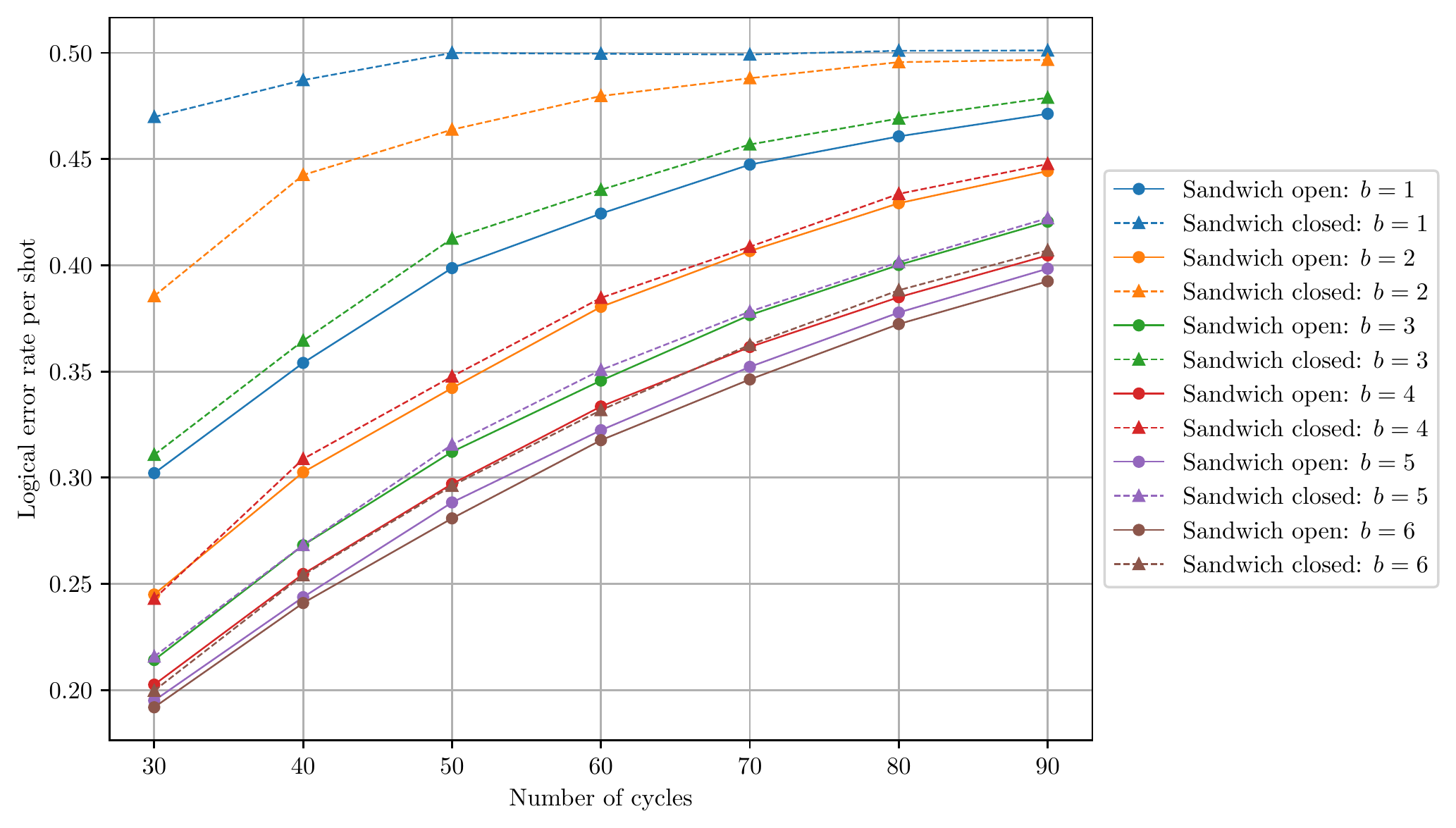}}}

    \subfigure[\label{fig:sandwich_closed_avg}]{
    \raisebox{.2cm}{\includegraphics[scale=.47]{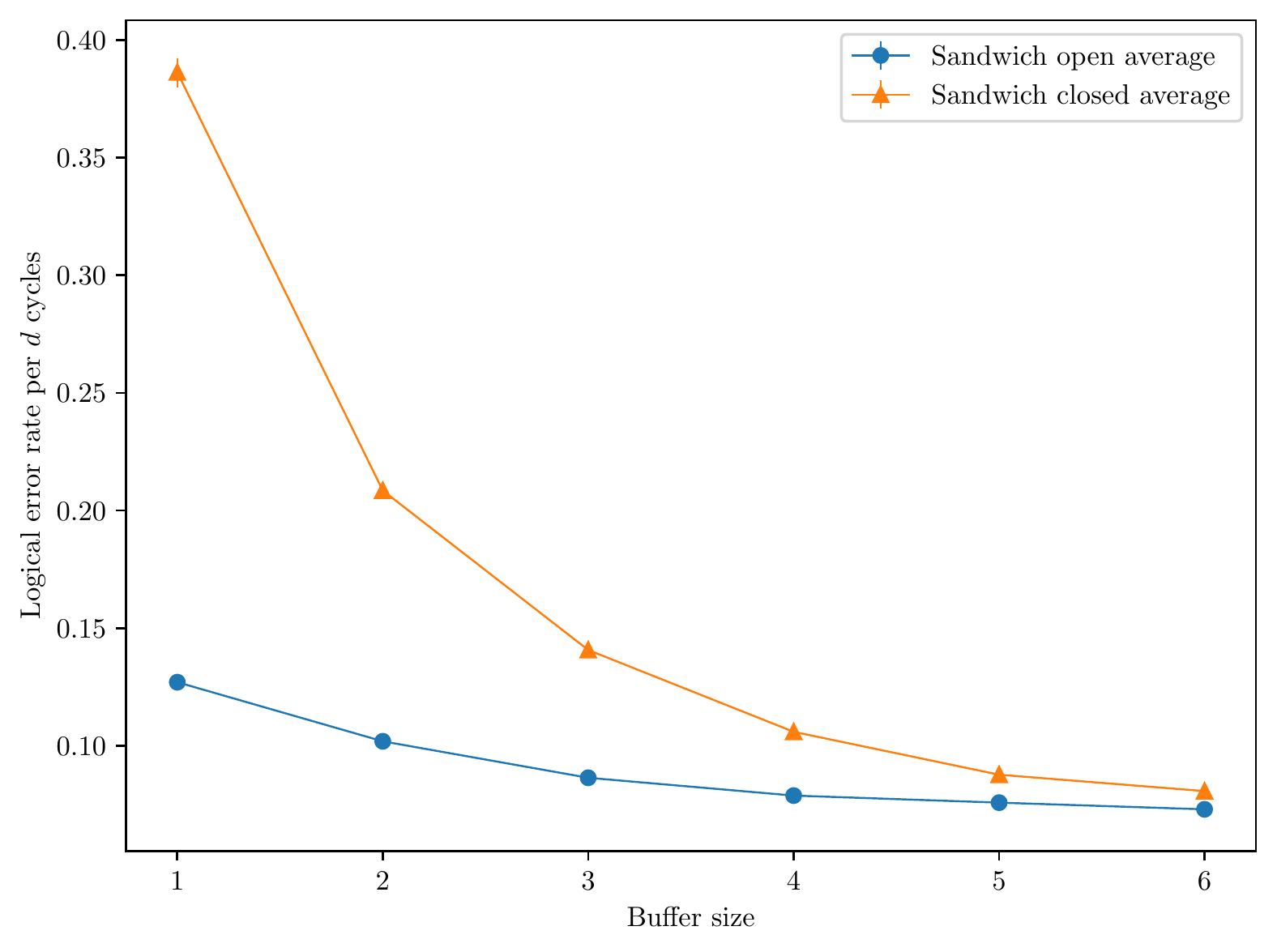}}}
    \end{tabular}
    \caption{Comparison between open and closed artificial time boundaries for forward windows. We fix $d=9$, $p=0.005$, and $s=4$. }
    \label{fig:closed}
\end{figure}

\subsection{Seam offset for inconsistent corrections}
\begin{figure}[!ht]
\begin{tabular}{cc}
    \subfigure[\label{fig:vary_seam_offset}]{
    \raisebox{.2cm}{\includegraphics[scale=.51]{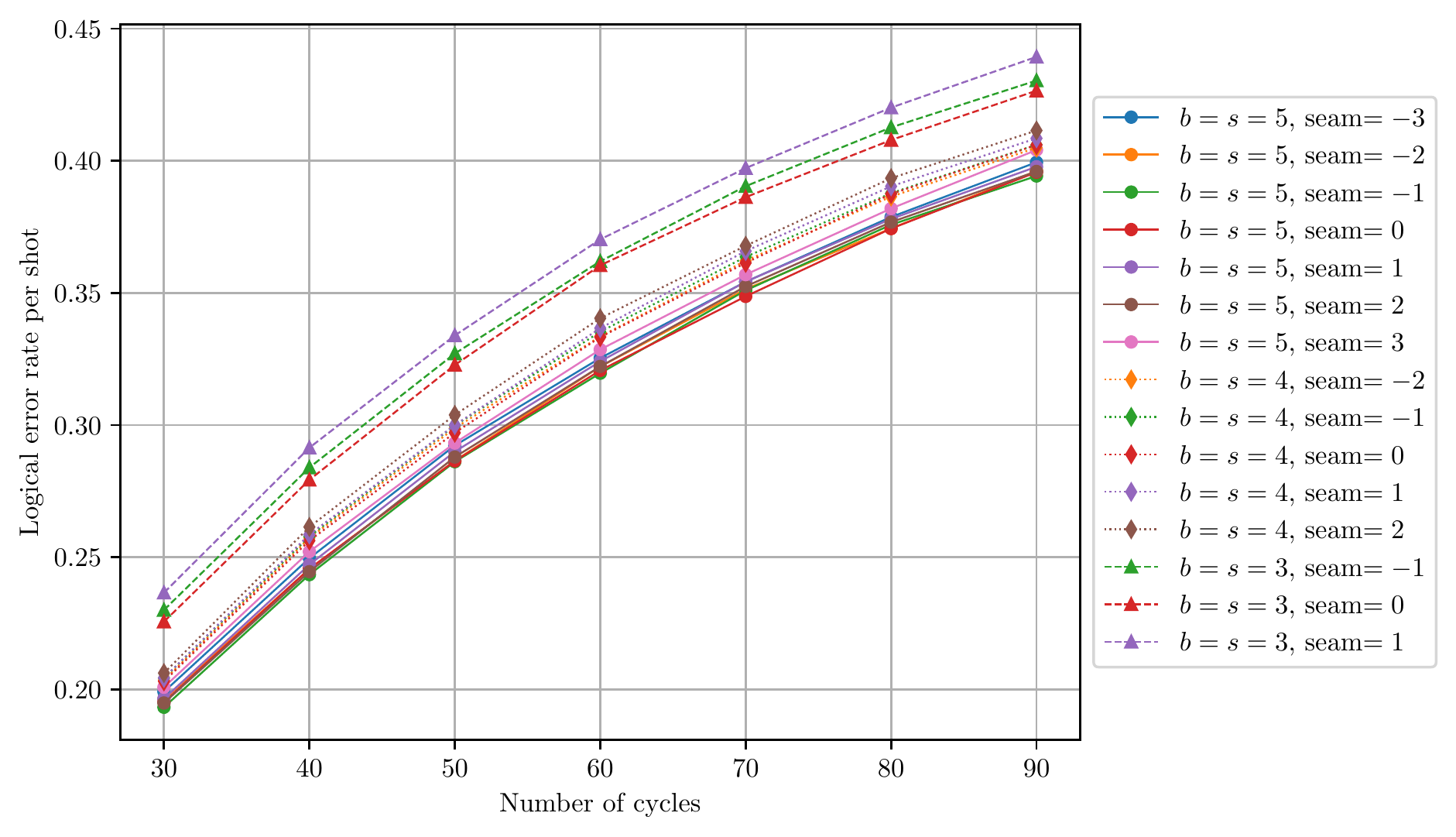}}}
    \subfigure[\label{fig:vary_seam_offset_avg}]{
    \raisebox{.2cm}{\includegraphics[scale=.51]{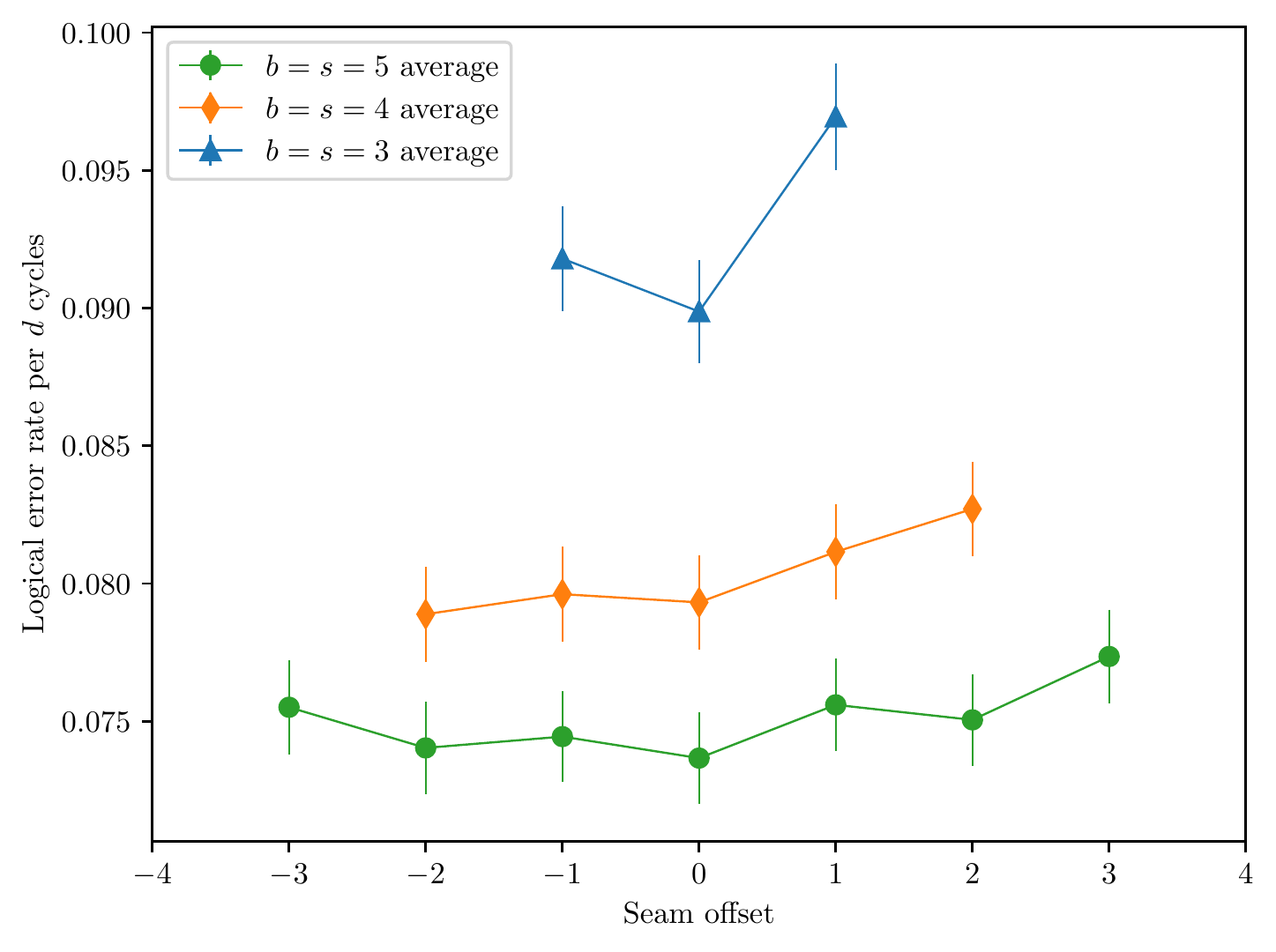}}}
\end{tabular}
\caption{Comparison between different seam offsets for sandwich decoders. We fix $d=9$ and $p=0.005$.}
\label{fig:seam_offset}
\end{figure}

In \figref{fig:seam_offset}, we study the effect of having different seam offsets (\figref{fig:seam_offset_ref}). The results seem to vaguely indicate that our choice of seam offset $0$ is actually the best or close to the best value, although the difference is small and the data is far from conclusive.

\subsection{Real-world data from Google QEC experiments}

\begin{figure}[!ht]
    \centering
    \includegraphics[width=.5\textwidth]{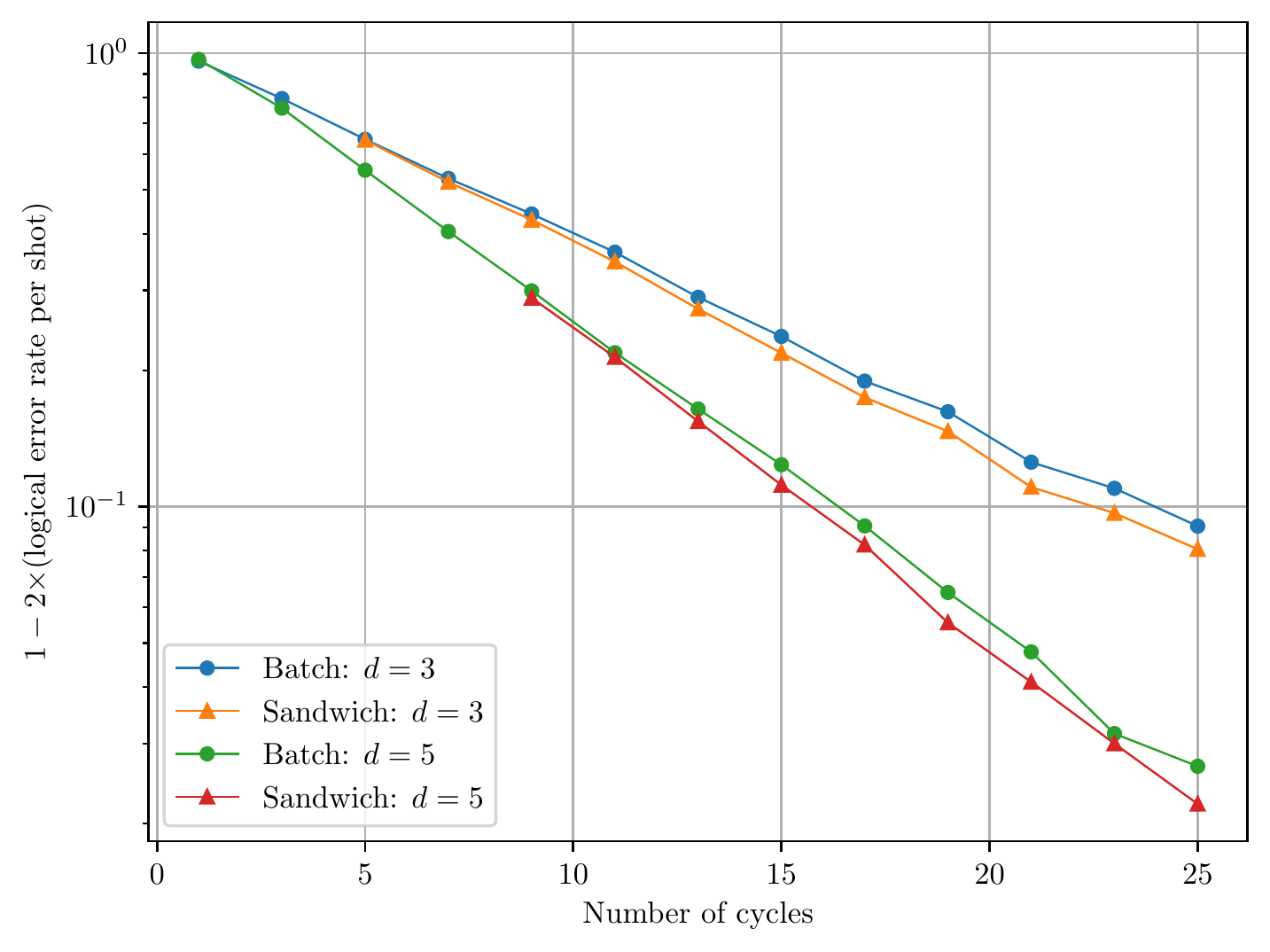}
    \caption{Comparison between the UF sandwich decoder and the UF batch decoder on Google QEC data. For $d=3$, each point is the average of $4$ configurations of the surface codes as specified in \cite{sm_acharya2022google}. There is only one configuration when $d=5$. Each configuration has $50,000$ shots. The sandwich decoder we choose has a step size of $(d+1)/2$ and a window size of $3(d+1)/2$. We plot in a style similar to Figure 3 in the main text of \cite{sm_acharya2022google}, where the $y$-axis is the logical fidelity $1-2\PL(n)$ and $n$ is the number of cycles. If follows from \eqnref{eq:logical_error_rate_recursion} that the slope indicates the logical error rate per cycle. }
    \label{fig:google_qec}
\end{figure}

We also evaluate the performance of the sandwich (UF) decoders on real-world data provided in \cite{sm_acharya2022google} for $d=3$ and $d=5$. As shown in \figref{fig:google_qec}, the sandwich decoders only have slightly larger logical error rates compared with that of using the batch (UF) decoders. Notice that \cite{sm_acharya2022google} used tensor network and belief decoders, which inherently have higher accuracy than UF decoders, to obtain the suppression of errors from $d=3$ to $d=5$.

\section{Future work}
\label{sec:future}
\subsection{Generalizations}
\label{sec:generalizations}
\paragraph{A general parallel divide-and-conquer algorithm}
Fix a stabilizer code and its syndrome-extraction circuit. Denote by $\mathcal V$ the set of detectors.
Then a stochastic Pauli noise model induces a (hyper)graph $(\mathcal V, \mathcal E)$ with
\begin{align}
\mathcal E=\left\{e\subseteq \mathcal V \,:\,\textrm{there is a fault that flips exactly the detectors in }e\right\}.
\end{align}
Consider the $\mathbb F_2$-linear map from the edge space to the vertex space $\partial: \mathbb F_2^{|\mathcal E|} \rightarrow \mathbb F_2^{|\mathcal V|}$,
\begin{align}
\partial E:=\sum_{e\in E}\sum_{v\in e}v,
\end{align}
where the vector addition corresponds to symmetric difference. 
For each $V\subseteq\mathcal{V}$ and $E\subseteq\mathcal{E}$, define
\begin{align}
\Delta(E, V) := \left\{e\in E\, :\, e\textrm{ incident to a vertex in }V\right\}.
\end{align}
Then, our {\em Generalized Sandwich} algorithm $\GS$ takes as the input the graph $(\mathcal{V}, \mathcal{E})$ and a set $D\subseteq\mathcal{V}$ of defects, outputs a set $K\subseteq\mathcal{E}$ of corrections, so that $\partial K =  D$. 
See Algorithm~\ref{alg:GS}.

The algorithm assumes that there is an ``inner decoder'', which is applied to instances of a small enough size.
Given fixed defects $D$, we refer to corrections $K$ as ``valid'' if $\partial K=D$.
If the inner decoder fails to find valid corrections at any instance, the decoder terminates and declares failure.
The algorithm also assumes a ``partition method'' (\emph{e.g.}, see Section~\ref{subsec:sandwich}) for decomposing the input graph in Step~\ref{step:core}. Note that each execution of the step does not necessarily partition the vertex set, but the ``cores'' from all the recursive steps put together do for the initial input graph.  
Algorithm~\ref{alg:GS} provides one generalization of the sandwich decoder regarding disjoint core regions across windows (such as having non-negative seam offset). One can easily construct similar variant for overlapping core regions. 
Let $\bigsqcup$ denote the union of disjoint sets. 

\begin{algorithm}[H]
\caption{Generalized Sandwich Decoder $\GS(V, E, D)$}\label{alg:GS}
\begin{algorithmic}[1]
\If{$(V, E)$ consists of disconnected subgraphs, each of a small enough size}
    \State Apply the inner decoder to each subgraph in parallel, \Return the union of the outputs
\EndIf
\State\label{step:core}Apply the partition method to choose ``cores'' $\left\{C_i\subseteq   V\right\}_i$ with disjoint $\{\Delta(E, C_i)\}_i$, each of a small enough size
\State\label{step:solve_cores}Apply the inner decoder in parallel to calculate corrections $K_i\subseteq \Delta(E, C_i)$, for all $i$, with $\partial K_i\cap C_i=D\cap C_i$
\State\label{step:update}$ V' \leftarrow  V \left \backslash \bigsqcup_i C_i \right. ,  \quad
 E'  \leftarrow  E \left \backslash \bigsqcup_i \Delta(E,C_i) \right. , \quad
D' \leftarrow D+\partial \left( \bigsqcup_i K_i   \right)$
\State \Return $\GS( V',  E', D') \sqcup \bigsqcup_i K_i$
\end{algorithmic}
\end{algorithm}

Algorithm~\ref{alg:GS} always terminates after finite recursions since $|\bar{V}|$ is strictly increasing. 
The emphasis of the algorithm is on how to parallelize the computation. It does not address the chance of failure, or the chance of logical error when valid corrections are output. Thus, its success and accuracy requires additional theoretical or empirical justifications. When it does not fail, however, the output is valid. To see this, first observe that Step~\ref{step:update} defines a valid input instance, i.e., all vertices involved in $E'$ are in $V'$, and $D'\subseteq  V'$. The former follows from the definition of $V'$ and $E'$. To see the latter, note
\begin{equation}
D':=D+\partial\left(\bigsqcup_i K_i\right) = D+ \sum_i\partial (K_i) 
=  V'\cap D + \sum_i\left[\left(D\cap C_i\right) + C_i\cap \partial  (K_i) + \bar C_i\cap\partial (K_i) \right],
\end{equation}
where ${\bar C_i}$ denotes the set complement of $C_i$. It then follows from $\partial K_i\cap C_i=D\cap C_i$ as required in Step~\ref{step:solve_cores} that
\begin{equation}
     V'\cap D + \sum_i \left(\bar C_i\cap \partial(K_i)\right).
\end{equation}
Note that for all $i$, $\bar C_i\cap\partial(K_i)\subseteq  V'$, for otherwise there will be a $j$, $j\ne i$, such that $\Delta(E,C_i)\cap \Delta(E,C_j)$ has an element in $\partial (K_i)$, a contradiction to the disjointness of $\left\{ \Delta(E,C_i)\right\}_i$. Therefore $D'\subseteq  V'$.

By an inductive argument,
\begin{equation}
\partial\left( \GS( V',  E', D') \sqcup \bigsqcup_i K_i\right) = 
\partial\left(\GS( V',  E', D')\right) + \partial\left(\bigsqcup_i  K_i\right) = D' + \partial\left(\bigsqcup_i K_i\right) = D.
\end{equation} 
Thus, the algorithm always outputs valid corrections, provided it does not fail.

The algorithm also leaves significant freedom in choosing the inner decoder and the partition method. Our sandwich decoder partitions the input graph along the time direction, which disconnects the graph, resulting in a depth-2 recursion. It also guarantees success based on the graph properties of the windows. An alternative inner decoder to what we experiment with is pre-computed lookup tables, when one sets the base input size to be small enough. It would be interesting to explore and evaluate the many design choices.

\paragraph{Stability experiment}
One omission from this paper is the \emph{stability experiment} as described in \cite{sm_gidney2022stability}. A technical problem we have encountered when trying to apply our scheme to the stability experiment is that, since the surface code patch used in the stability experiment has closed space boundaries on all sides, the decoder graph for a type-$2$ window would not have any open boundary at all. Yet it can still get an odd number of defects as the input if the two adjacent windows yield completely different corrections. Furthermore, it takes only $O(w)$ errors to cause such an irreconcilable consistency, whereas the stability experiment is supposed to be able to tolerate any $O(n)$ errors (remember that in our notations, $n$ is the total number of cycles in the experiment and $w$ is the number of cycles in a window).

One may argue that this is probably an inherent disadvantage of sliding-window schemes: The forward-window scheme does not have a decoder graph without any open boundary, but $O(w)$ errors can cause a logical error instead. However, the real and also more interesting question is whether it is meaningful to divide the stability experiment into different windows \emph{in the direction of time}. 

We note that one of the main motivations for the stability experiment is to emulate the ``space-like parts'' that arise in various useful logical operations with lattice surgery, such as moving a qubit or doing a two-qubit parity measurement. 
Ideally, each of those ``space-like part'' should last only for $O(d)$ surface code cycles, since adding more cycles has diminishing returns for suppressing time-like logical errors and is detrimental for suppressing space-like logical errors (of the opposite X/Z type). 
Therefore, it makes less sense to divide the stability experiment into windows by time, as opposed to the memory experiment, which in practical scenarios can last much more than $O(d)$ cycles (depending on the number of logical operations applied on a logical qubit). 
On the other hand, it is more plausible that the spatial span of a ``space-like part'' is significantly larger than $d$, depending on the physical distance on the surface code lattice between the qubits involved. 
Thus, it may make more sense to consider stability experiments on an elongated rectangular code patch and divide it into windows in a spatial direction instead. Such a sliding-window decoder would probably be much easier to formulate too, as it is not fundamentally different from a sliding-window decoder for the memory experiment, only with the roles of the time and one spatial dimension switched.

\paragraph{Lattice surgery}\label{para:lattice_surgery}
By the same token, it is not a stretch to straightforwardly generalize our sliding-window decoders to some more useful operations in lattice surgery, like the aforementioned qubit movements and two-qubit parity measurements. For example, the two-qubit parity measurement has an overall ``H''-shaped decoder graph, as opposed to the rectangular box-shaped decoder graph for the memory experiment (where the box is elongated in the temporal direction) or the stability experiment (where the box is elongated in a spatial direction), but it is still straightforward to divide the graph into 3D windows each with dimensions $O(d)\times O(d)\times O(d)$, as shown in \figref{fig:lattice_surgery}.
Two ``T''-shaped windows simply need to propagate seam syndromes in three directions instead of two. 
Other lattice surgery operations may add more complexity to the scheme---for example, a twist defect may require combining the $X$ decoder graph and the $Z$ decoder graph in some way---but it seems that the same principle should be able to handle everything.

\newcommand{\centered}[1]{\begin{tabular}{l} #1 \end{tabular}}

\begin{figure}[!ht]
\begin{tabular}[c]{ccc}
    \centered{\subfigure[\label{fig:lattice_surgery_1}]{
    \raisebox{.2cm}{\includegraphics[scale=.16]{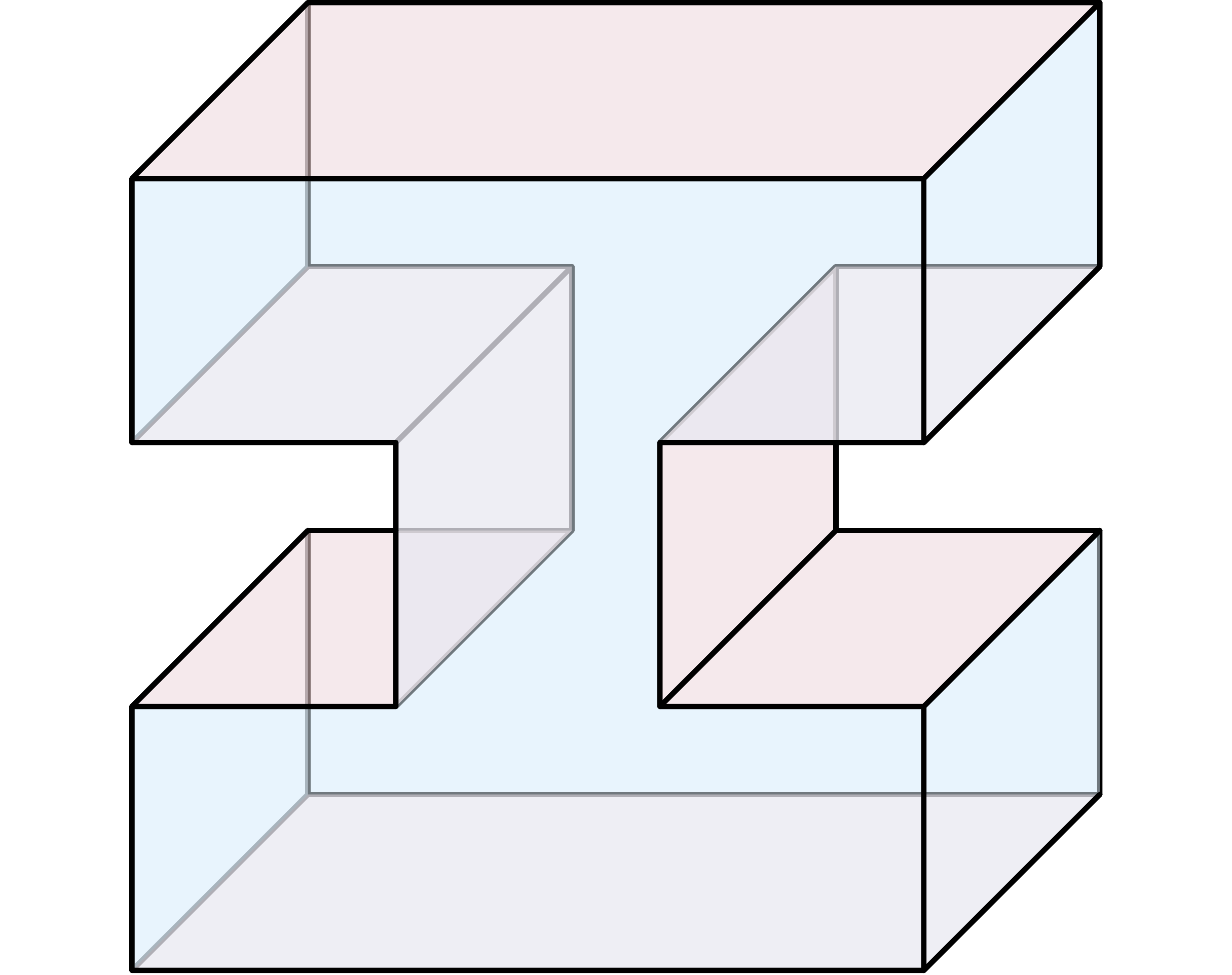}}}}
    \quad
    &
    \centered{\LARGE $\rightarrow$}
    &
\quad
    \centered{\subfigure[\label{fig:lattice_surgery_2}]{
    \raisebox{.2cm}{\includegraphics[scale=.16]{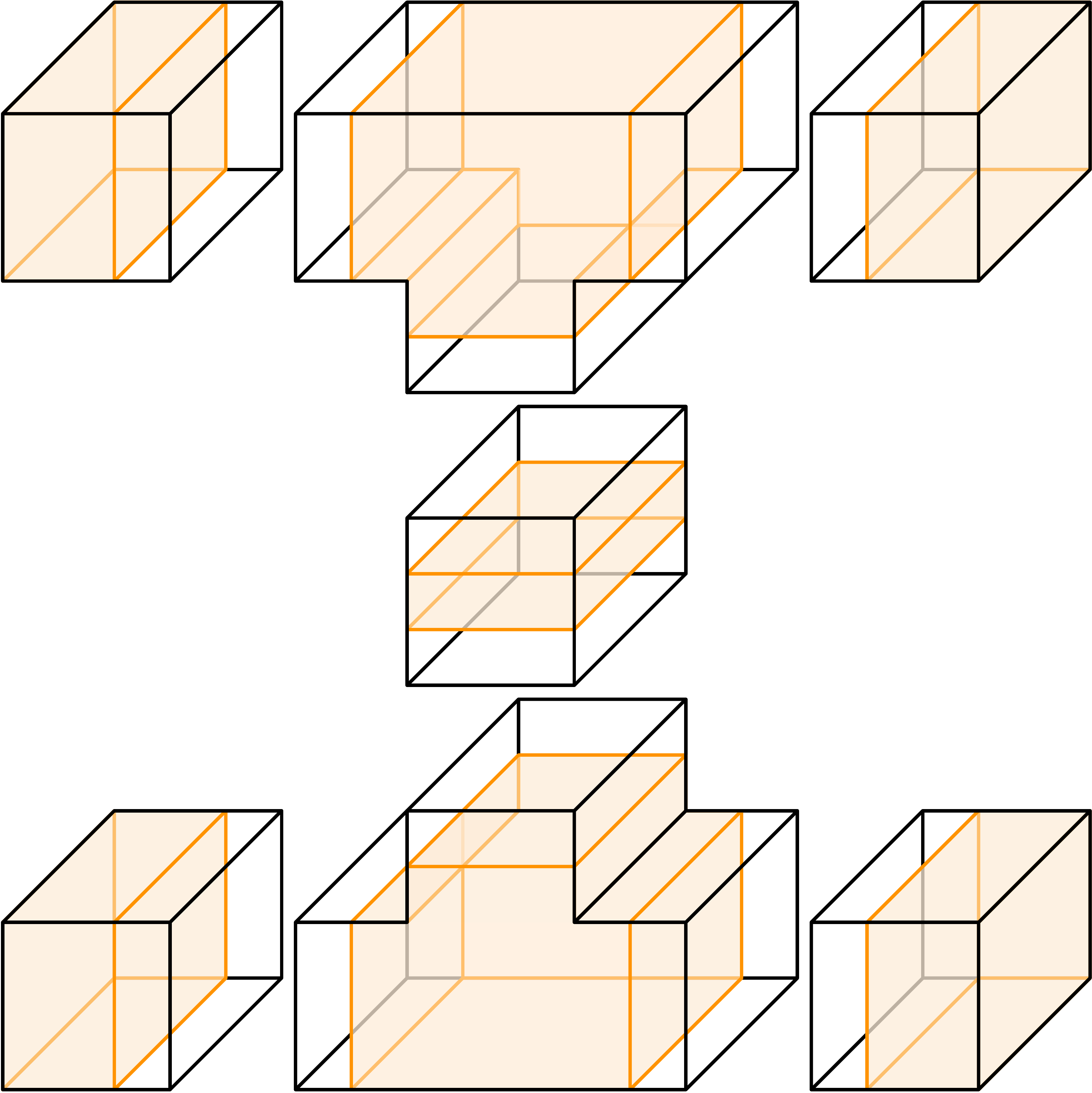}}}}
\end{tabular}
\caption{Illustration of how our scheme can be potentially extended to lattice surgery operations such as the two-qubit measurement. (a) The shape of the 3D decoder graph for an experiment containing a two-qubit measurement. (b) A way to divide this decoder graph into windows, with the core and buffer regions for each window illustrated. Each window has size $O(d)\times O(d)\times O(d)$.}
\label{fig:lattice_surgery}
\end{figure}

\subsection{Theoretical analysis}
This paper analyzes sliding-window decoders mainly via simulation experiments accompanied by some intuitive justifications for some of our design choices. Many of those choices have been made because the alternative would cause the \emph{effective code distance} to be less than $d$. For example, without any form of syndrome propagation, and assuming that the underlying decoder is a black box MWPM decoder which may output any correction with minimum weight, a constant-weight error would be enough to make the correction inconsistent and flip any logical operator representative. Similarly, without an adequate buffer region between windows, there exists an error with weight $d/4+O(1)$ that causes a logical error, making the effective code distance $d/2+O(1)$ instead of $d$.

However, we have not proven that the effective code distance of our current proposed scheme is indeed $d$. A rigorous proof would not only further justify our scheme, but also potentially uncover some alternative choices that could have been made without breaking the effective code distance guarantee.

Of course, the effective code distance is only half of the story, and arguably it is the less realistic half since surface codes are not usually supposed to work in the regime where the total number of faults is expected to be less than half of the code distance. 
Instead, the main strength of surface codes is their ability to correct most of the possible error configurations even when the total number of faults is much larger than the code distance. This can be well captured by the code threshold. A theoretical lower bound of the threshold of our scheme, even a very loose one, would be an interesting result.


\begin{thebibliography}{37}%
\makeatletter
\providecommand \@ifxundefined [1]{%
 \@ifx{#1\undefined}
}%
\providecommand \@ifnum [1]{%
 \ifnum #1\expandafter \@firstoftwo
 \else \expandafter \@secondoftwo
 \fi
}%
\providecommand \@ifx [1]{%
 \ifx #1\expandafter \@firstoftwo
 \else \expandafter \@secondoftwo
 \fi
}%
\providecommand \natexlab [1]{#1}%
\providecommand \enquote  [1]{``#1''}%
\providecommand \bibnamefont  [1]{#1}%
\providecommand \bibfnamefont [1]{#1}%
\providecommand \citenamefont [1]{#1}%
\providecommand \href@noop [0]{\@secondoftwo}%
\providecommand \href [0]{\begingroup \@sanitize@url \@href}%
\providecommand \@href[1]{\@@startlink{#1}\@@href}%
\providecommand \@@href[1]{\endgroup#1\@@endlink}%
\providecommand \@sanitize@url [0]{\catcode `\\12\catcode `\$12\catcode
  `\&12\catcode `\#12\catcode `\^12\catcode `\_12\catcode `\%12\relax}%
\providecommand \@@startlink[1]{}%
\providecommand \@@endlink[0]{}%
\providecommand \url  [0]{\begingroup\@sanitize@url \@url }%
\providecommand \@url [1]{\endgroup\@href {#1}{\urlprefix }}%
\providecommand \urlprefix  [0]{URL }%
\providecommand \Eprint [0]{\href }%
\providecommand \doibase [0]{https://doi.org/}%
\providecommand \selectlanguage [0]{\@gobble}%
\providecommand \bibinfo  [0]{\@secondoftwo}%
\providecommand \bibfield  [0]{\@secondoftwo}%
\providecommand \translation [1]{[#1]}%
\providecommand \BibitemOpen [0]{}%
\providecommand \bibitemStop [0]{}%
\providecommand \bibitemNoStop [0]{.\EOS\space}%
\providecommand \EOS [0]{\spacefactor3000\relax}%
\providecommand \BibitemShut  [1]{\csname bibitem#1\endcsname}%
\let\auto@bib@innerbib\@empty
\bibitem [{\citenamefont {Aharonov}\ and\ \citenamefont
  {Ben-Or}(2008)}]{aharonov2008fault}%
  \BibitemOpen
  \bibfield  {author} {\bibinfo {author} {\bibfnamefont {D.}~\bibnamefont
  {Aharonov}}\ and\ \bibinfo {author} {\bibfnamefont {M.}~\bibnamefont
  {Ben-Or}},\ }\bibfield  {title} {\bibinfo {title} {Fault-tolerant quantum
  computation with constant error rate},\ }\href
  {https://doi.org/10.1137/S0097539799359385} {\bibfield  {journal} {\bibinfo
  {journal} {{SIAM} J. Comput.}\ }\textbf {\bibinfo {volume} {38}},\ \bibinfo
  {pages} {1207} (\bibinfo {year} {2008})},\ \Eprint
  {https://arxiv.org/abs/arXiv:quant-ph/9906129} {arXiv:quant-ph/9906129}
  \BibitemShut {NoStop}%
\bibitem [{\citenamefont {Knill}(2005)}]{knill2005quantum}%
  \BibitemOpen
  \bibfield  {author} {\bibinfo {author} {\bibfnamefont {E.}~\bibnamefont
  {Knill}},\ }\bibfield  {title} {\bibinfo {title} {Quantum computing with
  realistically noisy devices},\ }\href {https://doi.org/10.1038/nature03350}
  {\bibfield  {journal} {\bibinfo  {journal} {Nature}\ }\textbf {\bibinfo
  {volume} {434}},\ \bibinfo {pages} {39} (\bibinfo {year} {2005})},\ \Eprint
  {https://arxiv.org/abs/arXiv:quant-ph/0410199} {arXiv:quant-ph/0410199}
  \BibitemShut {NoStop}%
\bibitem [{\citenamefont {Aliferis}\ \emph {et~al.}(2006)\citenamefont
  {Aliferis}, \citenamefont {Gottesman},\ and\ \citenamefont
  {Preskill}}]{aliferis2005quantum}%
  \BibitemOpen
  \bibfield  {author} {\bibinfo {author} {\bibfnamefont {P.}~\bibnamefont
  {Aliferis}}, \bibinfo {author} {\bibfnamefont {D.}~\bibnamefont
  {Gottesman}},\ and\ \bibinfo {author} {\bibfnamefont {J.}~\bibnamefont
  {Preskill}},\ }\bibfield  {title} {\bibinfo {title} {Quantum accuracy
  threshold for concatenated distance-3 codes},\ }\href
  {https://doi.org/10.26421/qic6.2-1} {\bibfield  {journal} {\bibinfo
  {journal} {Quantum Info. Comput.}\ }\textbf {\bibinfo {volume} {6}},\
  \bibinfo {pages} {97} (\bibinfo {year} {2006})},\ \Eprint
  {https://arxiv.org/abs/arXiv:quant-ph/0504218} {arXiv:quant-ph/0504218}
  \BibitemShut {NoStop}%
\bibitem [{\citenamefont {Gottesman}(2014)}]{gottesman2013fault}%
  \BibitemOpen
  \bibfield  {author} {\bibinfo {author} {\bibfnamefont {D.}~\bibnamefont
  {Gottesman}},\ }\bibfield  {title} {\bibinfo {title} {Fault-tolerant quantum
  computation with constant overhead},\ }\href
  {https://doi.org/10.26421/qic14.15-16-5} {\bibfield  {journal} {\bibinfo
  {journal} {Quantum Info. Comput.}\ }\textbf {\bibinfo {volume} {14}},\
  \bibinfo {pages} {1338} (\bibinfo {year} {2014})},\ \Eprint
  {https://arxiv.org/abs/arXiv:1310.2984} {arXiv:1310.2984} \BibitemShut
  {NoStop}%
\bibitem [{\citenamefont {Dennis}\ \emph {et~al.}(2002)\citenamefont {Dennis},
  \citenamefont {Kitaev}, \citenamefont {Landahl},\ and\ \citenamefont
  {Preskill}}]{dennis2002topological}%
  \BibitemOpen
  \bibfield  {author} {\bibinfo {author} {\bibfnamefont {E.}~\bibnamefont
  {Dennis}}, \bibinfo {author} {\bibfnamefont {A.}~\bibnamefont {Kitaev}},
  \bibinfo {author} {\bibfnamefont {A.}~\bibnamefont {Landahl}},\ and\ \bibinfo
  {author} {\bibfnamefont {J.}~\bibnamefont {Preskill}},\ }\bibfield  {title}
  {\bibinfo {title} {Topological quantum memory},\ }\href
  {https://doi.org/10.1063/1.1499754} {\bibfield  {journal} {\bibinfo
  {journal} {J. Math. Phys.}\ }\textbf {\bibinfo {volume} {43}},\ \bibinfo
  {pages} {4452} (\bibinfo {year} {2002})},\ \Eprint
  {https://arxiv.org/abs/arXiv:quant-ph/0110143} {arXiv:quant-ph/0110143}
  \BibitemShut {NoStop}%
\bibitem [{\citenamefont {Fowler}\ \emph
  {et~al.}(2012{\natexlab{a}})\citenamefont {Fowler}, \citenamefont
  {Mariantoni}, \citenamefont {Martinis},\ and\ \citenamefont
  {Cleland}}]{fowler2012surface}%
  \BibitemOpen
  \bibfield  {author} {\bibinfo {author} {\bibfnamefont {A.~G.}\ \bibnamefont
  {Fowler}}, \bibinfo {author} {\bibfnamefont {M.}~\bibnamefont {Mariantoni}},
  \bibinfo {author} {\bibfnamefont {J.~M.}\ \bibnamefont {Martinis}},\ and\
  \bibinfo {author} {\bibfnamefont {A.~N.}\ \bibnamefont {Cleland}},\
  }\bibfield  {title} {\bibinfo {title} {Surface codes: Towards practical
  large-scale quantum computation},\ }\href
  {https://doi.org/10.1103/PhysRevA.86.032324} {\bibfield  {journal} {\bibinfo
  {journal} {Phys. Rev. A}\ }\textbf {\bibinfo {volume} {86}},\ \bibinfo
  {pages} {032324} (\bibinfo {year} {2012}{\natexlab{a}})},\ \Eprint
  {https://arxiv.org/abs/arXiv:1208.0928} {arXiv:1208.0928} \BibitemShut
  {NoStop}%
\bibitem [{\citenamefont {Kitaev}(2003)}]{kitaev2003fault}%
  \BibitemOpen
  \bibfield  {author} {\bibinfo {author} {\bibfnamefont {A.}~\bibnamefont
  {Kitaev}},\ }\bibfield  {title} {\bibinfo {title} {Fault-tolerant quantum
  computation by anyons},\ }\href
  {https://doi.org/https://doi.org/10.1016/S0003-4916(02)00018-0} {\bibfield
  {journal} {\bibinfo  {journal} {Ann. Phys.}\ }\textbf {\bibinfo {volume}
  {303}},\ \bibinfo {pages} {2} (\bibinfo {year} {2003})}\BibitemShut {NoStop}%
\bibitem [{\citenamefont {Bravyi}\ and\ \citenamefont
  {Kitaev}(2005)}]{bravyi2005universal}%
  \BibitemOpen
  \bibfield  {author} {\bibinfo {author} {\bibfnamefont {S.}~\bibnamefont
  {Bravyi}}\ and\ \bibinfo {author} {\bibfnamefont {A.}~\bibnamefont
  {Kitaev}},\ }\bibfield  {title} {\bibinfo {title} {Universal quantum
  computation with ideal {C}lifford gates and noisy ancillas},\ }\href
  {https://doi.org/10.1103/PhysRevA.71.022316} {\bibfield  {journal} {\bibinfo
  {journal} {Phys. Rev. A}\ }\textbf {\bibinfo {volume} {71}},\ \bibinfo
  {pages} {022316} (\bibinfo {year} {2005})},\ \Eprint
  {https://arxiv.org/abs/arXiv:quant-ph/0403025} {arXiv:quant-ph/0403025}
  \BibitemShut {NoStop}%
\bibitem [{\citenamefont {Terhal}(2015)}]{terhal2015quantum}%
  \BibitemOpen
  \bibfield  {author} {\bibinfo {author} {\bibfnamefont {B.~M.}\ \bibnamefont
  {Terhal}},\ }\bibfield  {title} {\bibinfo {title} {Quantum error correction
  for quantum memories},\ }\href {https://doi.org/10.1103/RevModPhys.87.307}
  {\bibfield  {journal} {\bibinfo  {journal} {Rev. Mod. Phys.}\ }\textbf
  {\bibinfo {volume} {87}},\ \bibinfo {pages} {307} (\bibinfo {year} {2015})},\
  \Eprint {https://arxiv.org/abs/arXiv:1302.3428} {arXiv:1302.3428}
  \BibitemShut {NoStop}%
\bibitem [{\citenamefont {Raussendorf}\ and\ \citenamefont
  {Harrington}(2007)}]{raussendorf2007fault}%
  \BibitemOpen
  \bibfield  {author} {\bibinfo {author} {\bibfnamefont {R.}~\bibnamefont
  {Raussendorf}}\ and\ \bibinfo {author} {\bibfnamefont {J.}~\bibnamefont
  {Harrington}},\ }\bibfield  {title} {\bibinfo {title} {Fault-tolerant quantum
  computation with high threshold in two dimensions},\ }\href
  {https://doi.org/10.1103/PhysRevLett.98.190504} {\bibfield  {journal}
  {\bibinfo  {journal} {Phys. Rev. Lett.}\ }\textbf {\bibinfo {volume} {98}},\
  \bibinfo {pages} {190504} (\bibinfo {year} {2007})},\ \Eprint
  {https://arxiv.org/abs/arXiv:quant-ph/0610082} {arXiv:quant-ph/0610082}
  \BibitemShut {NoStop}%
\bibitem [{\citenamefont {Fowler}\ \emph {et~al.}(2009)\citenamefont {Fowler},
  \citenamefont {Stephens},\ and\ \citenamefont
  {Groszkowski}}]{fowler2009high}%
  \BibitemOpen
  \bibfield  {author} {\bibinfo {author} {\bibfnamefont {A.~G.}\ \bibnamefont
  {Fowler}}, \bibinfo {author} {\bibfnamefont {A.~M.}\ \bibnamefont
  {Stephens}},\ and\ \bibinfo {author} {\bibfnamefont {P.}~\bibnamefont
  {Groszkowski}},\ }\bibfield  {title} {\bibinfo {title} {High-threshold
  universal quantum computation on the surface code},\ }\href
  {https://doi.org/10.1103/PhysRevA.80.052312} {\bibfield  {journal} {\bibinfo
  {journal} {Phys. Rev. A}\ }\textbf {\bibinfo {volume} {80}},\ \bibinfo
  {pages} {052312} (\bibinfo {year} {2009})},\ \Eprint
  {https://arxiv.org/abs/arXiv:0803.0272} {arXiv:0803.0272} \BibitemShut
  {NoStop}%
\bibitem [{\citenamefont {Harrington}(2004)}]{harrington2004analysis}%
  \BibitemOpen
  \bibfield  {author} {\bibinfo {author} {\bibfnamefont {J.~W.}\ \bibnamefont
  {Harrington}},\ }\emph {\bibinfo {title} {Analysis of quantum
  error-correcting codes: Symplectic lattice codes and toric codes}},\ \href
  {https://doi.org/10.7907/AHMQ-EG82} {Ph.D. thesis},\ \bibinfo  {school}
  {California Institute of Technology} (\bibinfo {year} {2004})\BibitemShut
  {NoStop}%
\bibitem [{\citenamefont {Duclos-Cianci}\ and\ \citenamefont
  {Poulin}(2010)}]{duclos2010fast}%
  \BibitemOpen
  \bibfield  {author} {\bibinfo {author} {\bibfnamefont {G.}~\bibnamefont
  {Duclos-Cianci}}\ and\ \bibinfo {author} {\bibfnamefont {D.}~\bibnamefont
  {Poulin}},\ }\bibfield  {title} {\bibinfo {title} {Fast decoders for
  topological quantum codes},\ }\href
  {https://doi.org/10.1103/PhysRevLett.104.050504} {\bibfield  {journal}
  {\bibinfo  {journal} {Phys. Rev. Lett.}\ }\textbf {\bibinfo {volume} {104}},\
  \bibinfo {pages} {050504} (\bibinfo {year} {2010})},\ \Eprint
  {https://arxiv.org/abs/arXiv:0911.0581} {arXiv:0911.0581} \BibitemShut
  {NoStop}%
\bibitem [{\citenamefont {Bravyi}\ and\ \citenamefont
  {Haah}(2013)}]{bravyi2013quantum}%
  \BibitemOpen
  \bibfield  {author} {\bibinfo {author} {\bibfnamefont {S.}~\bibnamefont
  {Bravyi}}\ and\ \bibinfo {author} {\bibfnamefont {J.}~\bibnamefont {Haah}},\
  }\bibfield  {title} {\bibinfo {title} {Quantum self-correction in the 3{D}
  cubic code model},\ }\href {https://doi.org/10.1103/PhysRevLett.111.200501}
  {\bibfield  {journal} {\bibinfo  {journal} {Phys. Rev. Lett.}\ }\textbf
  {\bibinfo {volume} {111}},\ \bibinfo {pages} {200501} (\bibinfo {year}
  {2013})},\ \Eprint {https://arxiv.org/abs/arXiv:1112.3252} {arXiv:1112.3252}
  \BibitemShut {NoStop}%
\bibitem [{\citenamefont {Fujii}\ \emph {et~al.}(2014)\citenamefont {Fujii},
  \citenamefont {Negoro}, \citenamefont {Imoto},\ and\ \citenamefont
  {Kitagawa}}]{fujii2014measurement}%
  \BibitemOpen
  \bibfield  {author} {\bibinfo {author} {\bibfnamefont {K.}~\bibnamefont
  {Fujii}}, \bibinfo {author} {\bibfnamefont {M.}~\bibnamefont {Negoro}},
  \bibinfo {author} {\bibfnamefont {N.}~\bibnamefont {Imoto}},\ and\ \bibinfo
  {author} {\bibfnamefont {M.}~\bibnamefont {Kitagawa}},\ }\bibfield  {title}
  {\bibinfo {title} {Measurement-free topological protection using dissipative
  feedback},\ }\href {https://doi.org/10.1103/PhysRevX.4.041039} {\bibfield
  {journal} {\bibinfo  {journal} {Phys. Rev. X}\ }\textbf {\bibinfo {volume}
  {4}},\ \bibinfo {pages} {041039} (\bibinfo {year} {2014})},\ \Eprint
  {https://arxiv.org/abs/arXiv:1401.6350} {arXiv:1401.6350} \BibitemShut
  {NoStop}%
\bibitem [{\citenamefont {Herold}\ \emph {et~al.}(2015)\citenamefont {Herold},
  \citenamefont {Campbell}, \citenamefont {Eisert},\ and\ \citenamefont
  {Kastoryano}}]{herold2015cellular}%
  \BibitemOpen
  \bibfield  {author} {\bibinfo {author} {\bibfnamefont {M.}~\bibnamefont
  {Herold}}, \bibinfo {author} {\bibfnamefont {E.~T.}\ \bibnamefont
  {Campbell}}, \bibinfo {author} {\bibfnamefont {J.}~\bibnamefont {Eisert}},\
  and\ \bibinfo {author} {\bibfnamefont {M.~J.}\ \bibnamefont {Kastoryano}},\
  }\bibfield  {title} {\bibinfo {title} {Cellular-automaton decoders for
  topological quantum memories},\ }\href
  {https://doi.org/10.1038/npjqi.2015.10} {\bibfield  {journal} {\bibinfo
  {journal} {npj Quantum Inf.}\ }\textbf {\bibinfo {volume} {1}},\ \bibinfo
  {pages} {1} (\bibinfo {year} {2015})},\ \Eprint
  {https://arxiv.org/abs/arXiv:1406.2338} {arXiv:1406.2338} \BibitemShut
  {NoStop}%
\bibitem [{\citenamefont {Torlai}\ and\ \citenamefont
  {Melko}(2017)}]{torlai2017neural}%
  \BibitemOpen
  \bibfield  {author} {\bibinfo {author} {\bibfnamefont {G.}~\bibnamefont
  {Torlai}}\ and\ \bibinfo {author} {\bibfnamefont {R.~G.}\ \bibnamefont
  {Melko}},\ }\bibfield  {title} {\bibinfo {title} {Neural decoder for
  topological codes},\ }\href {https://doi.org/10.1103/PhysRevLett.119.030501}
  {\bibfield  {journal} {\bibinfo  {journal} {Phys. Rev. Lett.}\ }\textbf
  {\bibinfo {volume} {119}},\ \bibinfo {pages} {030501} (\bibinfo {year}
  {2017})},\ \Eprint {https://arxiv.org/abs/arXiv:1610.04238}
  {arXiv:1610.04238} \BibitemShut {NoStop}%
\bibitem [{\citenamefont {Holmes}\ \emph {et~al.}(2020)\citenamefont {Holmes},
  \citenamefont {Jokar}, \citenamefont {Pasandi}, \citenamefont {Ding},
  \citenamefont {Pedram},\ and\ \citenamefont {Chong}}]{holmes2020nisq+}%
  \BibitemOpen
  \bibfield  {author} {\bibinfo {author} {\bibfnamefont {A.}~\bibnamefont
  {Holmes}}, \bibinfo {author} {\bibfnamefont {M.~R.}\ \bibnamefont {Jokar}},
  \bibinfo {author} {\bibfnamefont {G.}~\bibnamefont {Pasandi}}, \bibinfo
  {author} {\bibfnamefont {Y.}~\bibnamefont {Ding}}, \bibinfo {author}
  {\bibfnamefont {M.}~\bibnamefont {Pedram}},\ and\ \bibinfo {author}
  {\bibfnamefont {F.~T.}\ \bibnamefont {Chong}},\ }\bibfield  {title} {\bibinfo
  {title} {{NISQ}+: Boosting quantum computing power by approximating quantum
  error correction},\ }in\ \href {https://doi.org/10.1109/ISCA45697.2020.00053}
  {\emph {\bibinfo {booktitle} {Proc. 47th ACM/IEEE Int. Symp. C. (ISCA)}}}\
  (\bibinfo  {publisher} {IEEE Press},\ \bibinfo {year} {2020})\ pp.\ \bibinfo
  {pages} {556--569},\ \Eprint {https://arxiv.org/abs/arXiv:2004.04794}
  {arXiv:2004.04794} \BibitemShut {NoStop}%
\bibitem [{\citenamefont {Ueno}\ \emph
  {et~al.}(2022{\natexlab{a}})\citenamefont {Ueno}, \citenamefont {Kondo},
  \citenamefont {Tanaka}, \citenamefont {Suzuki},\ and\ \citenamefont
  {Tabuchi}}]{ueno2022qulatis}%
  \BibitemOpen
  \bibfield  {author} {\bibinfo {author} {\bibfnamefont {Y.}~\bibnamefont
  {Ueno}}, \bibinfo {author} {\bibfnamefont {M.}~\bibnamefont {Kondo}},
  \bibinfo {author} {\bibfnamefont {M.}~\bibnamefont {Tanaka}}, \bibinfo
  {author} {\bibfnamefont {Y.}~\bibnamefont {Suzuki}},\ and\ \bibinfo {author}
  {\bibfnamefont {Y.}~\bibnamefont {Tabuchi}},\ }\bibfield  {title} {\bibinfo
  {title} {{QULATIS}: A quantum error correction methodology toward lattice
  surgery},\ }in\ \href {https://doi.org/10.1109/HPCA53966.2022.00028} {\emph
  {\bibinfo {booktitle} {2022 IEEE Int. S. High Perf. Comp. (HPCA)}}}\
  (\bibinfo {organization} {IEEE},\ \bibinfo {year} {2022})\ pp.\ \bibinfo
  {pages} {274--287}\BibitemShut {NoStop}%
\bibitem [{\citenamefont {Delfosse}(2020)}]{delfosse2020hierarchical}%
  \BibitemOpen
  \bibfield  {author} {\bibinfo {author} {\bibfnamefont {N.}~\bibnamefont
  {Delfosse}},\ }\href@noop {} {\bibinfo {title} {Hierarchical decoding to
  reduce hardware requirements for quantum computing},} \bibinfo {year}
  {2020},\ \Eprint {https://arxiv.org/abs/arXiv:2001.11427} {arXiv:2001.11427}
  \BibitemShut {NoStop}%
\bibitem [{\citenamefont {Meinerz}\ \emph {et~al.}(2022)\citenamefont
  {Meinerz}, \citenamefont {Park},\ and\ \citenamefont
  {Trebst}}]{meinerz2022scalable}%
  \BibitemOpen
  \bibfield  {author} {\bibinfo {author} {\bibfnamefont {K.}~\bibnamefont
  {Meinerz}}, \bibinfo {author} {\bibfnamefont {C.-Y.}\ \bibnamefont {Park}},\
  and\ \bibinfo {author} {\bibfnamefont {S.}~\bibnamefont {Trebst}},\
  }\bibfield  {title} {\bibinfo {title} {Scalable neural decoder for
  topological surface codes},\ }\href
  {https://doi.org/10.1103/PhysRevLett.128.080505} {\bibfield  {journal}
  {\bibinfo  {journal} {Phys. Rev. Lett.}\ }\textbf {\bibinfo {volume} {128}},\
  \bibinfo {pages} {080505} (\bibinfo {year} {2022})},\ \Eprint
  {https://arxiv.org/abs/arXiv:2101.07285} {arXiv:2101.07285} \BibitemShut
  {NoStop}%
\bibitem [{\citenamefont {Gicev}\ \emph {et~al.}(2021)\citenamefont {Gicev},
  \citenamefont {Hollenberg},\ and\ \citenamefont {Usman}}]{gicev2021scalable}%
  \BibitemOpen
  \bibfield  {author} {\bibinfo {author} {\bibfnamefont {S.}~\bibnamefont
  {Gicev}}, \bibinfo {author} {\bibfnamefont {L.~C.}\ \bibnamefont
  {Hollenberg}},\ and\ \bibinfo {author} {\bibfnamefont {M.}~\bibnamefont
  {Usman}},\ }\href@noop {} {\bibinfo {title} {A scalable and fast artificial
  neural network syndrome decoder for surface codes},} \bibinfo {year}
  {2021},\ \Eprint {https://arxiv.org/abs/arXiv:2110.05854} {arXiv:2110.05854}
  \BibitemShut {NoStop}%
\bibitem [{\citenamefont {Chamberland}\ \emph {et~al.}(2022)\citenamefont
  {Chamberland}, \citenamefont {Goncalves}, \citenamefont {Sivarajah},
  \citenamefont {Peterson},\ and\ \citenamefont
  {Grimberg}}]{chamberland2022techniques}%
  \BibitemOpen
  \bibfield  {author} {\bibinfo {author} {\bibfnamefont {C.}~\bibnamefont
  {Chamberland}}, \bibinfo {author} {\bibfnamefont {L.}~\bibnamefont
  {Goncalves}}, \bibinfo {author} {\bibfnamefont {P.}~\bibnamefont
  {Sivarajah}}, \bibinfo {author} {\bibfnamefont {E.}~\bibnamefont
  {Peterson}},\ and\ \bibinfo {author} {\bibfnamefont {S.}~\bibnamefont
  {Grimberg}},\ }\href@noop {} {\bibinfo {title} {Techniques for combining fast
  local decoders with global decoders under circuit-level noise},} \bibinfo
  {year} {2022},\ \Eprint {https://arxiv.org/abs/arXiv:2208.01178}
  {arXiv:2208.01178} \BibitemShut {NoStop}%
\bibitem [{\citenamefont {Smith}\ \emph {et~al.}(2022)\citenamefont {Smith},
  \citenamefont {Brown},\ and\ \citenamefont {Bartlett}}]{smith2022local}%
  \BibitemOpen
  \bibfield  {author} {\bibinfo {author} {\bibfnamefont {S.~C.}\ \bibnamefont
  {Smith}}, \bibinfo {author} {\bibfnamefont {B.~J.}\ \bibnamefont {Brown}},\
  and\ \bibinfo {author} {\bibfnamefont {S.~D.}\ \bibnamefont {Bartlett}},\
  }\href@noop {} {\bibinfo {title} {A local pre-decoder to reduce the bandwidth
  and latency of quantum error correction},} \bibinfo {year} {2022},\ \Eprint
  {https://arxiv.org/abs/arXiv:2208.04660} {arXiv:2208.04660} \BibitemShut
  {NoStop}%
\bibitem [{\citenamefont {Ueno}\ \emph
  {et~al.}(2022{\natexlab{b}})\citenamefont {Ueno}, \citenamefont {Kondo},
  \citenamefont {Tanaka}, \citenamefont {Suzuki},\ and\ \citenamefont
  {Tabuchi}}]{ueno2022neo}%
  \BibitemOpen
  \bibfield  {author} {\bibinfo {author} {\bibfnamefont {Y.}~\bibnamefont
  {Ueno}}, \bibinfo {author} {\bibfnamefont {M.}~\bibnamefont {Kondo}},
  \bibinfo {author} {\bibfnamefont {M.}~\bibnamefont {Tanaka}}, \bibinfo
  {author} {\bibfnamefont {Y.}~\bibnamefont {Suzuki}},\ and\ \bibinfo {author}
  {\bibfnamefont {Y.}~\bibnamefont {Tabuchi}},\ }\href@noop {} {\bibinfo
  {title} {{NEO-QEC}: Neural network enhanced online superconducting decoder
  for surface codes},} \bibinfo {year} {2022}{\natexlab{b}},\ \Eprint
  {https://arxiv.org/abs/arXiv:2208.05758} {arXiv:2208.05758} \BibitemShut
  {NoStop}%
\bibitem [{\citenamefont {Fowler}\ \emph
  {et~al.}(2012{\natexlab{b}})\citenamefont {Fowler}, \citenamefont
  {Whiteside},\ and\ \citenamefont {Hollenberg}}]{fowler2012towards}%
  \BibitemOpen
  \bibfield  {author} {\bibinfo {author} {\bibfnamefont {A.~G.}\ \bibnamefont
  {Fowler}}, \bibinfo {author} {\bibfnamefont {A.~C.}\ \bibnamefont
  {Whiteside}},\ and\ \bibinfo {author} {\bibfnamefont {L.~C.~L.}\ \bibnamefont
  {Hollenberg}},\ }\bibfield  {title} {\bibinfo {title} {Towards practical
  classical processing for the surface code},\ }\href
  {https://doi.org/10.1103/PhysRevLett.108.180501} {\bibfield  {journal}
  {\bibinfo  {journal} {Phys. Rev. Lett.}\ }\textbf {\bibinfo {volume} {108}},\
  \bibinfo {pages} {180501} (\bibinfo {year} {2012}{\natexlab{b}})},\ \Eprint
  {https://arxiv.org/abs/arXiv:1110.5133} {arXiv:1110.5133} \BibitemShut
  {NoStop}%
\bibitem [{\citenamefont {Fowler}(2015)}]{fowler2013minimum}%
  \BibitemOpen
  \bibfield  {author} {\bibinfo {author} {\bibfnamefont {A.~G.}\ \bibnamefont
  {Fowler}},\ }\bibfield  {title} {\bibinfo {title} {Minimum weight perfect
  matching of fault-tolerant topological quantum error correction in average
  {O}(1) parallel time},\ }\href {https://doi.org/10.26421/qic15.1-2-9}
  {\bibfield  {journal} {\bibinfo  {journal} {Quantum Info. Comput.}\ }\textbf
  {\bibinfo {volume} {15}},\ \bibinfo {pages} {145} (\bibinfo {year} {2015})},\
  \Eprint {https://arxiv.org/abs/arXiv:1307.1740} {arXiv:1307.1740}
  \BibitemShut {NoStop}%
\bibitem [{\citenamefont {Das}\ \emph {et~al.}(2022{\natexlab{a}})\citenamefont
  {Das}, \citenamefont {Pattison}, \citenamefont {Manne}, \citenamefont
  {Carmean}, \citenamefont {Svore}, \citenamefont {Qureshi},\ and\
  \citenamefont {Delfosse}}]{das2022afs}%
  \BibitemOpen
  \bibfield  {author} {\bibinfo {author} {\bibfnamefont {P.}~\bibnamefont
  {Das}}, \bibinfo {author} {\bibfnamefont {C.~A.}\ \bibnamefont {Pattison}},
  \bibinfo {author} {\bibfnamefont {S.}~\bibnamefont {Manne}}, \bibinfo
  {author} {\bibfnamefont {D.~M.}\ \bibnamefont {Carmean}}, \bibinfo {author}
  {\bibfnamefont {K.~M.}\ \bibnamefont {Svore}}, \bibinfo {author}
  {\bibfnamefont {M.}~\bibnamefont {Qureshi}},\ and\ \bibinfo {author}
  {\bibfnamefont {N.}~\bibnamefont {Delfosse}},\ }\bibfield  {title} {\bibinfo
  {title} {{AFS}: Accurate, fast, and scalable error-decoding for
  fault-tolerant quantum computers},\ }in\ \href
  {https://doi.org/10.1109/HPCA53966.2022.00027} {\emph {\bibinfo {booktitle}
  {2022 IEEE Int. S. High Perf. Comp. (HPCA)}}}\ (\bibinfo  {publisher}
  {IEEE},\ \bibinfo {year} {2022})\ pp.\ \bibinfo {pages} {259--273},\ \Eprint
  {https://arxiv.org/abs/arXiv:2001.06598} {arXiv:2001.06598} \BibitemShut
  {NoStop}%
\bibitem [{\citenamefont {Overwater}\ \emph {et~al.}(2022)\citenamefont
  {Overwater}, \citenamefont {Babaie},\ and\ \citenamefont
  {Sebastiano}}]{overwater2022neural}%
  \BibitemOpen
  \bibfield  {author} {\bibinfo {author} {\bibfnamefont {R.~W.}\ \bibnamefont
  {Overwater}}, \bibinfo {author} {\bibfnamefont {M.}~\bibnamefont {Babaie}},\
  and\ \bibinfo {author} {\bibfnamefont {F.}~\bibnamefont {Sebastiano}},\
  }\bibfield  {title} {\bibinfo {title} {Neural-network decoders for quantum
  error correction using surface codes: A space exploration of the hardware
  cost-performance tradeoffs},\ }\href
  {https://doi.org/10.1109/TQE.2022.3174017} {\bibfield  {journal} {\bibinfo
  {journal} {IEEE Transactions on Quantum Engineering}\ }\textbf {\bibinfo
  {volume} {3}},\ \bibinfo {pages} {1} (\bibinfo {year} {2022})},\ \Eprint
  {https://arxiv.org/abs/arXiv:2202.05741} {arXiv:2202.05741} \BibitemShut
  {NoStop}%
\bibitem [{\citenamefont {Das}\ \emph {et~al.}(2022{\natexlab{b}})\citenamefont
  {Das}, \citenamefont {Locharla},\ and\ \citenamefont
  {Jones}}]{das2021lilliput}%
  \BibitemOpen
  \bibfield  {author} {\bibinfo {author} {\bibfnamefont {P.}~\bibnamefont
  {Das}}, \bibinfo {author} {\bibfnamefont {A.}~\bibnamefont {Locharla}},\ and\
  \bibinfo {author} {\bibfnamefont {C.}~\bibnamefont {Jones}},\ }\bibfield
  {title} {\bibinfo {title} {{LILLIPUT}: A lightweight low-latency lookup-table
  decoder for near-term quantum error correction},\ }in\ \href
  {https://doi.org/10.1145/3503222.3507707} {\emph {\bibinfo {booktitle} {Proc.
  27th ACM ASPLOS}}}\ (\bibinfo  {publisher} {ACM},\ \bibinfo {address} {New
  York, NY, USA},\ \bibinfo {year} {2022})\ pp.\ \bibinfo {pages} {541--553},\
  \Eprint {https://arxiv.org/abs/arXiv:2108.06569} {arXiv:2108.06569}
  \BibitemShut {NoStop}%
\bibitem [{\citenamefont {Bartolucci}\ \emph {et~al.}(2021)\citenamefont
  {Bartolucci}, \citenamefont {Birchall}, \citenamefont {Bombin}, \citenamefont
  {Cable}, \citenamefont {Dawson}, \citenamefont {Gimeno-Segovia},
  \citenamefont {Johnston}, \citenamefont {Kieling}, \citenamefont {Nickerson},
  \citenamefont {Pant} \emph {et~al.}}]{bartolucci2021fusion}%
  \BibitemOpen
  \bibfield  {author} {\bibinfo {author} {\bibfnamefont {S.}~\bibnamefont
  {Bartolucci}}, \bibinfo {author} {\bibfnamefont {P.}~\bibnamefont
  {Birchall}}, \bibinfo {author} {\bibfnamefont {H.}~\bibnamefont {Bombin}},
  \bibinfo {author} {\bibfnamefont {H.}~\bibnamefont {Cable}}, \bibinfo
  {author} {\bibfnamefont {C.}~\bibnamefont {Dawson}}, \bibinfo {author}
  {\bibfnamefont {M.}~\bibnamefont {Gimeno-Segovia}}, \bibinfo {author}
  {\bibfnamefont {E.}~\bibnamefont {Johnston}}, \bibinfo {author}
  {\bibfnamefont {K.}~\bibnamefont {Kieling}}, \bibinfo {author} {\bibfnamefont
  {N.}~\bibnamefont {Nickerson}}, \bibinfo {author} {\bibfnamefont
  {M.}~\bibnamefont {Pant}}, \emph {et~al.},\ }\href@noop {} {\bibinfo {title}
  {Fusion-based quantum computation},} \bibinfo {year} {2021},\ \Eprint
  {https://arxiv.org/abs/arXiv:2101.09310} {arXiv:2101.09310} \BibitemShut
  {NoStop}%
\bibitem [{\citenamefont {Tomita}\ and\ \citenamefont
  {Svore}(2014)}]{tomitasvore}%
  \BibitemOpen
  \bibfield  {author} {\bibinfo {author} {\bibfnamefont {Y.}~\bibnamefont
  {Tomita}}\ and\ \bibinfo {author} {\bibfnamefont {K.~M.}\ \bibnamefont
  {Svore}},\ }\bibfield  {title} {\bibinfo {title} {Low-distance surface codes
  under realistic quantum noise},\ }\href
  {https://doi.org/10.1103/PhysRevA.90.062320} {\bibfield  {journal} {\bibinfo
  {journal} {Phys. Rev. A}\ }\textbf {\bibinfo {volume} {90}},\ \bibinfo
  {pages} {062320} (\bibinfo {year} {2014})},\ \Eprint
  {https://arxiv.org/abs/arXiv:1404.3747} {arXiv:1404.3747} \BibitemShut
  {NoStop}%
\bibitem [{\citenamefont {Delfosse}\ and\ \citenamefont
  {Nickerson}(2021)}]{delfosse2021almost}%
  \BibitemOpen
  \bibfield  {author} {\bibinfo {author} {\bibfnamefont {N.}~\bibnamefont
  {Delfosse}}\ and\ \bibinfo {author} {\bibfnamefont {N.~H.}\ \bibnamefont
  {Nickerson}},\ }\bibfield  {title} {\bibinfo {title} {Almost-linear time
  decoding algorithm for topological codes},\ }\href
  {https://doi.org/10.22331/q-2021-12-02-595} {\bibfield  {journal} {\bibinfo
  {journal} {Quantum}\ }\textbf {\bibinfo {volume} {5}},\ \bibinfo {pages}
  {595} (\bibinfo {year} {2021})},\ \Eprint
  {https://arxiv.org/abs/arXiv:1709.06218} {arXiv:1709.06218} \BibitemShut
  {NoStop}%
\bibitem [{sm()}]{sm}%
  \BibitemOpen
  \href@noop {} {}\bibinfo {note} {See Supplementary Material for (i) the
  syndrome-extraction circuit; (ii) generalizations of our sandwich decoder;
  and (iii) simulation details.}\BibitemShut {Stop}%
\bibitem [{\citenamefont {Gidney}(2022)}]{gidney2022stability}%
  \BibitemOpen
  \bibfield  {author} {\bibinfo {author} {\bibfnamefont {C.}~\bibnamefont
  {Gidney}},\ }\href@noop {} {\bibinfo {title} {Stability experiments: The
  overlooked dual of memory experiments},} \bibinfo {year} {2022},\ \Eprint
  {https://arxiv.org/abs/arXiv:2204.13834} {arXiv:2204.13834} \BibitemShut
  {NoStop}%
\bibitem [{\citenamefont {Chamberland}\ and\ \citenamefont
  {Campbell}(2022)}]{chamberland2022universal}%
  \BibitemOpen
  \bibfield  {author} {\bibinfo {author} {\bibfnamefont {C.}~\bibnamefont
  {Chamberland}}\ and\ \bibinfo {author} {\bibfnamefont {E.~T.}\ \bibnamefont
  {Campbell}},\ }\bibfield  {title} {\bibinfo {title} {Universal quantum
  computing with twist-free and temporally encoded lattice surgery},\ }\href
  {https://doi.org/10.1103/PRXQuantum.3.010331} {\bibfield  {journal} {\bibinfo
   {journal} {PRX Quantum}\ }\textbf {\bibinfo {volume} {3}},\ \bibinfo {pages}
  {010331} (\bibinfo {year} {2022})},\ \Eprint
  {https://arxiv.org/abs/arXiv:2109.02746} {arXiv:2109.02746} \BibitemShut
  {NoStop}%
\bibitem [{\citenamefont {Skoric}\ \emph {et~al.}(2022)\citenamefont {Skoric},
  \citenamefont {Browne}, \citenamefont {Barnes}, \citenamefont {Gillespie},\
  and\ \citenamefont {Campbell}}]{skoric2022parallel}%
  \BibitemOpen
  \bibfield  {author} {\bibinfo {author} {\bibfnamefont {L.}~\bibnamefont
  {Skoric}}, \bibinfo {author} {\bibfnamefont {D.~E.}\ \bibnamefont {Browne}},
  \bibinfo {author} {\bibfnamefont {K.~M.}\ \bibnamefont {Barnes}}, \bibinfo
  {author} {\bibfnamefont {N.~I.}\ \bibnamefont {Gillespie}},\ and\ \bibinfo
  {author} {\bibfnamefont {E.~T.}\ \bibnamefont {Campbell}},\ }\href@noop {}
  {\bibinfo {title} {Parallel window decoding enables scalable fault tolerant
  quantum computation},} \bibinfo {year} {2022},\ \Eprint
  {https://arxiv.org/abs/arXiv:2209.08552} {arXiv:2209.08552} \BibitemShut
  {NoStop}%
\end{thebibliography}

\begin{thebibliography}{6}%
\makeatletter
\providecommand \@ifxundefined [1]{%
 \@ifx{#1\undefined}
}%
\providecommand \@ifnum [1]{%
 \ifnum #1\expandafter \@firstoftwo
 \else \expandafter \@secondoftwo
 \fi
}%
\providecommand \@ifx [1]{%
 \ifx #1\expandafter \@firstoftwo
 \else \expandafter \@secondoftwo
 \fi
}%
\providecommand \natexlab [1]{#1}%
\providecommand \enquote  [1]{``#1''}%
\providecommand \bibnamefont  [1]{#1}%
\providecommand \bibfnamefont [1]{#1}%
\providecommand \citenamefont [1]{#1}%
\providecommand \href@noop [0]{\@secondoftwo}%
\providecommand \href [0]{\begingroup \@sanitize@url \@href}%
\providecommand \@href[1]{\@@startlink{#1}\@@href}%
\providecommand \@@href[1]{\endgroup#1\@@endlink}%
\providecommand \@sanitize@url [0]{\catcode `\\12\catcode `\$12\catcode
  `\&12\catcode `\#12\catcode `\^12\catcode `\_12\catcode `\%12\relax}%
\providecommand \@@startlink[1]{}%
\providecommand \@@endlink[0]{}%
\providecommand \url  [0]{\begingroup\@sanitize@url \@url }%
\providecommand \@url [1]{\endgroup\@href {#1}{\urlprefix }}%
\providecommand \urlprefix  [0]{URL }%
\providecommand \Eprint [0]{\href }%
\providecommand \doibase [0]{https://doi.org/}%
\providecommand \selectlanguage [0]{\@gobble}%
\providecommand \bibinfo  [0]{\@secondoftwo}%
\providecommand \bibfield  [0]{\@secondoftwo}%
\providecommand \translation [1]{[#1]}%
\providecommand \BibitemOpen [0]{}%
\providecommand \bibitemStop [0]{}%
\providecommand \bibitemNoStop [0]{.\EOS\space}%
\providecommand \EOS [0]{\spacefactor3000\relax}%
\providecommand \BibitemShut  [1]{\csname bibitem#1\endcsname}%
\let\auto@bib@innerbib\@empty
\bibitem [{\citenamefont {Tomita}\ and\ \citenamefont
  {Svore}(2014)}]{sm_tomitasvore}%
  \BibitemOpen
  \bibfield  {author} {\bibinfo {author} {\bibfnamefont {Y.}~\bibnamefont
  {Tomita}}\ and\ \bibinfo {author} {\bibfnamefont {K.~M.}\ \bibnamefont
  {Svore}},\ }\bibfield  {title} {\bibinfo {title} {Low-distance surface codes
  under realistic quantum noise},\ }\href
  {https://doi.org/10.1103/PhysRevA.90.062320} {\bibfield  {journal} {\bibinfo
  {journal} {Phys. Rev. A}\ }\textbf {\bibinfo {volume} {90}},\ \bibinfo
  {pages} {062320} (\bibinfo {year} {2014})},\ \Eprint
  {https://arxiv.org/abs/arXiv:1404.3747} {arXiv:1404.3747} \BibitemShut
  {NoStop}%
\bibitem [{\citenamefont {Das}\ \emph {et~al.}(2022)\citenamefont {Das},
  \citenamefont {Locharla},\ and\ \citenamefont {Jones}}]{sm_das2021lilliput}%
  \BibitemOpen
  \bibfield  {author} {\bibinfo {author} {\bibfnamefont {P.}~\bibnamefont
  {Das}}, \bibinfo {author} {\bibfnamefont {A.}~\bibnamefont {Locharla}},\ and\
  \bibinfo {author} {\bibfnamefont {C.}~\bibnamefont {Jones}},\ }\bibfield
  {title} {\bibinfo {title} {{LILLIPUT}: A lightweight low-latency lookup-table
  decoder for near-term quantum error correction},\ }in\ \href
  {https://doi.org/10.1145/3503222.3507707} {\emph {\bibinfo {booktitle} {Proc.
  27th ACM ASPLOS}}}\ (\bibinfo  {publisher} {ACM},\ \bibinfo {address} {New
  York, NY, USA},\ \bibinfo {year} {2022})\ pp.\ \bibinfo {pages} {541--553},\
  \Eprint {https://arxiv.org/abs/arXiv:2108.06569} {arXiv:2108.06569}
  \BibitemShut {NoStop}%
\bibitem [{\citenamefont {Dennis}\ \emph {et~al.}(2002)\citenamefont {Dennis},
  \citenamefont {Kitaev}, \citenamefont {Landahl},\ and\ \citenamefont
  {Preskill}}]{sm_dennis2002topological}%
  \BibitemOpen
  \bibfield  {author} {\bibinfo {author} {\bibfnamefont {E.}~\bibnamefont
  {Dennis}}, \bibinfo {author} {\bibfnamefont {A.}~\bibnamefont {Kitaev}},
  \bibinfo {author} {\bibfnamefont {A.}~\bibnamefont {Landahl}},\ and\ \bibinfo
  {author} {\bibfnamefont {J.}~\bibnamefont {Preskill}},\ }\bibfield  {title}
  {\bibinfo {title} {Topological quantum memory},\ }\href
  {https://doi.org/10.1063/1.1499754} {\bibfield  {journal} {\bibinfo
  {journal} {J. Math. Phys.}\ }\textbf {\bibinfo {volume} {43}},\ \bibinfo
  {pages} {4452} (\bibinfo {year} {2002})},\ \Eprint
  {https://arxiv.org/abs/arXiv:quant-ph/0110143} {arXiv:quant-ph/0110143}
  \BibitemShut {NoStop}%
\bibitem [{\citenamefont {Delfosse}\ and\ \citenamefont
  {Nickerson}(2021)}]{sm_delfosse2021almost}%
  \BibitemOpen
  \bibfield  {author} {\bibinfo {author} {\bibfnamefont {N.}~\bibnamefont
  {Delfosse}}\ and\ \bibinfo {author} {\bibfnamefont {N.~H.}\ \bibnamefont
  {Nickerson}},\ }\bibfield  {title} {\bibinfo {title} {Almost-linear time
  decoding algorithm for topological codes},\ }\href
  {https://doi.org/10.22331/q-2021-12-02-595} {\bibfield  {journal} {\bibinfo
  {journal} {Quantum}\ }\textbf {\bibinfo {volume} {5}},\ \bibinfo {pages}
  {595} (\bibinfo {year} {2021})},\ \Eprint
  {https://arxiv.org/abs/arXiv:1709.06218} {arXiv:1709.06218} \BibitemShut
  {NoStop}%
\bibitem [{\citenamefont {{Google Quantum AI}}(2022)}]{sm_acharya2022google}%
  \BibitemOpen
  \bibfield  {author} {\bibinfo {author} {\bibnamefont {{Google Quantum AI}}},\
  }\href@noop {} {\bibinfo {title} {Suppressing quantum errors by scaling a
  surface code logical qubit},} \bibinfo {year} {2022},\ \Eprint
  {https://arxiv.org/abs/arXiv:2207.06431} {arXiv:2207.06431} \BibitemShut
  {NoStop}%
\bibitem [{\citenamefont {Gidney}(2022)}]{sm_gidney2022stability}%
  \BibitemOpen
  \bibfield  {author} {\bibinfo {author} {\bibfnamefont {C.}~\bibnamefont
  {Gidney}},\ }\href@noop {} {\bibinfo {title} {Stability experiments: The
  overlooked dual of memory experiments},} \bibinfo {year} {2022},\ \Eprint
  {https://arxiv.org/abs/arXiv:2204.13834} {arXiv:2204.13834} \BibitemShut
  {NoStop}%
\end{thebibliography}
\end{document}